\documentclass[11pt]{article}
\usepackage{axodraw2}
\usepackage{comment}
\usepackage{epsfig}
\usepackage{amsfonts}
\usepackage{amsmath,amssymb}
\usepackage{bbm,bm}
\usepackage{cite}
\allowdisplaybreaks[1]

\newcommand\newtext[1]{{\color{black}#1}}
\newcommand\noveltext[1]{{\color{black}#1}}
\usepackage[normalem]{ulem}
\usepackage{enumerate}
\newcounter{dummy}

\usepackage{graphicx}
\usepackage{tikz}
\usetikzlibrary{shapes,arrows,shadows,automata,positioning}
\newdimen\nodeDist
\usepackage[
  colorlinks=true,
  linkcolor={red!50!black},
  citecolor={blue!50!black},
  filecolor=black,
  urlcolor=black,
  breaklinks=true
  ]{hyperref}

\renewcommand{\vec}[1]{{\bf #1}}
\newcommand{\nn}{\nonumber \\}
\newcommand{\rmi}[1]{{\mbox{\scriptsize #1}}}
\newcommand{\rmii}[1]{{\mbox{\tiny\rm{#1}}}}

\newcommand{\bmu}{\bar{\Lambda}}

\newcommand{\gammaE}{{\gamma_\rmii{E}}}
\newcommand{\alphas}{\alpha_{\rm s}}

\newcommand{\re}{\mathop{\mbox{Re}}}
\newcommand{\im}{\mathop{\mbox{Im}}}
\newcommand{\T}{\rmii{$T$}}

\newcommand\MSbar{$\overline{\rm MS}$}

\newcommand{\nF}{n_\rmii{F}}
\newcommand{\nFt}{\tilde{n}_\rmii{F}}
\newcommand{\nB}{n_\rmii{B}}

\newcommand{\Tint}[1]{{\hbox{$\sum$}\!\!\!\!\!\!\!\int\,}_{\!\!\!\!\raise-0.9ex\hbox{$\scriptstyle{#1}$}}}
\newcommand{\Tinti}[1]{{{\Sigma}\!\!\!\!\raise0.3ex\hbox{$\int$}_\rmii{${#1}$}}}
\newcommand{\Tintip}[1]{{{\Sigma'}\!\!\!\!\!\raise0.3ex\hbox{$\int$}_\rmii{${#1}$}}}
\newcommand\varpm{\mathbin{\vcenter{\hbox{%
  \oalign{\hfil{$+$}\hfil\cr
          \noalign{\kern-.3ex}
          \rmii{$(-)$}\cr}%
}}}}

\def\Lwidth{1}

\def\Agl(#1,#2)(#3,#4,#5){\PhotonArc(#1,#2)(#3,#4,#5){\Lwidth}
  {6.283 #3 mul 360 div #4 #5 sub #4 #5 sub mul sqrt mul Ldensity mul}}
\def\Lgl(#1,#2)(#3,#4){\Photon(#1,#2)(#3,#4){\Lwidth}
  {#1 #3 sub #1 #3 sub mul #2 #4 sub #2 #4 sub mul add sqrt Ldensity mul}}
\def\Aqu(#1,#2)(#3,#4,#5){\ArrowArc(#1,#2)(#3,#4,#5)}
\def\Auq(#1,#2)(#3,#4,#5){\ArrowArcn(#1,#2)(#3,#5,#4)}
\def\Lqu(#1,#2)(#3,#4){\ArrowLine(#1,#2)(#3,#4)}
\def\Luq(#1,#2)(#3,#4){\ArrowLine(#3,#4)(#1,#2)}
\def\Aqq(#1,#2)(#3,#4,#5){\CArc(#1,#2)(#3,#4,#5)}
\def\Lqq(#1,#2)(#3,#4){\Line(#1,#2)(#3,#4)}

\def\Textt(#1,#2,#3){\Text(#1,#2)[t]{{$\scriptstyle #3$}}}
\def\Textb(#1,#2,#3){\Text(#1,#2)[b]{{$\scriptstyle #3$}}}
\def\Textl(#1,#2,#3){\Text(#1,#2)[l]{{$\scriptstyle #3$}}}
\def\Textr(#1,#2,#3){\Text(#1,#2)[r]{{$\scriptstyle #3$}}}
\def\Texttl(#1,#2,#3){\Text(#1,#2)[tl]{{$\scriptstyle #3$}}}
\def\Textbl(#1,#2,#3){\Text(#1,#2)[bl]{{$\scriptstyle #3$}}}
\def\Texttr(#1,#2,#3){\Text(#1,#2)[tr]{{$\scriptstyle #3$}}}
\def\Textbr(#1,#2,#3){\Text(#1,#2)[br]{{$\scriptstyle #3$}}}

\def\TopoVR(#1){\pic{%
  #1(15,15)(15,0,180)%
  #1(15,15)(15,180,360)%
}}
\def\ToptVS(#1,#2,#3){\pic{
  #1(15,15)(15,0,180)%
  #2(15,15)(15,180,360)%
  #3(30,15)(0,15)%
}}
\def\ToptVSn(#1,#2,#3,#4,#5,#6){\pic{
  #1(15,15)(15,0,180)
  #2(15,15)(15,180,360)%
  #3(30,15)(0,15)
  \Textb(15,32,#4)%
  \Textb(15,2,#5)%
  \Textb(15,16,#6)%
}}
\def\ToptVE(#1,#2){\picc{
  #1(15,15)(15,0,360)
  #2(45,15)(15,-180,180)}}
\def\TopoSBtxt(#1,#2,#3,#4,#5,#6,#7){\picb{
  #1(0,15)(7.5,15)
  #1(45,15)(37.5,15)
  #2(22.5,15)(15,0,180)%
  #3(22.5,15)(15,180,360)
  \Textr(0,15,#4)%
  \Textl(45,15,#5)%
  \Textb(22.5,32,#6)
  \Textt(22.5,-2,#7)
  }}

\def\scfc{0.7}  
\def\phgt{21}   
\def\pwc{21}    
\def\pwcb{31.5} 
\def\pwcc{42} 
\newcommand{\PIC}[4]{\;\parbox[c]{#2 pt}{\begin{picture}(#2,#3)(0,0)
\SetWidth{1.0}\SetScale{#4} #1 \end{picture}}\;}
\newcommand{\pic}[1]{\PIC{#1}{\pwc}{\phgt}{\scfc}}
\newcommand{\picb}[1]{\PIC{#1}{\pwcb}{\phgt}{\scfc}}
\newcommand{\picc}[1]{\PIC{#1}{\pwcc}{\phgt}{\scfc}}

\makeatletter \@addtoreset{equation}{section} \makeatother
\renewcommand{\theequation}{\arabic{section}.\arabic{equation}}
\makeatletter
\renewcommand\section{\@startsection{section}{1}{\z@}%
  {-5.5ex \@plus -1ex \@minus -.2ex}
  {2.3ex \@plus.2ex}%
  {\normalfont\large\bfseries}}
\renewcommand\subsection{\@startsection{subsection}{2}{\z@}%
  {-3.25ex\@plus -1ex \@minus -.2ex}%
  {1.5ex \@plus .2ex}%
  {\normalfont\normalsize\bfseries}}
\renewcommand\thesection{\@arabic\c@section}
\renewcommand\thesubsection{\thesection.\@arabic\c@subsection}
\renewcommand{\@seccntformat}[1]{%
  \csname the#1\endcsname.\hspace{1.0em}}
\makeatother

\begin{document}

\flushbottom

\begin{titlepage}

\begin{flushright}
HIP-2023-7/TH
\end{flushright}
\begin{centering}

\vfill

{\Large{\bf
Integrating by parts at finite density
}}

\vspace{0.8cm}

\renewcommand{\thefootnote}{\fnsymbol{footnote}}
Juuso {\"O}sterman$^{\rm a,}$%
\footnote{juuso.s.osterman@helsinki.fi},
Philipp Schicho$^{\rm b,a,}$%
\footnote{schicho@itp.uni-frankfurt.de},  and
Aleksi Vuorinen$^{\rm a,}$%
\footnote{aleksi.vuorinen@helsinki.fi}

\vspace{0.8cm}

$^\rmi{a}$%
{\em
Department of Physics and Helsinki Institute of Physics,\\
P.O.~Box 64, FI-00014 University of Helsinki, Finland
}

\vspace{0.3cm}

$^\rmi{b}$%
{\em
Institute for Theoretical Physics, Goethe Universit{\"a}t Frankfurt, 60438 Frankfurt, Germany
}

\vspace*{0.8cm}

\mbox{\bf Abstract}

\end{centering}

\vspace*{0.3cm}

\noindent
Both nonzero temperature and chemical potentials break the Lorentz symmetry present in vacuum quantum field theory by singling out the rest frame of the heat bath. This leads to complications in the application of thermal perturbation theory, including the appearance of novel infrared divergences in loop integrals and an apparent absence of four-dimensional integration-by-parts (IBP) identities, vital for high-order computations. Here, we propose a new strategy that enables the use of IBP techniques in the evaluation of \newtext{Feynman integrals, in particular vacuum or bubble diagrams,} in the limit of vanishing temperature $T$ but nonzero chemical potentials $\mu$. The central elements of the new setup include a contour representation for the temporal momentum integral, the use of a small but nonzero $T$ as an \newtext{IR} regulator, and the systematic application of both temporal and spatial differential operators in the generation of \newtext{linear} relations \newtext{among the loop integrals of interest}. The relations we derive contain novel inhomogeneous terms featuring differentiated Fermi-Dirac distribution functions, which severely complicate calculations at nonzero temperature, but are shown to reduce to solvable lower-dimensional objects as $T$ tends to zero. Pedagogical example computations are kept at the one- and two-loop levels, but the application of the new method to higher-order calculations is discussed in some detail.

\vfill
\end{titlepage}

{\hypersetup{hidelinks}
\tableofcontents
}
\clearpage

\renewcommand{\thefootnote}{\arabic{footnote}}
\setcounter{footnote}{0}

%
\section{Introduction}

Perturbation theory is undoubtedly the single most powerful technique for making quantitative predictions in quantum field theory~\cite{Peskin:1995ev,Collins:1984xc,Weinberg:1995mt}. Importantly, its application is not limited to problems in vacuum, but perturbative methods are equally applicable to the study of extended systems in and even out of thermal equilibrium~\cite{Kapusta:2006pm,Bellac:2011kqa,Laine:2016hma,Ghiglieri:2020dpq}. In the latter context, the physical systems of relevance range from the early Universe to the extremely dense cores of neutron stars and the hot fireball created in heavy-ion collisions. In such applications, Quantum Chromodynamics (QCD) has proven a particularly challenging theory \cite{Brambilla:2014jmp}  due to a combination of complicated phenomenological problems and the slow convergence of weak-coupling expansions. For this reason, thermal perturbation theory calculations in QCD need to be pushed to particularly high loop orders, and, e.g., the equation of state (EoS) of deconfined QCD matter is currently known to partial four-loop level both at high temperatures $T$ and small or vanishing chemical potentials $\mu$ \cite{Kajantie:2002wa,Vuorinen:2003fs,Laine:2006cp,Andersen:2011sf,Mogliacci:2013mca,Haque:2014rua} as well as
at high $\mu$ and small or vanishing $T$  \cite{Kurkela:2009gj,Kurkela:2016was,Gorda:2021znl,Gorda:2021kme}.

In high-loop-order perturbative calculations, be that in vacuum or a thermal setting, various methods of automation become indispensable. In vacuum, one of the most crucial tools enabling, e.g., the determination of the five-loop running coupling constant of QCD~\cite{Baikov:2016tgj,Herzog:2017ohr}, is the integration-by-parts (IBP) technique~\cite{Chetyrkin:1981qh,Bender:1976pw}. It allows for the derivation of linear relations between different Feynman integrals, reducing the number of independent {\em master} integrals in need of explicit analytic or numerical evaluation. These identities are typically generated by multiplying the integrand of a Feynman integral, schematically $\int_P g(P)$, by a momentum-dependent function $f_\nu(P)$ and using the fact that integrals over total derivatives vanish, i.e.~$\int_P \partial_\nu \big( f_\nu (P)g(P)\big)=0$.

Unfortunately, the derivation of IBP relations relies heavily on Lorentz symmetry, which is broken by both a nonzero $T$ and $\mu$. In the imaginary-time formalism of thermal field theory, applicable in thermal equilibrium, the spacetime metric becomes Euclidean with integrals over the temporal momentum component being replaced by discrete sums over the so-called Matsubara frequencies~\cite{Matsubara:1955ws}. The latter take the form $p_0 = \omega_n^{\rmii{B/F}}$ with $\omega_n^\rmii{B} = 2 n\pi T$ for bosons and $\omega_n^\rmii{F} = (2n+1)\pi T+i\mu$ for fermions, $n\in \mathbb{Z}$, where an imaginary shift appears for nonzero chemical potentials. This implies that total derivatives in the temporal direction no longer lead to vanishing momentum integrals, which prevents the closing of traditional IBP relations. In existing literature, this issue has typically been resolved by only considering spatial derivative operators and integrations in the derivation of IBP relations~\cite{Nishimura:2012ee,Laporta:2000dsw}, which however severely limits their power in practical computations.\footnote{\newtext{Note also that while spatial IBP operators are always applicable, the presence of chemical potential changes the structure of the corresponding IBP relations derived at $\mu=0$ as demonstrated in appendix \ref{app:spatialcheck}.}}

In a largely unrelated development, it was recently pointed out in~\cite{Gorda:2022yex} that even the simplest loop calculations involving chemical potentials require particular care in the low-temperature limit. A naive application of the residue theorem may lead to the missing of important contributions to physical quantities. This issue appears whenever fermionic propagators are raised to (integer) powers higher than 1 and can be most conveniently circumvented through a consistent use of a contour-integral formulation for the temporal integral, introduced in~\cite{Gorda:2022yex} and reviewed in sec.~\ref{sec:formalism} below. In the strict $T=0$ limit, this procedure gives rise to $\delta$-function terms easily missed in standard residue calculations that utilize a single linear contour along the real axis,\footnote{Somewhat analogous $\delta$-function terms are known to arise from the temporal integration in the context of the real-time formalism of thermal field theory~\cite{Gorbar:2013upa,Laine:2016hma} when using a generalized Cauchy principal value.} which also highlights the fact that thermal field theory calculations performed at $T=0,\; \mu\neq 0$ should always be thought of as the $T\to 0$ limit of a finite-temperature computation. As detailed in~\cite{Gorda:2022yex}, even if infinitesimally small, the parameter $T$ plays an important regulatory role in loop integrals.

Motivated by the phenomenological need to determine the EoS of cold and ultradense quark matter, stemming from attempts to determine the EoS of neutron-star matter in a model-agnostic manner (see e.g., \cite{Annala:2021gom,Gorda:2022jvk,Annala:2023cwx} for recent studies), our aim in this paper is to generalize IBP methods to be \newtext{at least partially available} in the limit of nonzero chemical potentials but vanishing temperatures. Crucially, we are interested in IBPs in all $D\equiv 4-2\epsilon$ dimensions, and to this end insist on including also temporal derivatives in the operators generating linear relations. As we will detail below, the use of the contour-integral prescription of \cite{Gorda:2022yex} will give rise to extra {\em boundary terms} or {\em inhomogeneities} in IBP relations that often feature derivatives of the Fermi-Dirac distribution function. While highly problematic at nonzero temperatures, in the strict $T\to 0$ limit these derivative terms simplify considerably, leading to \newtext{more} easily computable entities that close the \newtext{extended} IBP relations \newtext{although some differences to standard vacuum IBP relations  are seen to  arise}.\footnote{\newtext{Unlike in the standard $T=\mu=0$ IBP framework, a consecutive application of a given temporal derivative operator does not provide additional information. Furthermore, each individual temporal derivative introduces a new closed group operation for linear algebra, characterized by the number of differentiated distribution functions in a loop integral.}}

In the practical derivation of \newtext{the new} IBP relations, our strategy is to use temporal derivatives to write the boundary terms in a form, where \newtext{ideally} all fermionic distribution functions have been differentiated, while all other terms are of a form familiar from vacuum-type IBP relations. While the latter can be further simplified using spatial IBP relations as discussed in sec.~\ref{sec:IBPintro} below, the integrals with differentiated distribution functions need to be evaluated explicitly. The techniques needed in such computations are outlined in sec.~\ref{sec:formalism}, where we also introduce our notation and general formalism. In secs.~\ref{sec:1-loopformulae} and \ref{sec:2-loop}, we then provide explicit examples of this procedure at the one- and two-loop levels, wherein we also describe the required regularization due to the complex nature of fermionic propagators at nonzero $\mu$ \cite{Gorda:2022yex}. As a byproduct of our calculations, we end up generalizing the factorization of the vacuum sunset diagram \newtext{with collinear scales} to finite $\mu$, motivated by an observed factorization from our IBP relations; this topic is discussed in detail in \newtext{appendix}~\ref{sec:factorizationof2loop}. The final objective of our program is to create an automatable algorithm that, in combination with other tools such as the cutting rules of~\cite{Ghisoiu:2016swa}, will facilitate computations at high orders of perturbation theory. \newtext{Although this important extension of our framework is largely left to future work, some related aspects are} discussed in the concluding sec.~\ref{sec:concl}.

%
\section{Formalism and setup}
\label{sec:formalism}

We work in the imaginary-time formalism of thermal field theory. Here, the starting point of any computation at nonzero temperature $T$ involves discrete sums over Matsubara frequencies~\cite{Matsubara:1955ws}, amounting to $\omega_n^\rmii{B} = 2 n\pi T$ for bosons and $\omega_n^\rmii{F} = (2 n+1)\pi T+i\mu$ for fermions at a nonzero chemical potential $\mu$, with $n\in\mathbb{Z}$.
As demonstrated in numerous textbooks~\cite{Kapusta:2006pm,Bellac:2011kqa,Laine:2016hma}, sums over these discrete frequencies can be easily converted to continuous integrals using the analyticity properties of the Bose-Einstein and Fermi-Dirac distribution functions,
\begin{align}
  \nF (x) = \frac{1}{e^x+1}
  \,, \qquad
  \nB (x) = \frac{1}{e^x-1}
  \,.
\end{align}
Concretely, we may write the (sum-)integration measures appearing in thermal Feynman integrals in the forms
\begin{eqnarray}
\label{eq:bosonsum}
  \Tint{P} f(\omega_n^\rmii{B},\vec{p})
  =
  T \sum_{\omega_n^\rmii{B}} \int_{\vec{p}} f(\omega_n^\rmii{B},\vec{p}) &=&
  \oint_{P}^b
  \nB(i\beta p_0) f(p_0,\vec{p})
  \, , \\
\label{eq:fermionsum}
  \Tint{\{P\}} f(\omega_n^\rmii{F},\vec{p})
  =
  T \sum_{\omega_n^\rmii{F}} \int_{\vec{p}} f(\omega_n^\rmii{F},\vec{p}) &=&
  \oint_{P}^f
  \nF \left[i\beta (p_0-i\mu) \right]f(p_0,\vec{p})
    \, ,
\end{eqnarray}
where $P = (p_0,\vec{p})$ stands for a momentum in $D=d+1$ dimensions,
$\vec{p}$ is a $d$-dimensional spatial vector, and $p \equiv |\vec{p}|$. Sum-integrals are denoted by
the shorthand $\Tinti{P} = T\sum_{p_n}\int_{\vec{p}}$, where $p_n^{ } = \omega_n^{\rmii{B/F}}$ and curly brackets are used to indicate the fermionic nature of the loop momentum.

Above, we have also introduced a shorthand notation for
the $(d+1)$-dimensional integral and
its temporal integration contours
\begin{align}
\label{eq:oint:f}
    \oint_{P}^{b/f}
    &\equiv
    \oint_{p_0}^{b/f}\!\!\!
    \int_{\vec{p}}
      \,,\;\;\; &
    \oint_{p_0}^{b}
    &\equiv \biggl[
          \int_{-\infty - i\eta}^{\infty  - i\eta}
        + \int_{\infty  + i\eta}^{-\infty + i\eta} \biggr] \frac{{\rm d}p_0}{2\pi}
      \,,\;\; &
    \oint_{p_0}^f &\equiv \biggl[
          \int_{-\infty + i\mu + i\eta}^{\infty + i \mu + i\eta}
        + \int_{\infty + i\mu - i\eta}^{-\infty + i \mu - i\eta} \biggr] \frac{{\rm d} p_0}{2\pi}
    \, ,
\end{align}
where the integration contour runs (anti-)clockwise for fermions (bosons) around the real axis of the complex $p_0$-plane. For the dimensionally regulated (\MSbar) spatial integrals in $d=3-2\epsilon$ dimensions, we finally introduce the measure
\begin{eqnarray}
  \int_{\vec{p}} &\equiv&
  \biggl( \frac{e^\gammaE \bmu^2}{4 \pi} \biggr)^{\frac{3-d}{2}}
  \int \frac{{\rm d}^d \vec{p} }{(2\pi)^d}
  \, ,
\end{eqnarray}
where $\bmu$ is the corresponding renormalization scale and
$\gammaE$ is the Euler-Mascheroni constant.
Unless otherwise stated, we will provide results for general $d$ dimensions whenever possible.

Below, we will briefly discuss two technical issues that are both frequently used in our forthcoming presentation. The first issue has to do with subtleties in taking the zero-temperature limit in the above (sum-)integration measures; see sec.~\ref{sec:smallTreg}.
The second introduces two convenient deformations of the fermionic $p_0$-integration contour defined above, which turn out to greatly simplify practical calculations; see sec.~\ref{sec:contour:def}.

\subsection{Small-temperature limit and regularization}
\label{sec:smallTreg}

As discussed in detail in~\cite{Gorda:2022yex}, taking the zero-temperature limit in the integrals defined above can be surprisingly non-trivial, and in particular, changing the order of the temporal and spatial integrations must be handled with care. To prepare for these subtleties, we identify the real-valued parameters $x \equiv \re (p_0)$ and
$y \equiv \im (p_0)-\mu$ for the Fermi-Dirac distribution as well as $x \equiv \re (p_0)$ and
$y \equiv \im (p_0)$ for the Bose-Einstein distribution, whereby the low-temperature limits
of the corresponding distribution functions become
\begin{eqnarray}
\label{eq:nfbtheta}
  n_{\rmii{F(B)}} \left[i \beta x-\beta y \right] &=&
      \varpm \frac{1}{2}\biggl[1 + \frac{\sinh(\beta y)}{\cosh(\beta y) \varpm \cos(\beta x)}\biggr]
      - \frac{i}{2}\frac{\sin(\beta y)}{\cosh(\beta y) \varpm \cos(\beta x)}
    \nn
    &\stackrel{T < |y|}{\longrightarrow}&
    \varpm \theta_{\T}^{\rmii{F(B)}} (y)
    - i 0_{\T \rmii{1}}^{\rmii{F(B)}}
    \, .
\end{eqnarray}
In this result, the bosonic case is always shown in parenthesis, and the chosen notation for the two functions defined on the lower line corresponds to the respective behaviors of the real and imaginary parts of the distribution functions at small but non-vanishing $T$: $\lim_{T\to 0} \theta_{\T}^{\rmii{F(B)}} (y) = \theta(y)$ and $\lim_{T\to 0} 0_{\T \rmii{1}}^{\rmii{F(B)}} = 0$.%
\footnote{
  Upper indices in eq.~\eqref{eq:nfbtheta} indicate the particular $\theta$-function and $0$-sequences defined by this equation.
}

The derivative of the \newtext{real-valued} fermionic distribution function finally gives rise to a well-known \newtext{nascent delta distribution, referred to as the $\delta$- sequence below}:
\begin{eqnarray}
\label{nFprimeT0}
    -\beta n_{\rmii{F}}'[-\beta y] &=& \frac{\beta}{2}\frac{1}{[\cosh(\beta y) + 1]}
    \stackrel{T < |y|}{\longrightarrow}
    \delta_{\T}(y)
    \, .
    \end{eqnarray}
    Similarly, the complex generalizations of the differentiated Fermi-Dirac and Bose-Einstein distributions (often referred to as {\em primed} distribution functions) yield
    \begin{eqnarray}
\label{eq:nfb_complexdelta}
    -\beta n_{\rmii{F(B)}}' \left[i \beta x - \beta y \right] &=&
      \frac{ \beta}{2} \frac{\cosh(\beta y) \cos(\beta x) \varpm 1}{[\cosh(\beta y) \varpm \cos(\beta x)]^2}
    \varpm \frac{i\beta}{2} \frac{\sinh(\beta y) \sin(\beta x)}{[\cosh(\beta y) \varpm \cos(\beta x) ]^2}
    \nn
    &\stackrel{T < |y|}{\longrightarrow}&
    \varpm \delta_{\T}^{\rmii{F(B)}}(y)
    \varpm i 0_{\T \rmii{2}}^{\rmii{F(B)}}
    \, ,
\end{eqnarray}
where we introduced two more functions with the respective limits of $\lim_{T\to 0} \delta_{\T}^{\rmii{F(B)}} (y) = \delta(y)$ and $\lim_{T\to 0} 0_{\T \rmii{2}}^{\rmii{F(B)}} = 0$.

The simplest example of an integrand, for which the use of $T$ as a regulatory parameter (through the above relations) becomes important%
\footnote{\newtext{Keeping $T$ nonzero prevents the temporal components of fermionic momenta $p_0^f=\omega_n^\rmii{F} = (2 n+1)\pi T+i\mu$ from vanishing, which in turn protects integrals with high powers of fermionic propagators from problematic divergences from the region $p=\mu$, $\re (p_0)=0$.}}, consists of a massless fermionic propagator raised to an arbitrary real-valued power $\alpha$. Here, a naive way to proceed would be to
first drop the lower part of the $p_0$-contour in eq.~\eqref{eq:fermionsum} due to its exponential suppression at small $T$ and then to set $\nF\left[i \beta (p_0-i\mu) \right] \to 1$, resulting in
\begin{equation}
\label{eq:naiveleadingmapping}
  \oint_{p_0}^f
    \int_{\vec{p}}
    \frac{\nF\left[i \beta (p_0-i\mu) \right]}{[p_0^2+p^2]^\alpha} \mapsto
    \int_{\vec{p}} \int_{-\infty}^{\infty}\frac{{\rm d}p_0}{2\pi}
    \frac{1}{[(p_0+i\mu)^2+p^2]^\alpha}
    \, .
\end{equation}
This strategy indeed yields a $p_0$-integral that can be easily evaluated using the residue theorem, but the result turns out to be the physically correct one only for $1/2 < \alpha \leq 1$; see appendix~\ref{app:singleprop} and \cite{Gorda:2022yex}. For $\alpha>1$, an explicit divergence occurs at $p = \mu$, $p_0 \sim 0$ 
\begin{equation}
\label{eq:Iapole}
    \int_{-\epsilon}^\epsilon \frac{{\rm d}p_0}{2 \pi} \frac{1}{[p_0 + i\mu + ip]^\alpha [p_0 + i\mu - ip]^\alpha} =
    \int_{-\epsilon}^\epsilon \frac{{\rm d}p_0}{2 \pi} \frac{1}{[p_0 + 2i\mu ]^\alpha [p_0]^\alpha}\, ,
\end{equation}
which explains the fact that different integration orders are seen to yield differing results
\begin{equation}
\label{eq:Ianeq}
  \int_{\vec{p}} \int_{-\infty}^\infty \frac{{\rm d}p_0}{2 \pi} \frac{1}{[(p_0+i\mu)^2+p^2]^{\alpha}} \neq
  \int_{-\infty}^\infty \frac{{\rm d}p_0}{2 \pi}  \int_{\vec{p}}\frac{1}{[(p_0+i\mu)^2+p^2]^{\alpha}}\, .
\end{equation}
As first pointed out in \cite{Gorda:2022yex} (see also app.~\ref{app:singleprop}), this discrepancy is correctly treated through the Fermi-Dirac distribution and its derivative, which induces an additional $\delta$-function-type contribution when $p_0$ crosses the imaginary axis along the contour of eq.~\eqref{eq:fermionsum}.

Similar issues are absent for purely bosonic integrals, given the lack of scales like chemical potentials. However, graphs mixing bosonic and fermionic momenta may again feature nontrivial low-temperature limits. Like in the fermionic case, keeping $T$ small but nonzero in eqs.~\eqref{eq:nfbtheta} and \eqref{eq:nfb_complexdelta} and utilizing (simplifications of) the bosonic \newtext{$\delta$}-sequences ensure that the order of the temporal and spatial integrations are always interchangeable.

\subsection{Contour deformations}
\label{sec:contour:def}

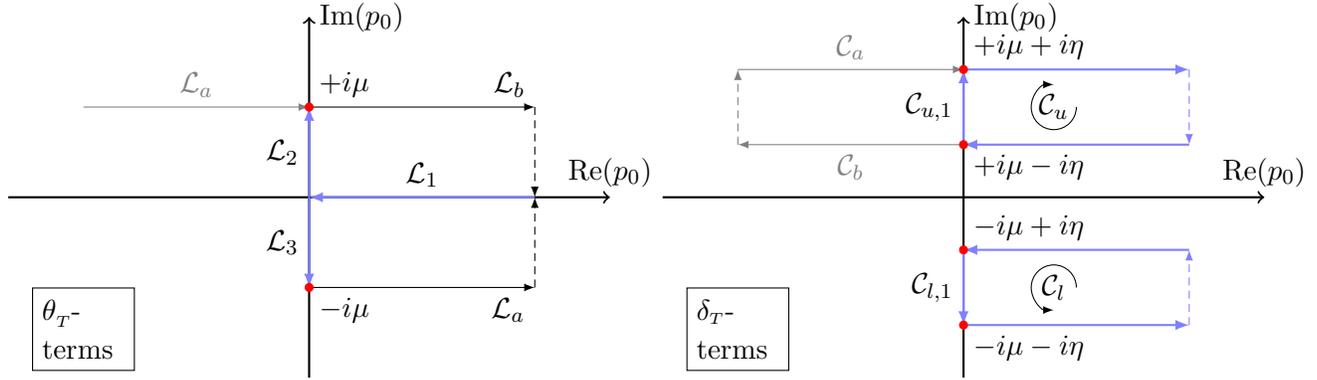
\begin{figure}[t]
\centering
\begin{tikzpicture}[scale=1]
\def\blockh{1.2}
\def\blockl{3.0}
  \draw[->,thick] (-\blockl-1,0) -- (\blockl+1,0) node[above] {$\re(p_0)$};
  \draw[->,thick] (0,-2*\blockh) -- (0,2*\blockh) node[right] {$\im(p_0)$};

  \draw[-latex,gray] (-\blockl,\blockh) -- (0,\blockh) node[midway,above] {$\mathcal{L}_{a}$};
  \draw[-latex] (0,\blockh) -- (\blockl,\blockh) node[above left] {$\mathcal{L}_{b}$};
  \draw[-latex] (0,-\blockh) -- (\blockl,-\blockh) node[below left] {$\mathcal{L}_{a}$};
  \draw[-latex,densely dashed] (\blockl,+\blockh) -- (\blockl,0) node[right] {};
  \draw[-latex,densely dashed] (\blockl,-\blockh) -- (\blockl,0) node[right] {};
  \draw[-latex,thick,blue!50] (\blockl,0) -- (0,0) node[black,midway,above] {$\mathcal{L}_{1}$};
  \draw[-latex,thick,blue!50] (0,0) -- (0,+\blockh) node[black,midway,left] {$\mathcal{L}_{2}$};
  \draw[-latex,thick,blue!50] (0,0) -- (0,-\blockh) node[black,midway,left] {$\mathcal{L}_{3}$};

  \draw[red,fill] (0,+\blockh) circle [radius=1.5pt] node[black,above right] {$+i\mu$};
  \draw[red,fill] (0,-\blockh) circle [radius=1.5pt] node[black,below right] {$-i\mu$};

  \draw[] (-\blockl,-\blockh) node[draw,below,text width=3\blockh,minimum height=3\blockh]
    {$\theta_\T^{ }$-\\terms};
\end{tikzpicture}%
\begin{tikzpicture}[scale=1]
\def\blockh{1.2}
\def\blockl{3.0}
  \draw[->,thick] (-\blockl-1,0) -- (\blockl+1,0) node[above] {$\re(p_0)$};
  \draw[->,thick] (0,-2*\blockh) -- (0,2*\blockh) node[right] {$\im(p_0)$};

  \draw[-latex,gray] (-\blockl,\blockh+0.5) -- (0,\blockh+0.5) node[midway,above] {$\mathcal{C}_{a}$};
  \draw[-latex,gray] (0,\blockh-0.5) -- (-\blockl,\blockh-0.5) node[midway,below] {$\mathcal{C}_{b}$};
  \draw[-latex,thick,blue!50] (0,\blockh+0.5) -- (\blockl,\blockh+0.5) node[black, above left] {};
  \draw[-latex,densely dashed,gray] (-\blockl,+\blockh-0.5) -- (-\blockl,+\blockh+0.5) node[right] {};
  \draw[-latex,densely dashed,blue!50] (\blockl,+\blockh+0.5) -- (\blockl,+\blockh-0.5) node[right] {};
  \draw[-latex,thick,blue!50] (0,+\blockh-0.5) -- (0,+\blockh+0.5) node[black,midway,left] {$\mathcal{C}_{u,1}$};
  \draw[-latex,thick,blue!50] (\blockl,\blockh-0.5) -- (0,\blockh-0.5) node[black, above left] {};

  \draw[-latex,thick,blue!50] (+\blockl,-\blockh+0.5) -- (0,-\blockh+0.5) node[above left] {};
  \draw[-latex,densely dashed,blue!50] (\blockl,-\blockh-0.5) -- (\blockl,-\blockh+0.5) node[right] {};
  \draw[-latex,thick,blue!50] (0,-\blockh+0.5) -- (0,-\blockh-0.5) node[black,midway,left] {$\mathcal{C}_{l,1}$};
  \draw[-latex,thick,blue!50] (0,-\blockh-0.5) -- (\blockl,-\blockh-0.5) node[black,above left] {};

  \draw[-latex] (\blockl/2,+\blockh) arc (0:-270:\blockh/4) node[black,below] {$\mathcal{C}_{u}$};
  \draw[red,fill] (0,+\blockh+0.5) circle [radius=1.5pt] node[black,above right] {$+i\mu+i\eta$};
  \draw[red,fill] (0,+\blockh-0.5) circle [radius=1.5pt] node[black,below right] {$+i\mu-i\eta$};

  \draw[-latex] (\blockl/2,-\blockh) arc (0:+270:\blockh/4) node[black,above] {$\mathcal{C}_{l}$};
  \draw[red,fill] (0,-\blockh+0.5) circle [radius=1.5pt] node[black,above right] {$-i\mu+i\eta$};
  \draw[red,fill] (0,-\blockh-0.5) circle [radius=1.5pt] node[black,below right] {$-i\mu-i\eta$};

  \draw[] (-\blockl,-\blockh) node[draw,below,text width=3\blockh,minimum height=3\blockh]
    {$\delta_\T$-\\terms};
\end{tikzpicture}
\caption[]{%
  Left: deformation of the line integral~\eqref{eq:cauchy1} for integrals involving $\nF$-terms, i.e.\ ($\mathcal{L}_a,\mathcal{L}_b$) to  ($\mathcal{L}_1,\mathcal{L}_2,\mathcal{L}_3$). The contours $\mathcal{L}_a$ and $\mathcal{L}_b$ indicate the original line integral parallel to the real axis. By mapping $\mathcal{L}_a$ from
  $(-\infty + i \mu, i \mu)$ to $(-i \mu, \infty -i \mu)$ and changing the sign of the integrand $f(p_0)\mapsto \newtext{-} f(-p_0)$, the rectangular contour of interest can be closed, while the dashed contours can be neglected. In turn, the integral is evaluated in terms of the thick blue lines $\mathcal{L}_1$, $\mathcal{L}_2$ as well as $\mathcal{L}_1$, $\mathcal{L}_3$.  The lines $\mathcal{L}_2$ and $\mathcal{L}_3$ are oriented in opposite direction, while $\mathcal{L}_1$ is used twice in this construction.
  Right:   deformation of the fermionic contour~\eqref{eq:fermionsum} for integrals involving $\nF'$-terms,  i.e.\  ($\mathcal{C}_a,\mathcal{C}_b$) to  ($\mathcal{C}_u,\mathcal{C}_l$).
  The initial contour, which is closed by two vertical segments at
  $\re(p_0) = \pm \infty$,  is first split into two semi-infinite closed contours by introducing
  the vertical segment $\mathcal{C}_{u,1}$ and its inverse. The left of these closed contours is then mapped to the fourth quadrant of the complex plane upon setting $p_0\mapsto -p_0$, whereby the inverse of $\mathcal{C}_{u,1}$ becomes $\mathcal{C}_{l,1}$ and we end up with the final blue contours  $\mathcal{C}_u$ and $\mathcal{C}_l$.
  }
  \label{fig:contoursplit}
\end{figure}

To no surprise, the technical complications that we encountered in the $T\to 0$ limit of the one-loop integral  \eqref{eq:naiveleadingmapping} and that culminated in the isolation of a novel divergence in eqs.~\eqref{eq:Iapole} and \eqref{eq:Ianeq} continue to be present at higher loop orders. Although one may often bypass these issues by analytically continuing results derived for convergent values of parameters such as the dimension $d$ and the exponent $\alpha$ above, in the application of IBP relations we will again encounter integrals containing differentiated Fermi-Dirac distribution functions that give rise to extra contributions from the momentum region that makes the argument of the $\nF'$ vanish, namely $p = \mu$, $p_0 \sim 0$. It turns out that for the two classes of $p_0$-integrals featuring either undifferentiated or differentiated $\nF$ functions the optimal computational strategies differ somewhat and in particular feature slightly different ways of deforming the original $p_0$-contour \eqref{eq:oint:f}. Below, we briefly introduce these two deformations that are respectively summarized in fig.~\ref{fig:contoursplit}~(left) and (right).

The former class of integrals is \newtext{initially} characterized by such parameter values that allow one to take the step-function limit of the Fermi-Dirac distribution function without introducing new divergences,  \newtext{leading to special function solutions analytically continued to the relevant parameter values}.
Akin to eq.~\eqref{eq:naiveleadingmapping}, it is then sufficient to only consider the first line integral in eq.~\eqref{eq:oint:f}, and $\eta$ can furthermore be taken to 0. The practical evaluation of such integrals can be simplified by deforming the remaining integral in a fashion illustrated in fig.~\ref{fig:contoursplit}~(left) and detailed in~\cite{Gorda:2022yex}. Using the Cauchy theorem, we can namely replace the infinite integral ($\mathcal{L}_a$, $\mathcal{L}_b$) parallel to the real axis by three separate line integrals ($\mathcal{L}_1$, $\mathcal{L}_2$, $\mathcal{L}_3$), of which $\mathcal{L}_2$ and $\mathcal{L}_3$ are of finite length and run along the imaginary axis while $\mathcal{L}_1$ runs along the real axis, giving
\begin{eqnarray}
\label{eq:cauchy1}
      \int_{-\infty+i \mu}^{\infty + i \mu} \frac{{\rm d}p_0}{2\pi} f(p_0) =
      \int_{-\infty}^\infty \frac{{\rm d}p_0}{2\pi} f(p_0)
    - \int_{0}^{+i\mu} \frac{{\rm d}p_0}{2\pi} f(p_0)
    - \int_{0}^{-i\mu} \frac{{\rm d}p_0}{2\pi} f(-p_0)
    \, .
\end{eqnarray}
The validity of this construction, detailed in appendix~\ref{app:singleprop}, requires that
the integrated function $f(p_0)$ be regular inside the closed Cauchy contour. This can typically be achieved for certain ranges of propagator exponents, resulting in expressions that can be later analytically continued to the divergent parameter regions.

For the second class of integrals, a different strategy involving the full fermionic contour of eq.~\eqref{eq:fermionsum} is needed, owing to the sensitivity of the differentiated Fermi-Dirac distribution to those momentum regions where its argument vanishes. This can be understood by recognizing that eq.~\eqref{eq:nfb_complexdelta} describes a two-sided $\delta$-distribution, which results in both line integrals of the fermionic contour \eqref{eq:oint:f} contributing to the results at $T \to 0$. Assuming, as usual, that $f(p_0)$ is regular within the fermionic contour, we can again split the original integrals, denoted now by $\mathcal{C}_{a}$ and $\mathcal{C}_{b}$, at the imaginary axis as shown in fig.~\ref{fig:contoursplit} (right). The contour on the left-hand side of the imaginary axis is then mapped to the fourth quadrant of the complex plane, so that $(\mathcal{C}_{a} + \mathcal{C}_{b}) \mapsto (\mathcal{C}_{u} + \mathcal{C}_{l})$ as detailed in the figure.

Considering now the $\nF'$-terms integrated over the contours $\mathcal{C}_{u}$ and $\mathcal{C}_{l}$, we obtain
\begin{align}
  \oint_{p_0}^f f(p_0) \Bigl\{i \beta \nF' \left[i \beta (p_0-i\mu) \right] \Bigr\} &=
    \oint_{p_0}^{\mathcal{C}_u} f(p_0) \Bigl\{i \beta \nF' \left[i \beta ( p_0-i\mu) \right] \Bigr\}
  \nn &
  + \oint_{p_0}^{\mathcal{C}_l} f(-p_0)\Bigl\{i \beta \nF' \left[i \beta (-p_0-i\mu) \right] \Bigr\}
  \, ,
\end{align}
where the only nonzero contributions originate from $\mathcal{C}_{u,1}$ and $\mathcal{C}_{l,1}$ of fig.~\ref{fig:contoursplit} (right). The leading low-temperature behavior of the fermionic contour integral then becomes
\begin{align}
\label{eq:primecontoursimple}
  \oint_{p_0}^f f(p_0) \Bigl\{i \beta \nF' \left[i \beta (p_0-i\mu) \right] \Bigr\}
  &\to
  \int_{i\mu-i\eta}^{i\mu+i\eta} \frac{{\rm d}p_0}{2\pi} f(p_0)\Bigl\{i\beta \nF' \left[i\beta (p_0-i\mu) \right] \Bigr\}
  \nn
  &+
  \int_{-i\mu+i\eta}^{-i\mu-i\eta} \frac{{\rm d}p_0}{2\pi} f(-p_0) \Bigl\{i\beta \nF' \left[i\beta (-p_0-i\mu) \right] \Bigr\}
  \, ,
\end{align}
which demonstrates that the only nonzero contributions to the original integral originate from the neighborhood of $p_0 = i\mu$ of eq.~\eqref{eq:oint:f}. The deformed contours defined here greatly simplify analytic computations involving the $\delta$-sequences but are nevertheless able to capture the novel thermal corrections from the differentiated distribution functions. This resembles, to some extent, the role of the finite line integrals in eq.~\eqref{eq:cauchy1}.

%
\section{Integration-by-parts relations}
\label{sec:IBPintro}

Moving on from the details of integration contours to the derivation of the desired IBP relations, let us first specify the precise form of the vacuum-type Feynman integrals we will be studying in this paper. To this end, we abbreviate the Fermi-Dirac distribution functions as
\begin{equation}
\nFt(p_0) \equiv \nF \left[i \beta (p_0-i\mu) \right]
\, ,\qquad
\nFt'(p_0) \equiv i \beta \nF' \left[i \beta (p_0-i\mu) \right]
\, ,
\end{equation}
and assume \newtext{a scalarized} numerator structure in the \newtext{Feynman} integrals to be considered.
A generic vacuum-type Feynman integral then takes the form
\begin{equation}
\label{eq:genericZ}
  \mathcal{Z}_{\{\alpha\},\{\gamma\},\{\phi\}}^{\{s\},\{t\}}(\mu,T) =
    \biggl\{
    \prod_{k=1}^{N^{\rm f}_{\ell}}
    \oint_{P_{k}}^f
    \frac{p_{k,0}^{s_k}\nFt(p_{k,0})}{[P_k^2]^{\alpha_k}}
    \biggr\}
    \biggl\{
     \prod_{l=1}^{N^{\rm b}_{\ell}}
    \oint_{Q_{l}}^b
    \frac{q_{l,0}^{t_l}\nB[i \beta q_{l,0}]}{[Q_l^2]^{\gamma_l}}
    \biggr\}
  \prod_{R \in \newtext{\mathcal{R}^{\pm}_{\geq 2}}(P_k, Q_l)} \frac{1}{[R^2]^{\phi_{R}}}
  \, ,
\end{equation}
where we again denote momenta in $d+1$ dimensions by $P = (p_0,\vec{p})$. In this expression, $N^{\rm f}_{\ell}$ stands for the number of fermionic and $N^{\rm b}_{\ell}$ for the number of bosonic loops, with $P_k = (p_{k,0},\vec{p}_k)$ denoting fermionic and $Q_l$ bosonic loop momenta. Finally, the momenta $R$ (no indices) are linear combinations of the $P_k$ and $Q_l$, picked from the sets $\newtext{\mathcal{R}^{\pm}_{\geq 2}} (P_k, Q_l)$.%
\footnote{
  We denote by $\newtext{\mathcal{R}_{\geq 2}} (P_k, Q_l)$ the set of distinct loop momenta with two or more constituents, while $\newtext{\mathcal{R}^{\pm}_{\geq 2}} (P_k, Q_l)$  further includes a unique sign signature for each subset corresponding to the fermionic flow in the corresponding diagram.
}

As two explicit examples of the (scalarized) integrals to be considered, the one-loop fermion bubble and the two-loop fermionic sunset are defined as
\begin{align}
\label{eq:I:1l}
  \mathcal{I}_{\alpha}^{s} (\mu,T) \equiv
  \TopoVR(\Aqu) &=
    \oint_{p_0}^f
    p_0^s\,\nFt(p_0)\,
    I_{\alpha}(p_0)
    \, ,\\[2mm]
\label{eq:sunset:fb}
 \newtext{\mathcal{S}_{\alpha_1 \alpha_2 \alpha_3}^{s_1 s_2}(\mu, T) =}
  \noveltext{\ToptVS(\Aqu,\Aqu,\Lgl)} &=
  \oint_{p_0}^f \oint_{q_0}^b \newtext{p_0^{s_1} q_0^{s_2}}
  \nFt(p_0)
  \noveltext{\nFt (q_0)}\,
  S_{\alpha_1 \alpha_2 \alpha_3}(p_0,q_0,p_0-q_0)
  \, ,
\end{align}
where we henceforth use
calligraphic letters for in-medium and
Latin letters for \newtext{$d$-dimensional $T = 0$} 
loop integrals and also identify
$\mathcal{I}_{\alpha}^{ }(\mu,T) \equiv \mathcal{I}_{\alpha}^{0} (\mu,T)$ \newtext{as well as $\mathcal{S}_{\alpha_1 \alpha_2 \alpha_3}(\mu, T)  \equiv \mathcal{S}_{\alpha_1 \alpha_2 \alpha_3}^{00}(\mu, T) $ }.
In the graphical notation,
wiggly lines always stand for bosonic and
directed solid lines for fermionic propagators.
We also introduced the corresponding $d$-dimensional vacuum integrals for the
one-loop bubble and
two-loop sunset
\begin{align}
\label{eq:I:m}
  I_\alpha(m) &=
    \int_{\vec{p}}
    \frac{1}{[p^2 + m^2]^{\alpha}}
    =
    \biggl( \frac{e^\gammaE\bmu^2}{4\pi} \biggr)^\frac{3-d}{2}
    \frac{
    \Gamma\bigl(\alpha-\frac{d}{2}\bigr)}{
    \Gamma\bigl(\alpha\bigr)}
    \frac{[m^2]^{\frac{d}{2}-\alpha}}{(4\pi)^\frac{d}{2}}
    \,,
    \\[2mm]
\label{eq:S:m123}
  S_{\alpha_1\alpha_2\alpha_3}(m_1,m_2,m_3) &=
    \int_{\vec{p},\vec{q}}
    \frac{1}{
      [p^2 + m_1^2]^{\alpha_{1}}
      [q^2 + m_2^2]^{\alpha_{2}}
      [|\vec{p}-\vec{q}|^2 + m_3^2]^{\alpha_{3}}
    }
    \,,
\end{align}
while deferring special mass signatures of the latter to appendix~\ref{sec:vacuum:2l}.

In practice, \newtext{both spatial and temporal} IBP relations are derived starting from integral expressions that vanish as total derivatives. Assuming that the integration orders are interchangeable, for one loop momentum, $P_\mu$,
such a relation can be written as
\begin{align}
\label{differential}
    0 =
    \oint_{P}^f
    \frac{\partial}{\partial P^\mu} \bigg[ P^\nu\;
      \nFt(p_0) f(P)
    \bigg]
    &\equiv
    \frac{\partial}{\partial P^\mu} \circ P^\nu \bigg[
    \oint_{P}^f\!
    \nFt(p_0) f(P)
    \bigg]
    \nn &=
    \oint_{P}^f
    \biggl[
      \nFt^{ }(p_0)\Bigl(
        \delta_{\mu}^\nu
        + P^\nu \frac{\partial}{\partial P^\mu}
        \Bigr)f(P)
      + \delta_{\mu 0}^{ } P^\nu \nFt'(p_0) f(P)
    \biggr]
  \, ,
\end{align}
where the integrand $f(P)$ typically takes the form of a product of individual propagators and we have introduced a notation for the differential operator $\frac{\partial}{\partial P^\mu} \circ P^\nu$
acting on an integral. The last term on the last line of eq.~\eqref{differential} is seen to contain a derivative of the Fermi-Dirac distribution, which is specific to a thermal setting. The explicit distribution functions are essential for ensuring that the total derivatives vanish and that integration orders are interchangeable.

Considering next two distinct loop momenta $P$ and $Q$, the simplest differential operators that can be expected to give rise to non-trivial IBP relations are bilinear in both momenta and derivatives. They form two disjoint classes that are either diagonal or off-diagonal\footnote{\newtext{The off-diagonal operators lead to scalarized expressions only when combining an even number of them to a composite operator. While of no interest at the one-loop level, they can play a more pronounced role in multi-loop problems.}} in Lorentz indices,
\begin{align}
\label{eq:ibp:set}
\biggl\{
\frac{\partial}{\partial p_i} \circ p_i,
\frac{\partial}{\partial p_i} \circ q_i,
\frac{\partial}{\partial p_0} \circ p_0,
\frac{\partial}{\partial p_0} \circ q_0,\cdots
\biggr\}
& \qquad
\text{diagonal}
\, ,
\nn
\biggl\{
\frac{\partial}{\partial p_0} \circ p_i,
\frac{\partial}{\partial p_i} \circ q_0, \cdots
\biggr\}
& \qquad
\text{off-diagonal}
\, .
\end{align}
\newtext{When acting on the integrand apart from the distribution functions, the two classes are self-contained}, such that integrals generated by an operator of a given class can be algebraically related to integrals generated by other operators in the same class. This property can be seen in the one-loop expression
\begin{align}
\label{oneloopmanipulation}
    \oint_{P}^f\!
    \nFt(p_0)  \frac{\partial}{\partial p_0}
    \frac{p_0}{[p_0^2+p^2]^\alpha} &=
    \Bigl[
    (d+1-2 \alpha)
    -\frac{\partial}{\partial p_i} \circ p_i
    \Bigr] \mathcal{I}_\alpha^{ } (\mu,T)
    \,,
\end{align}
where the emergent quadratic numerator structure $p_0^2$ has been expressed in terms of
a scalar multiplier and
a total spatial derivative.

So far, only spatial \newtext{differential operators} have been considered in the derivation of IBP relations~\cite{Nishimura:2012ee,Laporta:2000dsw} for systems with broken Lorentz symmetry. \newtext{The extension we propose here} is to \newtext{generalize the} IBP relations \newtext{to} the full $(d+1)$-dimensional spacetime, \newtext{taking advantage of natural simplifications such as that present in eq.~\eqref{oneloopmanipulation}}. \newtext{A pedagogical example of this procedure at the one-loop level is obtained by acting on $\mathcal{I}_{\alpha}^{s}(\mu, T)$ with both spatial and temporal derivatives. This leads to
\setcounter{dummy}{\value{equation}}
\renewcommand{\theequation}{ibp.1}
\begin{subequations}
  \begin{alignat}{3}
\label{eq:ibp:1l:spatial}
    0 &=
    \Bigl(\frac{\partial}{\partial p_i} \circ p_i\Bigr)
    \mathcal{I}_\alpha^{s} (\mu,T)
    &&=
    (d-2 \alpha) \mathcal{I}_\alpha^{s} (\mu,T)
    +2 \alpha
    \mathcal{I}_{\alpha+1}^{s+2} (\mu,T)
    \, ,
    \\[2mm]
\label{eq:ibp:1l:temporal}
    0 &=
    \Bigl(\frac{\partial}{\partial p_0} \circ p_0\Bigr)
    \mathcal{I}_{\alpha}^{s} (\mu,T)
    &&=
    (1+s) \mathcal{I}_{\alpha}^{s} (\mu,T)
    -2\alpha\mathcal{I}_{\alpha+1}^{s+2} (\mu,T)
    +\oint_{P}^f
      \frac{
        p_0^{s+1}\,
        \nFt'(p_0)
      }{[p_0^2+p^2]^\alpha}
    \, ,
    \\[2mm]
\label{eq:ibp:1l:d+1}
    0 &=
    \Bigl(\frac{\partial}{\partial P_\mu} \circ P_\mu\Bigr)
    \mathcal{I}_\alpha^{s} (\mu,T)
    &&=
    (d+1-2 \alpha + s)
    \mathcal{I}_\alpha^{s} (\mu,T)
    +\oint_{P}^f
      \frac{
        p_0^{s+1}\,
        \nFt'(p_0)
      }{[p_0^2+p^2]^\alpha}
    \,,
\end{alignat}
\end{subequations}
\renewcommand{\theequation}{\arabic{section}.\arabic{equation}}%
\setcounter{equation}{\value{dummy}} 
where on the last line we combined spatial and temporal total derivatives into the $(d+1)$-dimensional bilinear operator
\begin{equation}
\frac{\partial}{\partial P_\mu} \circ P_\mu =
  \frac{\partial}{\partial p_0} \circ p_0
+ \frac{\partial}{\partial p_i} \circ p_i
  \,.
\end{equation}}

\newtext{In the relations \eqref{eq:ibp:1l:spatial}--\eqref{eq:ibp:1l:d+1}, each term notably involves either one differentiated or one non-differentiated distribution function. Proceeding to higher loop orders, the classification of integrals appearing in our generalized IBP identities will naturally become slightly more complicated, but all integrals nevertheless still fall into one of the following three disjoint classes:
\begin{enumerate}[\bf type~A:]
  \item
    \label{typeA}
    integrals with no differentiated distribution functions,
  \item
  \label{typeB}
    integrals with only differentiated distribution functions, and
      \item
    \label{typeC}
     integrals with at least one differentiated and one non-differentiated distribution function.
\end{enumerate}
Conventional} spatial IBP identities such as \eqref{eq:ibp:1l:spatial} \newtext{only involve} integrals of type \ref{typeA}, which \newtext{are also familiar} from vacuum computations~\cite{Chetyrkin:1981qh,Bender:1976pw}.
There they typically reduce the numerator and denominator powers in type \ref{typeA} integrals while
introducing rational functions of the dimension $d$. \newtext{In contrast, the novel terms of types} \newtext{\ref{typeB} and \ref{typeC} are generated by temporal derivatives in the thermal context, where they} play the role of \newtext{inhomogeneous terms} closing \newtext{the IBP relations} between type \ref{typeA} integrals. This can be seen already in the one-loop relation~\eqref{eq:ibp:1l:d+1}, where the presence of the last term allows writing $\mathcal{I}_\alpha^{ } (\mu,T)$ in terms of an integral of reduced dimensionality in the zero-temperature limit \newtext{due to the $\delta$-sequence of eq.~\eqref{nFprimeT0},} thus constituting a new representation for the original integral.

\newtext{At higher loop orders, the utility of the novel IBP relations similarly derives from a qualitative hierarchy between the computational workloads associated with evaluating type \ref{typeA}, \ref{typeB}, and \ref{typeC} integrals: type \ref{typeA} $>$ type \ref{typeC} $\gg$ type \ref{typeB}. This is due to the appearance of differentiated distribution functions in the type \ref{typeC} and \ref{typeB} integrals, which according to the $\delta$-sequence of eq.~\eqref{nFprimeT0} amounts to a reduction in their dimensionality.%
\footnote{\noveltext{Each occurring delta function trades a temporal integration for a linear combination of on-shell substitutions, simplifying the evaluation of the corresponding loop integral. This is somewhat analogous to how the cutting rules of \cite{Ghisoiu:2016swa} operate.}} As we will demonstrate in the following two sections of the article, results for type \ref{typeB} integrals are typically immediately available through a straightforward analytic continuation of known massive vacuum ($T=\mu=0$) integrals, but even type \ref{typeC} integrals are normally dramatically simpler to evaluate than the original type  \ref{typeA} entities. Accordingly, our general strategy will focus on deriving relations between type \ref{typeA} and \ref{typeB} terms, i.e.~integrals involving either only differentiated or non-differentiated distributions functions, with type \ref{typeC} terms remaining present only if absolutely necessary.}

%
\section{One-loop integrals: new strategy in action}
\label{sec:1-loopformulae}

Having laid out our general strategy above, we will now take a closer look at explicit IBP calculations at the one-loop level, where we first return to the well-known integral~\eqref{eq:I:1l} \noveltext{with $s=0$ (for $s \in \mathbb{N}$, see eq.~\eqref{eq:I:alpha:s})}. It was shown in~\cite{Gorda:2022yex} that in the limit $T\to 0$ and for any $\alpha\in \mathbb{R}$, the integral takes the form
\begin{align}
\label{eq:fermion1master_new}
  \mathcal{I}_\alpha^{ } (\mu) \equiv
  \lim_{T \to 0}
  \mathcal{I}_\alpha^{ } (\mu,T)
    &=
    -2\re \int_{\vec{p}} \theta(\mu-p)
      \int_0^{i \mu} \frac{{\rm d}p_0}{2\pi}
      \frac{1}{[p_0^2+p^2]^\alpha}
    \nn &=
    -\biggl( \frac{e^\gammaE\bmu^2}{4\pi} \biggr)^\frac{3-d}{2}
    \frac{1}{ (4\pi)^\frac{d}{2}
      \Gamma\bigl(\alpha\bigr)
      \Gamma\bigl(\frac{d}{2}+1-\alpha \bigr)}
    \frac{\mu^{d+1-2 \alpha}}{\bigl(d+1-2 \alpha\bigr)}
    \, ,
\end{align}
which is nothing but the $T\to 0$ limit of the $\widetilde{\mathcal{I}}_\alpha^0$ function defined in~\cite{Vuorinen:2003fs}, generalized to an arbitrary spatial dimensionality $d$. The integral converges for parameter values of $\alpha$ and $d$ inside the triangle-shaped region
$\{ 2 \alpha-d  > 1 \,| \, \frac{1}{2}<\alpha < 1 \, \wedge \, d > 0 \}$, but for strictly vanishing $T$ features a $d$-independent novel divergence in the neighborhood of $p = \mu$ for $\alpha > 1$ that  requires analytic continuation from the convergent region. The result in eq.~\eqref{eq:fermion1master_new} is easiest derived starting from the spatial integral and utilizing a complex-valued generalization of Euler's beta function,%
\footnote{
  Assuming that $\alpha_{2}> \alpha_{1} > 0$ and that either $c \neq 0$ or $y > 0$ for $c,y\in\mathbb{R}$, we can generalize Euler's beta function integral such that
  $\int_0^\infty \frac{{\rm d}x\,x^{\alpha_{1}-1}}{(x + y+ ic)^{\alpha_{2}}} =
  \frac{
    \Gamma (\alpha_{1})
    \Gamma (\alpha_{2}-\alpha_{1})}{
    \Gamma (\alpha_{2})}
  (y+ ic)^{\alpha_{1}-\alpha_{2}}
  $
  as used in~\cite{Gorda:2022yex}.
}
which ensures a consistent treatment of the problematic neighborhood in the low-temperature limit. A generalization of this result to non-vanishing masses is detailed in appendix~\ref{sec:massl} with additional context found in sec.~III of~\cite{Gorda:2022yex}.
Further consistency checks are relegated to appendix~\ref{sec:consistency}, including the verification of the fact that acting on the master integral
eq.~\eqref{eq:fermion1master_new} with spatial total derivatives $\frac{\partial}{\partial p_i} \circ p_i$ leads to a vanishing result as it should.

As alluded to in the previous section, acting on thermal integrals with temporal total derivative operators may give rise to non-trivial linear relations, where terms featuring differentiated distribution functions are expected to lead to simplifications in the low-temperature limit. To demonstrate this at the simplest possible level, we focus on the identity~\eqref{eq:ibp:1l:d+1}, which clearly allows the evaluation of the master integral $\mathcal{I}_\alpha(\mu)$ through a new representation
\begin{eqnarray}
\label{eq:1loopdeltaformula1}
  \mathcal{I}_\alpha^{ }(\mu)
  =-\frac{1}{(d+1-2\alpha)} \lim_{T\to 0}\,
  \oint_{p_0}^f
  p_0^{ }\,\nFt'(p_0)
  I_{\alpha}(p_0)
    \, .
\end{eqnarray}
To evaluate the right-hand side of this relation, we first apply a complex-valued generalization of the integral form of Euler's beta function, resulting in
\begin{eqnarray}
\label{eq:prime1d}
  \oint_{p_0}^f\!
  p_0^{ }\,\nFt'(p_0)
  I_{\alpha}(p_0)
  &=&\hspace{-3mm}
  -i
  \biggl(\frac{e^\gammaE \bmu^2}{4 \pi} \biggr)^\frac{3-d}{2}
  \frac{\Gamma \left(\alpha-\frac{d}{2} \right)}{(4\pi)^\frac{d}{2} \Gamma (\alpha)}
  \oint_{p_0}^f\!
  p_0 [p_0^2]^{\frac{d}{2}-\alpha}
  \Bigl\{
    -\beta \nF' \left[i\beta (p_0-i\mu)\right]
  \Bigr\}
  \nn
  & \stackrel{{\rm eq.\eqref{eq:nfb_complexdelta}}}{=}&\hspace{-3mm}
  -i
  \biggl(\frac{e^\gammaE \bmu^2}{4 \pi} \biggr)^\frac{3-d}{2}
  \frac{\Gamma \left(\alpha-\frac{d}{2} \right)}{(4\pi)^\frac{d}{2} \Gamma (\alpha)}
  \oint_{p_0}^f\!
  p_0 [p_0^2]^{\frac{d}{2}-\alpha}
  \Bigl\{
      \delta_{\T}^{\rmii{F}}\left[\mu-\im (p_0)\right]
    + i 0_{\T \rmii{2}}^{\rmii{F}}
  \Bigr\}
  \,.\hspace{0.9cm}
\end{eqnarray}
Subsequently, we utilize the contour prescription of fig.~\ref{fig:contoursplit}~(right), where nonzero contributions arise from the two finite line segments $\mathcal{C}_{u,1}$ and $\mathcal{C}_{l,1}$. Since these two terms are related to each other via complex conjugation, we easily obtain
\begin{align}
  \lim_{T\to 0}
  \oint_{p_0}^f\!
  p_0^{ }\,\nFt'(p_0)
  I_{\alpha}(p_0)
  &=
  - \biggl(\frac{e^\gammaE \bmu^2}{4 \pi} \biggr)^\frac{3-d}{2}
  \frac{\Gamma \left(\alpha-\frac{d}{2} \right)}{(4\pi)^\frac{d}{2} \Gamma (\alpha)}
  \frac{1}{\pi}
  \re \biggl\{i \int_{i\mu -i\eta}^{i\mu+ i\eta} \!{\rm d}p_0 \,
  p_0 [p_0^2]^{\frac{d}{2}-\alpha}
  \delta(\mu+ip_0)
  \biggr\}
  \, .
\end{align}
Finally, to ensure that the monomial structure resulting from the final $p_0$-integration is well-defined for arbitrary non-integer powers $\alpha$, we regulate the imaginary unit inside the integral by defining
\begin{equation}
\label{eq:i:reg}
  i \mapsto \Pi^+\equiv \exp \Bigl[ \frac{i \pi}{2}(1+ \kappa ) \Bigl]
  \, ,
\end{equation}
with $\kappa>0$ an infinitesimal positive real number.
This allows to extract a result in terms of trigonometric functions, swiftly leading to the same result reported in eq.~\eqref{eq:fermion1master_new} upon taking $\kappa \to 0$.

As demonstrated by the above example, the appearance of a differentiated distribution function reduces the number of integrations by one in the $T\to 0$ limit, while the remaining ones correspond to complex-valued generalizations of standard $d$-dimensional integrals. It is often operationally useful to first perform the temporal integral but leave the complex-valued substitution (with the regulated imaginary unit $\Pi^+$) to be performed after the spatial integrations. In the above example, this would imply writing
\begin{align}
\label{eq:cauchyex1}
  \lim_{T\to 0}
  \oint_{p_0}^f\!
  p_0^{ }\,\nFt'(p_0)
  I_{\alpha}(p_0)
  =& \frac{1}{2\pi}
  \Bigl\{
        p_0 I_{\alpha}(p_0) \bigr|_{p_0 \mapsto \mu\,\Pi^+}
      + p_0 I_{\alpha}(p_0) \bigr|_{p_0 \mapsto \mu/\Pi^+}
    \Bigr\}
  \, ,
\end{align}
where
the spatial integral can be identified as the standard one-loop vacuum integral from eq.~\eqref{eq:I:m}.

To conclude this section, a few general remarks are in order. First, it should be apparent from above that the procedure applied in eq.~\eqref{eq:cauchyex1} will greatly simplify the treatment of
the inhomogeneous (type~\ref{typeB}) terms at higher loop orders, which should optimally contain the maximal number of differentiated distribution functions. Second, we have relegated several additional example calculations to appendix~\ref{sec:consistency}; these include consistency checks of the one-loop IBP relation~\eqref{eq:1loopdeltaformula1}, an explicit computation involving linear $p_0$-structures within the integrand numerator, and  a brief discussion of the use of parametric $\mu$-differentiation.

%
\section{Proceeding to the two-loop level and beyond}
\label{sec:2-loop}

The value of the \newtext{proposed novel} IBP machinery only becomes apparent in multi-loop computations. While this paper is mostly devoted to developing the new formalism, we will therefore next discuss its application at the lowest nontrivial order of perturbation theory, i.e.~two loops, which amounts to the \newtext{leading} correction to the non-interacting limit for the QCD pressure. This order is somewhat special, as it allows the derivation of nontrivial IBP relations while keeping the practical computations at a pedagogical and easily tractable level. This is particularly true for integrals with unit numerators but sufficiently general denominators, so that the result does not trivially factorize \newtext{into one-loop structures}.%
\footnote{
For two-loop vacuum bubbles with integer-valued exponents, factorization will later be seen to emerge through IBP relations (see \newtext{appendix}~\ref{sec:factorizationof2loop} below).}

For the aforementioned reasons, we shall concentrate on the fermionic sunset (cf.\ eq.~\eqref{eq:sunset:fb})
\begin{align}
\label{eq:S:a123:s12}
  \ToptVSn(\Aqu,\Aqu,\Lgl,1,2,3) =
  \mathcal{S}_{\alpha_{1}\alpha_{2}\alpha_{3}}^{s_1 s_2} &=
    \oint_{P,Q}^f
    \frac{p_0^{s_1}\,\nFt(p_0)}{[p_0^2+p^2]^{\alpha_1}}
    \frac{q_0^{s_2}\,\nFt(q_0)}{[q_0^2+q^2]^{\alpha_2}}
    \frac{1}{[(p_0-q_0)^2+ |\mathbf{p}-\mathbf{q}|^2]^{\alpha_3}}
    \nn[1mm] &=
    \oint_{p_0,q_0}^f
    p_0^{s_1} q_0^{s_2}\,
    \nFt(q_0)
    \nFt(p_0)
    S_{\alpha_{1}\alpha_{2}\alpha_{3}}(p_0,q_0,p_0-q_0)
    \, ,
\end{align}
where
$
\mathcal{S}_{\alpha_{1}\alpha_{2}\alpha_{3}}^{00} \equiv
\mathcal{S}_{\alpha_{1}\alpha_{2}\alpha_{3}}^{}
$ and
the non-calligraphic $S_{\alpha_{1}\alpha_{2}\alpha_{3}}$ denoted the corresponding $d$-dimensional vacuum sunset from eq.~\eqref{eq:S:m123}. The denominator exponents $\{\alpha_1, \alpha_2, \alpha_3 \}$ can be chosen such that the $p_0$- and $q_0$-integrals are regular both in the ultraviolet (UV) and at possible individual poles. As detailed in sec.~\ref{sec:formalism} and appendix~\ref{sec:Cauchy_appendix}, this allows for using the Cauchy theorem (see e.g.~eq.~\eqref{eq:cauchy1}) to analytically continue the results to all integer-valued exponents. In addition, it allows choosing the integration order at will even in the low-temperature limit and conveniently isolates the novel boundary contributions to finite line integrals along the imaginary axis as summarized in fig.~\ref{fig:contoursplit} (left).

Finally, we reiterate that the main goal of our \newtext{new} IBP program is to find relations between Feynman integrals that are of the vacuum-type \ref{typeA} in the presence of additional non-vacuum type~\ref{typeB} terms that are needed to close the relations. The latter terms include a maximal number of differentiated distribution functions and are computationally less complex as demonstrated in the previous section at the one-loop level. At the two-loop level, scrutinized in the present section, we begin our discussion from these maximally primed terms in sec.~\ref{sec:maxprime}, derive and solve novel IBP relations in sec.~\ref{sec:ibp2loopsec}, and finally, dwell on the closely related factorization of $d$-dimensional two-loop vacuum integrals in the subsequent \newtext{appendix}~\ref{sec:factorizationof2loop}.
\newtext{While the implementation of the computational methods is explicitly discussed only at the two-loop level, all results are at least in principle straightforwardly generalizable to an arbitrary loop order.}

\subsection{Integrals with only differentiated distribution functions}
\label{sec:maxprime}

Let us first examine the evaluation of the boundary or inhomogeneous terms containing only primed distribution functions $\nF'$. Using the contour prescription of eq.~\eqref{eq:primecontoursimple}, we can perform the $\Pi^+$ substitutions akin to eq.~\eqref{eq:cauchyex1} independently for the $T\to 0$ limit of each temporal integral, which generates $2^n$ independent terms from the contour integrals of an $n$-loop diagram. The sign of each term is determined by the even/odd symmetry of the integrand as indicated by eq.~\eqref{eq:primecontoursimple}. Below, we detail this procedure at the two-loop level focusing on the sunset integral of eq.~\eqref{eq:S:a123:s12}.

Acting on $\mathcal{S}_{\alpha_1 \alpha_2 \alpha_3}$ with two temporal derivatives gives rise to a term with two primed distribution functions, i.e.~an integral of the maximally primed type~\ref{typeB}. \noveltext{For the purpose of abbreviating such expressions, we introduce $\nF$-differentiating operators ${\bf D}_{k}$ that only act on the distribution functions of the corresponding temporal momentum via
\begin{align}
\label{eq:Dp}
  {\bf D}_{k}\,
  \oint_{k_{0}}^f
  \!
  \nFt(k_{0})
  &=
  \oint_{k_{0}}^f
  \!
  \nFt'(k_{0})
  \, .
\end{align}}
\noindent
Consistently with the above discussion, we find altogether $2^2=4$ separate terms in the low-temperature limit, totaling
\begin{eqnarray}
\label{eq:2loopmlong}
\noveltext{{\bf D}_{p} {\bf D}_{q}}\,
  \mathcal{S}_{\alpha_1\alpha_2\alpha_3}^{s_1 s_2}
  &\noveltext{=}
  &
    \int_{\vec{p},\vec{q}}
    \oint_{p_0,q_0}^f
    \frac{p_0^{s_1}\,\nFt'(p_0)}{[p_0^2+p^2]^{\alpha_1}}
    \frac{q_0^{s_2}\,\nFt'(q_0)}{[q_0^2+q^2]^{\alpha_2}}
    \frac{1}{[(p_0-q_0)^2+|\vec{p}-\vec{q}|^2]^{\alpha_3}}
  \nn
  &\stackrel{T\to 0}{=}&
    \frac{1}{(2\pi)^2}
    \int_{\vec{p},\vec{q}}
    \frac{p_0^{s_1}q_0^{s_2}}{
    [p_0^2+p^2]^{\alpha_1}
    [q_0^2+q^2]^{\alpha_2}
    [(p_0-q_0)^2+|\vec{p}-\vec{q}|^2]^{\alpha_3}}
    \biggr|_{\scriptsize
      \begin{aligned}
        p_0 &\mapsto \mu\,\Pi^+\\[-2mm]
        q_0 &\mapsto \mu\,\Pi^+
      \end{aligned}
      }
  \nn
  &-& \frac{1}{(2\pi)^2}
    \int_{\vec{p},\vec{q}}
    \frac{p_0^{s_1}(-q_0)^{s_2}}{
    [p_0^2+p^2]^{\alpha_1}
    [q_0^2+q^2]^{\alpha_2}
    [(p_0+q_0)^2+|\vec{p}-\vec{q}|^2]^{\alpha_3}}
    \biggr|_{\scriptsize
      \begin{aligned}
        p_0 &\mapsto \mu\,\Pi^+\\[-2mm]
        q_0 &\mapsto \mu/\Pi^+
      \end{aligned}
      }
  \nn
  &-& \frac{1}{(2\pi)^2}
    \int_{\vec{p},\vec{q}}
    \frac{(-p_0)^{s_1}q_0^{s_2}}{
    [p_0^2+p^2]^{\alpha_1}
    [q_0^2+q^2]^{\alpha_2}
    [(p_0+q_0)^2+|\vec{p}-\vec{q}|^2]^{\alpha_3}}
    \biggr|_{\scriptsize
      \begin{aligned}
        p_0 &\mapsto \mu/\Pi^+\\[-2mm]
        q_0 &\mapsto \mu\,\Pi^+
      \end{aligned}
      }
  \nn
  &+& \frac{1}{(2\pi)^2}
    \int_{\vec{p},\vec{q}}
    \frac{(-p_0)^{s_1}(-q_0)^{s_2}}{
    [p_0^2+p^2]^{\alpha_1}
    [q_0^2+q^2]^{\alpha_2}
    [(p_0-q_0)^2+|\vec{p}-\vec{q}|^2]^{\alpha_3}}
    \biggr|_{\scriptsize
      \begin{aligned}
        p_0 &\mapsto \mu/\Pi^+\\[-2mm]
        q_0 &\mapsto \mu/\Pi^+
      \end{aligned}
      }
    \;.
    \hspace{0.9cm}
\end{eqnarray}
Each of the terms can be mapped into a form where the propagator masses of underlying vacuum sunset $S_{\alpha_1\alpha_2\alpha_3}(m_1,m_2,m_3)$ obey the linear relation $m_1 + m_2 = m_3$. This class of massive Feynman integrals is typically referred to as {\em collinear}~\cite{Davydychev:2022dcw}. For their evaluation, we use standard Feynman parametrization and carefully regulate the vanishing bosonic mass scale%
\footnote{
  From the perspective of a $d$-dimensional momentum integral,
  the temporal part of the propagator plays the role of a mass term, which explains this terminology.
}
related to $(p_0 \pm q_0) = \mathcal{O}(\kappa)$ with the $\kappa$-regulator introduced in eq.~\eqref{eq:i:reg}. One curious detail of this procedure is that for positive integer values of the exponent $\alpha_3$, associated to the bosonic scale, each arising hypergeometric integral from eq.~(\ref{eq:intermhypergeom}) can be shown to factorize into a product of one-loop vacuum integrals. This is apparent for all four integrals in eq.~\eqref{eq:2loopmlong} wherein the complex-valued scales $p_0$ and $q_0$ allow for extracting contributions proportional to $p_0^2-q_0^2 = \mathcal{O}(\kappa)$. This factorization property at the two-loop level will be addressed using a closed-form solution of the collinear vacuum sunset in \newtext{appendix}~\ref{sec:factorizationof2loop} and at the hypergeometric function level in appendix \ref{sec:vacuum:2l}.

Next, we demonstrate in detail, how the above maximally primed integrals of type~\ref{typeB} can be computed for $\mathcal{S}_{\alpha_1 \alpha_2 \alpha_3}^{s_1 s_2}$ using the strategy established in sec.~\ref{sec:1-loopformulae}. We do this by allowing the power of the temporal momentum components $s_1+s_2$ in the numerator to take arbitrary even or odd values. To this end, we first write eq.~\eqref{eq:2loopmlong} in the form
\begin{eqnarray}
\label{eq:resunset1}
\noveltext{{\bf D}_{p}{\bf D}_{q}}\,
  \mathcal{S}_{\alpha_1\alpha_3\alpha_2}^{s_1 s_2}
  &\noveltext{=}&
  \oint_{p_0, q_0}
  p_0^{s_1}
  q_0^{s_2}\,
  \nFt'(p_0)\nFt'(q_0) S_{\alpha_{1}\alpha_{3}\alpha_{2}} ( p_0, p_0-q_0, q_0)
  \nn
  &\stackrel{T\to 0}{=}&
    \frac{1+(-1)^{s_1+s_2}}{(2\pi)^2} \re \Bigl[ p_0^{s_1+s_2}
    S_{\alpha_{1}\alpha_{3}\alpha_{2}} ( p_0, \mathcal{O}(\kappa), p_0)  \Bigr]_{p_0 \mapsto \mu\,\Pi^+}
  \nn&+& \frac{1-(-1)^{s_1+s_2}}{(2\pi)^2} \Pi^+ \im \Bigl[ p_0^{s_1+s_2}
    S_{\alpha_{1}\alpha_{3}\alpha_{2}} ( p_0, \mathcal{O}(\kappa), p_0)  \Bigr]_{p_0 \mapsto \mu\,\Pi^+}
  \nn&-&
    \frac{(-1)^{s_1} + (-1)^{s_2}}{(2\pi)^2} \re \Bigl[p_0^{s_1} q_0^{s_2}\,
    S_{\alpha_{1}\alpha_{2}\alpha_{3}} ( p_0, q_0, \mathcal{O}(\kappa))  \Bigr]_{\scriptsize
      \begin{aligned}
        p_0 &\mapsto \mu\,\Pi^+\\[-2mm]
        q_0 &\mapsto \mu/\Pi^+
      \end{aligned}
      }\nn
      &+&\frac{(-1)^{s_1} -(-1)^{s_2}}{(2\pi)^2} \Pi^+ \im \Bigl[p_0^{s_1} q_0^{s_2}\,
    S_{\alpha_{1}\alpha_{2}\alpha_{3}} ( p_0, q_0, \mathcal{O}(\kappa))  \Bigr]_{\scriptsize
      \begin{aligned}
        p_0 &\mapsto \mu\,\Pi^+\\[-2mm]
        q_0 &\mapsto \mu/\Pi^+
      \end{aligned}}
  \,,
    \hspace{1cm}
\end{eqnarray}
where we have only used the fact that $s_1+s_2$ is an integer but note in addition that the term corresponding to the imaginary (real) part here clearly vanishes for even (odd) values of this parameter. Finally, the corresponding $d$-dimensional vacuum sunset integral is evaluated in eq.~\eqref{eq:final}.\footnote{\noveltext{Note that the ordering of the exponents $\{\alpha_1, \alpha_2, \alpha_3\}$ and the temporal momentum scales $\{p_0, q_0, q_0-p_0\}$ are tied to one another, so that e.g.~the replacement of $\alpha_3 \leftrightarrow \alpha_2$ and $q_0-p_0 \leftrightarrow q_0$ yields the spatial integral $S_{\alpha_1 \alpha_3 \alpha_2} (p_0, q_0-p_0, q_0)$ but does not change the value of the original $S_{\alpha_1 \alpha_2 \alpha_3} (p_0, q_0, q_0-p_0)$. The spatial integrals in eq.~\eqref{eq:resunset1} are written to make the collinearity of the scales explicit (e.g.~$p_0+(q_0-p_0) = q_0$) which allows the direct application of the results of~\cite{Davydychev:2022dcw}, further discussed in appendix \ref{sec:factorizationof2loop}.}}

For concreteness, let us next inspect the simplest nontrivial sunset $\mathcal{S}_{111}$. After directly evaluating the $d$-dimensional integral using~\eqref{eq:S111:pp:sol}, \noveltext{we observe that eq.~(\ref{eq:resunset1}) can be recast as a product} of two  one-loop integrals. The result reads
\begin{eqnarray}
\label{eq:primesunsetsol1}
\noveltext{{\bf D}_{p}{\bf D}_{q}}\,
  \mathcal{S}_{111}^{ }
  &\noveltext{=}&
  \oint_{p_0, q_0} \nFt'(p_0)\nFt'(q_0) S_{111}^{ } (p_0, p_0-q_0, q_0)
  \nn
  &\stackrel{T\to 0}{=}&
    \frac{2}{(2\pi)^2} \re\Bigl[S_{111}(\mu \Pi^+,\mu \Pi^+,0) \Bigr]
  - \frac{2}{(2\pi)^2} \re\Bigl[S_{111}(\mu \Pi^+,\mu / \Pi^+,0) \Bigr]
  \nn &=&
  - \frac{2(d-3)}{(d-2)} \mathcal{I}_2 (\mu) \mathcal{I}_2 (\mu)
    \,.
\end{eqnarray}
We note that in this expression, $S_{111} (\mu \Pi^+, \mu \Pi^+,0)$ resembles the standard result for the vacuum sunset with two real-valued mass scales~\cite{Vladimirov:1979zm}, continued to complex scales,\footnote{
The mass scales appearing in the propagators can be kept general in such a computation, with the only necessary assumption being that the squared mass scales are not strictly negative (i.e.\ that they are either complex-valued or positive and real). As shown in \cite{Gorda:2022yex}, this allows the necessary extraction of beta functions on the complex plane~\cite{Gorda:2022yex}.} while $S_{111} (\mu \Pi^+, \mu/ \Pi^+,0)$ needs to be evaluated with eq.~\eqref{eq:final}. \newtext{It is also worth pointing out that akin to eq.~\eqref{eq:primesunsetsol1}, all type \ref{typeB} terms originating from eq.~\eqref{eq:S:a123:s12} are seen to naturally factorize into one-loop master integrals.} For more details on these considerations and an independent calculation applying \noveltext{the factorization of vacuum bubble diagrams \cite{Davydychev:2022dcw}}, we refer the reader to appendix \ref{sec:vacuum:2l}.

\subsection{Integration-by-parts at two loops}
\label{sec:ibp2loopsec}

Armed with the knowledge of the explicit expressions of the maximally primed terms introduced above, we will next inspect the derivation of the $(d+1)$-dimensional IBP relations at the two-loop level. Here, we will again concentrate on the sunset integrals of eq.~\eqref{eq:S:a123:s12}, with our aim being to relate them to simpler structures. To achieve this, we find it convenient to study $\mathcal{S}_{111}$ in a slightly different manner from the strategy described in sec.~\ref{sec:IBPintro}, which involves an IBP relation \newtext{initially} containing integrals \newtext{of type \ref{typeC}, containing both differentiated and non-differentiated distribution functions.} While this leads to a simple factorizing result in this particular case, with more complicated propagator structures, we strongly recommend the use of IBP relations containing only terms of type \ref{typeA} and \ref{typeB}.

The first nontrivial IBP relation arises upon taking a diagonal total derivative along the temporal direction, which produces
\begin{align}
  0=
   \Bigl(\frac{\partial}{\partial p_0} \circ p_0 \Bigr)
   \mathcal{S}_{\alpha_1\alpha_2\alpha_3}^{s_1 s_2}
   & =
   (1+s_1)\mathcal{S}_{\alpha_1\alpha_2\alpha_3}^{s_1 s_2}
  +
   \oint_{P,Q}^f \frac{p_0^{s_1+1}q_0^{s_2}
   \,
    \nFt' (p_0)
    \nFt (q_0)}{
      [P^2]^{\alpha_1}
      [Q^2]^{\alpha_2}
      [(P-Q)^2]^{\alpha_3}
    }
  \nn[2mm] &
   -2
   \oint_{P,Q}^f  \biggl\{
     \frac{
        \alpha_{1}(P-Q)^2\,p_0^{2}
      + \alpha_{3}P^2\, p_0 (p_0-q_0)}{
        [P^2]^{\alpha_1+1}
        [Q^2]^{\alpha_2}
        [(P-Q)^2]^{\alpha_3+1}
      }
      \biggr\}
    p_0^{s_1}q_0^{s_2}
    \nFt (p_0)
    \nFt (q_0)
    \, .
\end{align}
Applying next a diagonal total derivative in the spatial direction \newtext{on the last term on the right-hand side, it} can be further recast into
\begin{align}
\label{eq:S111:2}
  -2\Bigl(
      \alpha_{1}\mathcal{S}_{\alpha_1+1,\alpha_2,\alpha_3}^{s_1+2,s_2}
    &+\alpha_{3}\mathcal{S}_{\alpha_1,\alpha_2,\alpha_3+1}^{s_1+2,s_2}
    + \alpha_{3}\mathcal{S}_{\alpha_1,\alpha_2,\alpha_3+1}^{s_1+1,s_2+1}
  \Bigr)
  \\[2mm] &=
   \Bigl[
     (d-2\alpha_{1} - \alpha_{3})
    -\frac{\partial}{\partial p_i} \circ p_i
    \Bigr]\mathcal{S}_{\alpha_1\alpha_2\alpha_3}
  + \alpha_{3}\Bigl(
    \mathcal{S}_{\alpha_1,\alpha_2-1,\alpha_3+1}^{s_1 s_2}
  - \mathcal{S}_{\alpha_1-1,\alpha_2,\alpha_3+1}^{s_1 s_2}
   \Bigr)
  \, ,\nonumber
\end{align}
where the spatial derivative term clearly vanishes.

Following the same strategy as for the diagonal temporal $p_0$-derivatives, all other purely temporal combinations from the bilinear set of eq.~\eqref{eq:ibp:set} generate a system of in total $n^2$ linear relations at the $n$-loop level.
Together with the $n^2$ spatial IBP relations \eqref{eq:ibp:11:spatial}--\eqref{eq:ibp:12:spatial} \newtext{in appendix \ref{app:spatialcheck}},
this gives rise to the following $(d+1)$-dimensional IBP relations for the two-loop fermionic sunset
\setcounter{dummy}{\value{equation}}
\renewcommand{\theequation}{ibp.2}
\begin{subequations}
\begin{align}
\label{eq:ibp:11}
\noveltext{
  (d+1-2\alpha_{1} - \alpha_{3} + s_1)
  + \alpha_{3} {\bf 3_{+}}({\bf 2_{-}} - {\bf 1_{-}})
  + {\bf 1^{+}}{\bf D}_{p}
  }
  &\equiv
  \noveltext{
  \Bigl(\frac{\partial}{\partial P_\mu}\circ P_\mu\Bigr)
  }
  \, ,
  \\
\label{eq:ibp:12}
\noveltext{
  (\alpha_{3} - \alpha_{1})
  + s_1 {\bf 1^{-}}{\bf 2^{+}}
  + \alpha_{1} {\bf 1_{+}}({\bf 3_{-}} - {\bf 2_{-}})
  + \alpha_{3} {\bf 3_{+}}({\bf 2_{-}} - {\bf 1_{-}})
  + {\bf 2^{+}}{\bf D}_{p}
  }
  &\equiv
  \noveltext{
  \Bigl(\frac{\partial}{\partial P_\mu}\circ Q_\mu\Bigr)
  }
  \,,
\end{align}%
\end{subequations}
\renewcommand{\theequation}{\arabic{section}.\arabic{equation}}%
\setcounter{equation}{\value{dummy}}%
which are written in a more compact operator form,
omitting $\mathcal{S}_{\alpha_1\alpha_2\alpha_3}^{s_1 s_2}$ onto which each operator acts.
\newtext{The two remaining bilinear combinations corresponding to $P \leftrightarrow Q$ can be found through substitutions
${\bf 1_{ }}\leftrightarrow {\bf 2_{ }}$.}

Here, we have employed standard IBP notation for the raising and lowering operators~\cite{Smirnov:2006ry} for the denominator and $\{p_0, q_0 \}$ numerator powers via
\begin{align}
\label{eq:n_pm}
  {\bf n_{\pm}^{ }}\, \mathcal{S}^{s_1 s_2}_{\alpha_1\dots \alpha_3}{} &=
  \mathcal{S}^{s_1 s_2}_{\dots \alpha_{n}\pm1\dots}{}
  \, ,
  \qquad
  \text{for}\;
  n = 1,\dots,3
  \, ,
  \\[2mm]
\label{eq:n^pm}
  {\bf n^{\pm}_{ }}\, \mathcal{S}^{s_1 s_2}_{\alpha_1\dots \alpha_3}{} &=
  \mathcal{S}^{\dots s_{n}\pm1\dots}_{\alpha_1\dots \alpha_3}{}
  \, ,
  \qquad
  \text{for}\;
  n = 1,\dots,2
  \, .
\end{align}

The relations~\eqref{eq:ibp:11}--\eqref{eq:ibp:12} represent the two-loop equivalent of the one-loop $(d+1)$-dimensional IBP relation~\eqref{eq:ibp:1l:d+1}. Independent of the number of loops, the system of IBP equations is always linear, underdetermined, and inhomogeneous, as the finite-$\mu$ relations are characterized by their right-hand sides containing the ${\bf D}_{k}$ differential operators.
While no general closed-form solution of such infinite systems of equations can be found with current methods, a typical approach to find a solution is to attempt solving for a finite number of master integrals~\cite{Smirnov:2010hn} using a finite set of starting integrals in the parameter space of $\{\alpha_1,\alpha_3,\alpha_3;s_1,s_2\}$ for $\mathcal{S}_{\alpha_1 \alpha_2 \alpha_3}^{s_1 s_2}$ and its $\nF'$-version. The solution is then found via Gaussian elimination implemented in a Laporta-type algorithm~\cite{Laporta:2000dsw}. While this type of a systematic approach, implemented using suitable programs for symbolic manipulation (e.g.\ {\tt FORM}~\cite{Ruijl:2017dtg}), typically quickly becomes indispensable at higher loop orders, below we approach the two-loop problem iteratively by hand.

For concreteness, let us now focus on the specific choice of indices $\alpha_1 = \alpha_2 = \alpha_3 = 1$,
$s_1 = s_2 = 0$, which constitutes the $\mathcal{S}_{111}$ integral. In this case, the IBP relation~\eqref{eq:ibp:11} gives rise to an aesthetically pleasing linear dependence between differentiated and non-differentiated integrals \newtext{after applying the $P \leftrightarrow Q$ symmetry between the loop momenta},
\begin{equation}
\label{eq:nFp:lhs}
    (d-2)\mathcal{S}_{111}
    =
  - \oint_{P,Q}^f \frac{ p_0
    \nFt' (p_0)
    \nFt (q_0)}{P^2 Q^2 (P-Q)^2 }
    \,.
\end{equation}
Next, we linearly combine IBP relations by operating on the above expression for $\mathcal{S}_{111}$ with a diagonal total derivative in both the $P_\mu$- and $(P_\mu-Q_\mu)$-directions. This results in the relation
\begin{align}
\label{eq:sunsetansatz}
  0 &= \Bigl[
    (d-2)\frac{\partial}{\partial P_\mu}\circ P_\mu
    - \frac{\partial}{\partial Q_\mu}\circ (Q_\mu - P_\mu)\,
      {\bf 1^{+}}{\bf D}_{p}
    \Bigr]
    \mathcal{S}_{111}
    \nn &=
    \Bigl[
        (d-2)^2
      + (d-2){\bf 3_{+}}({\bf 2_{-}} - {\bf 1_{-}})
      + {\bf 2_{+}}({\bf 3_{-}} - {\bf 1_{-}}){\bf 1^{+}}{\bf D}_{p}
      - {\bf 1^{+}}\bigl({\bf 2^{+}} - {\bf 1^{+}}\bigr){ \bf D}_{p}{\bf D}_{q}
    \Bigr]
    \mathcal{S}_{111}
    \,,
\end{align}
where the term
${\bf 3_{+}}({\bf 2_{-}} - {\bf 1_{-}})\mathcal{S}_{111} = \mathcal{S}_{102}-\mathcal{S}_{012}$ can be shown to cancel on the integrand level upon a relabeling of loop momenta. The integrals
$\mathcal{S}_{012}$ and $\mathcal{S}_{102}$ also vanish individually in the $T \to 0$ limit, since they factorize into a fermionic and a vanishing bosonic one-loop integral, as detailed in appendix~\ref{app:momentumshift}.

By explicitly writing the integrals in the above result, we arrive at the relation
\begin{align}
\label{eq:sunsetexplicitIBP}
  (d-2)^2 \mathcal{S}_{111}
  &=
  - \biggl[\oint_{P}^f \frac{p_0 \nFt' (p_0)}{P^2} \biggr]
    \biggl[\oint_{Q}^f \frac{\nFt (q_0)}{Q^4} \biggr]
  \nn
  &+
  \oint_{P,Q}^f
  \frac{p_0 \nFt'(p_0) \nFt (q_0)}{Q^4 (P-Q)^2}
  +
  \oint_{P,Q}^f
  \frac{p_0 (q_0-p_0) }{P^2 Q^2 (P-Q)^2}
  \nFt' (p_0)
  \nFt' (q_0)
  \, ,
\end{align}
where on the right-hand side  we identify a sum of three terms. Among them, the first integral clearly factorizes into two one-loop expressions, the second contains a shifted bosonic propagator with momentum $P-Q$, and the third is a linear combination of \newtext{type \ref{typeB} terms in eq.~\eqref{eq:S111typeBterms}} which vanishes in the $T\to 0$ limit. Upon closer inspection, detailed in appendix~\ref{app:momentumshift}, even the second integral can be shown to vanish as $T \to 0$, since it contains a $\mu$-independent loop integral.

By removing the vanishing integrals from eq.~\eqref{eq:sunsetexplicitIBP} and using the one-loop $(d+1)$-dimensional IBP relation \eqref{eq:ibp:1l:d+1}, the low-temperature limit of the sunset integral is seen to simplify to
\begin{align}
 \label{eq:ibp2loopexample}
      \mathcal{S}_{111}
      &\stackrel{T \to 0}{=}
      \frac{(d-1)}{(d-2)^2}\mathcal{I}_1 (\mu) \mathcal{I}_2 (\mu)
      \, .
\end{align}
This result is in agreement with computations of the low-temperature pressure of QCD (see e.g.~appendix~B of \cite{Vuorinen:2003fs}) and can moreover be computed directly using the cutting rules of~\cite{Ghisoiu:2016swa}, as we demonstrate in appendix~\ref{app:cuttingrules}. As far as we are aware, the factorization formula is, however, new. It is notably of different structure than the factorizing integrals of odd-valued numerator exponents in eq.~\eqref{eq:S111:T:cross}, which was derived using a set of spatial IBP operators listed in eqs.~\eqref{eq:ibp:11:spatial}--\eqref{eq:ibp:12:spatial} of appendix \ref{app:spatialcheck}.
As a consequence, we can relate the factorizing one-loop integrals with odd-valued numerator exponents to the even-valued ones. This is technically not a linear relation among one-loop integrals and curiously only gets generated at the two-loop level. At low temperatures, we find the relation
\begin{align}
\label{eq:S:a12:s12:odd:ibp}
    \mathcal{I}_{\alpha_1+1}^{2s_1+1}(\mu)\,
    \mathcal{I}_{\alpha_2}^{2s_2+1}(\mu)
    &=-
    \frac{(d-2\alpha_{1})}{2\alpha_{1}}
    \frac{
      (d + 1 - 2\alpha_{1} + 2s_{1})}{
      (d - 2\alpha_{1} + 2s_{1})}
    \frac{
      (d + 1 - 2\alpha_{2} + 2s_{2})}{
      (d + 2 - 2\alpha_{2} + 2s_{2})}
    \mathcal{I}_{\alpha_1}^{2s_1}(\mu)\,
    \mathcal{I}_{\alpha_2}^{2s_2}(\mu)
    \;,
\end{align}
which \newtext{represents the $T=0$ limit of a more complicated  relation valid at all $T$ and $\mu$. Note that in the opposing limit of $\mu=0$ but $T\neq 0$,} integrals with odd-valued numerator powers such as the left-hand side of~\eqref{eq:S:a12:s12:odd:ibp} vanish, \newtext{implying that the relation derived here will necessarily obtain corrections at nonzero $T$}.

While the above example led to a simple and aesthetically pleasing result, it was not perfectly in line with the strategy we outlined in sec.~\ref{sec:IBPintro}, where we emphasized the importance of the maximally primed  type~\ref{typeB} terms. A more systematic and more easily generalizable strategy in line with this idea is to consider exclusively differential operators that contain temporal derivatives with respect to all fermionic loop momenta. The vanishing of the total derivatives then leads to relations such as
\begin{align}
\label{eq:2loop1alternative}
    \Bigl(\frac{\partial}{\partial p_0} \circ p_0 \Bigr)
    \Bigl(\frac{\partial}{\partial q_0} \circ (p_0-q_0) \Bigr)\mathcal{S}_{111}
    &=
    \oint_{P,Q}
    \frac{\partial^2}{\partial p_0 \partial q_0}
    \biggl[
    \frac{p_0 (p_0-q_0)}{P^2 Q^2 (P-Q)^2}
    \biggr]
    \nFt(p_0) \nFt (q_0)
    \nn[2mm] &
    - \oint_{P,Q}
    \frac{p_0 (p_0-q_0)}{P^2 Q^2 (P-Q)^2}
    \nFt'(p_0) \nFt' (q_0)
    \;,
\end{align}
where only terms of type \ref{typeA} and \ref{typeB} are present.

To demonstrate the latter strategy in action, let us again consider the general fermionic sunset $S_{\alpha_1 \alpha_2 \alpha_3}$ with no numerators. While the full $(d+1)$-dimensional diagonal IBP relations~\eqref{eq:ibp:11}--\eqref{eq:ibp:12} already give rise to non-trivial relations, full IBP simplification is only possible via their combination with either the temporal or spatial IBP relations generated from the bilinear set~\eqref{eq:ibp:set}. By combining two consecutive diagonal $d+1$ and temporal derivatives, we indeed obtain
\begin{align}
\label{eq:2loopibp2first}
    0&=
    \Bigl(\frac{\partial}{\partial P_\mu} \circ P_\mu \Bigr)
    \Bigl(\frac{\partial}{\partial q_0} \circ q_0 \Bigr)
    \mathcal{S}_{\alpha_1\alpha_2\alpha_3}^{s_1-1,s_2-1}
    \nn[2mm] &
    = \oint_{P,Q}
    \frac{\partial^2}{\partial p_0 \partial q_0}
    \biggl[
      \frac{p_0^{s_1} q_0^{s_2}}{
      [P^2]^{\alpha_1}
      [Q^2]^{\alpha_2}
      [(P-Q)^2]^{\alpha_3}}
    \biggr]
    \nFt(p_0) \nFt (q_0)
    -
    {\bf D}_{p}
    {\bf D}_{q}\,
    \mathcal{S}_{\alpha_{1}\alpha_{2}\alpha_{3}}^{s_1 s_2}
    \,,
\end{align}
which is the only relation we obtain given its symmetry under the exchange of $P\leftrightarrow Q$. After using the relations~\eqref{eq:ibp:11} and the temporal version of \eqref{eq:ibp:12}, we carefully expand all derivatives on the left-hand side of this equation and again combine the corresponding spatial and temporal total derivatives following eq.~\eqref{eq:ibp:1l:d+1}. After multiple steps of straightforward algebra, the resulting expression \noveltext{\eqref{eq:2loopibp2first}} simplifies to a recurrence relation between various sunset integrals $\mathcal{S}^{s_1 s_2}_{\alpha_1\alpha_2\alpha_3}$ that can be dressed in the finite-$T$ form%
\footnote{\noveltext{This result arises by carefully removing explicit cancellations between terms arising from the initial total derivative.}}
\noveltext{
\begin{equation}
     \Bigl(\frac{\partial}{\partial P_\mu} \circ P_\mu \Bigr)
    \Bigl(\frac{\partial}{\partial q_0} \circ q_0 \Bigr)
    \mathcal{S}_{\alpha_1\alpha_2\alpha_3}^{s_1-1,s_2-1} = \noveltext{\mathcal{D}_{00} \, \mathcal{S}^{s_1 s_2}_{\alpha_1\alpha_2\alpha_3} } = 0\, ,
\end{equation}}
where 
\begin{align}
\label{eq:2loopibp:ops}
     \noveltext{\mathcal{D}_{00} } &\equiv 2 \alpha_{1} (\alpha_{2}-\alpha_{3}){\bf 1_{+}}
    + 2 \alpha_{2} (\alpha_{1}-\alpha_{3}){\bf 2_{+}}
    + 2 \alpha_{3} (d-1
                   -\alpha_{1} -\alpha_{2} -2\alpha_{3} + s_{1}) {\bf 3_{+}}
    \nn[1mm] &
    + 2 \alpha_{3}(
        \alpha_{1} {\bf 1_{+}} {\bf 2_{-}}
      + \alpha_{2} {\bf 2_{+}} {\bf 1_{-}}
      ) {\bf 3_{+}}
    - 2 \alpha_{1} \alpha_{2} {\bf 1_{+}} {\bf 2_{+}} {\bf 3_{-}}
    - 2 s_1 (\alpha_{2}{\bf 2_{+}} + \alpha_{3}{\bf 3_{+}}){\bf 1^{-}}{\bf 2^{+}}
    \nn[1mm] &
    + s_2 (d
      - 2\alpha_{1} - \alpha_{3} + s_{1}
      + \alpha_{3}({\bf 2_{-}}-{\bf 1_{-}}){\bf 3_{+}}
      ) {\bf 1^{-}}{\bf 2^{-}}
    -
    {\bf D}_{p}
    {\bf D}_{q}
    \,,
\end{align}
\noveltext{and we have} used the operator definitions of eqs.~\eqref{eq:Dp}, \eqref{eq:n_pm} and \eqref{eq:n^pm}. The right-hand side of this expression can be studied using the factorization formulae derived in appendix \ref{sec:vacuum:2l} or the collinear integrand formulae given in \cite{Davydychev:2022dcw} and their extensions discussed in \newtext{appendix}~\ref{sec:factorizationof2loop}.

As an interesting \noveltext{special case} of the above result, we note that by setting once again
$
\alpha_{1} =
\alpha_{2} =
\alpha_{3} = 1
$ and
$s_1=s_2=0$,
the identity reduces to
\begin{align}
\label{eq:S112temporalderivation}
    2(d-5) \mathcal{S}_{112}
    + 4 \mathcal{S}_{202}
    - 2 \mathcal{S}_{220}
    &=
    {\bf D}_{p}
    {\bf D}_{q}\,
    \mathcal{S}_{111}
    \, .
\end{align}
\noveltext{After removing the vanishing bosonic integral $\mathcal{S}_{202}$ and inserting the result of eq.~\eqref{eq:primecontoursimple}, the $T \to 0$ limit of eq.~\eqref{eq:S112temporalderivation} reads
\begin{equation}
\label{eq:s112check}
    2(d-5) \mathcal{S}_{112} \stackrel{T \to 0}{\rightarrow} 2 [\mathcal{I}_2 (\mu)]^2 -\frac{2(d-3)}{(d-2)} \left[\mathcal{I}_2(\mu) \right]^2\,,
\end{equation}
given again in terms of eq.~\eqref{eq:fermion1master_new}}. By independently studying the \noveltext{lengthy} spatial IBP relation \eqref{eq:ibp:2l:spatial}, \newtext{we find a factorization} into the one-loop integrals of eq.~\eqref{eq:I:1l} and recover a non-trivial result for general $\mu$ and $T$
\begin{align}
\label{eq:S112:T}
  \mathcal{S}_{112}
  &=
  \frac{1}{(d-2)(d-5)}\Bigl(
      \Bigl[\mathcal{I}_2 (\mu, T)\Bigr]^{2}
      -2 \mathcal{I}_2 (\mu, T) \mathcal{I}_2^b(T)
    \Bigr)
     \, ,
\end{align}
which also agrees with the $T \neq 0$, $\mu = 0$ result of~\cite{Laine:2019uua}\footnote{\noveltext{This somewhat surprising result can be seen to stem from the $d$-dimensional integral involved: according to \cite{Davydychev:2022dcw}, the massive two-loop integral $S_{\alpha_1 \alpha_2 \alpha_3 } (m_1, m_2, m_3)$ from  eq.~(\ref{eq:S:m123}) factorizes into a linear combination of products of two one-loop integrals $I_{\alpha_i} (m_i) I_{\alpha_j} (m_j)$ as long as the mass scales involved satisfy the ``collinearity'' condition $m_1+m_2 =m_3$ and $\alpha_k \in \mathbb{N}$. The implications of this property, which crucially holds even for complex-valued masses $m_k$, is further discussed in appendix \ref{sec:factorizationof2loop}.}} and wherein we \noveltext{use the superscript $b$ to} denote the bosonic finite-temperature integral \noveltext{$\mathcal{I}_\alpha^b (T)$.} \noveltext{It is also straightforward to take the $T \to 0$ limit from  eq.~\eqref{eq:S112:T}, giving}
\begin{align}
\label{eq:S112:T0:fac}
  \mathcal{S}_{112}
  \stackrel{T \to 0}{=}
  \frac{1}{(d-2)(d-5)} \Bigl[\mathcal{I}_2 (\mu)\Bigr]^{2}
    \, ,
\end{align}
\noveltext{which is in agreement with eq.~\eqref{eq:s112check}. We emphasize, though, that the derivation leading to the more general spatial IBP relation requires a lengthy ansatz with 10 bilinear operators [see eq.~\eqref{eq:ibp:2l:spatial}] as opposed to the more compact one in eq.~\eqref{eq:2loopibp2first}. We find this a very promising observation from the point of view of the future evaluation $T=0$ Feynman integrals at higher loop levels.}

%
\section{Conclusions and outlook}
\label{sec:concl}

As recently pointed out in \cite{Gorda:2023usm}, to improve from the present uncertainties in the model-agnostic determination of the neutron-star-matter equation of state \cite{Annala:2021gom,Gorda:2022jvk,Annala:2023cwx}, the thermodynamic properties of cold and dense quark matter need to be known at unprecedented precision. In the absence of non-perturbative tools, perturbation theory plays a pronounced role in efforts to reach this goal, with the most prominent individual challenge being the extension of the weak-coupling expansion of the pressure of cold and dense unpaired quark matter. While the current state of the art in this problem lies at an incomplete $\mathcal{O}(\alphas^3)$ order~\cite{Gorda:2021kme,Gorda:2021znl,Gorda:2022zyc}, all qualitative issues hindering the completion of this full order have been resolved by now, so that only one numerical coefficient is lacking. This is the contribution of the hard momentum scale $\mu_{\rmii{$B$}}$ to the $\mathcal{O}(\alphas^3)$ pressure, encoded in the sum of all four-loop vacuum, or bubble, diagrams of full QCD.

While conceptually a straightforward task, the practical evaluation of four-loop vacuum diagrams in a thermal setting presents an extremely complicated technical challenge. Only a handful of simple individual diagrams have been successfully completed by now~\cite{Gynther:2009qf,Gynther:2007bw,Navarrete:2022adz}, and they all correspond either to a scalar field theory or to very specific gauge-theory topologies that greatly simplify the evaluation of the diagrams. The particular case of dense fermionic matter in the $T=0$ limit can, however, be seen to exhibit some simplifications, including the vanishing of diagrams with no fermion loops and the presence of a convenient computational tool, the so-called cutting rules of~\cite{Ghisoiu:2016swa}. It is, however, evident that a successful evaluation of all four-loop vacuum diagrams in the theory necessitates automated tools of computation, both in the identification and eventual evaluation of the master integrals. \newtext{Given the complexity of the task at hand, even minor advances, perhaps leading to the successful evaluation of a handful of four-loop master integrals, would be extremely valuable.}

In the paper at hand, we have taken first \newtext{preliminary} steps towards generalizing one of the most powerful methods of vacuum ($T=\mu=0$) perturbation theory to the context of nonzero chemical potentials: integration-by-parts, or IBP, techniques. They provide linear relations between master integrals, which are typically derived using the vanishing of (loop) integrals over total derivatives and lead to a reduction in the overall number of masters. In the thermal context, a problem can be seen to arise from additional boundary contributions that are generated due to the breaking of Lorentz invariance and are related to the special nature of the temporal direction. Our novel idea, presented in this article, is to generate the full $(d+1)$-dimensional IBP identities starting from a contour integral formulation of thermal sums. This leads to the extra boundary terms being tractable in the small-$T$ limit due to the simple limiting behaviors of the bosonic and fermionic distribution functions and their derivatives. Another key element of our approach is the use of the temperature $T$ as a natural regulator for specific types of infrared divergences that are encountered in the strict $T=0$ limit.

As part of our new framework, we introduce two deformations of the usual complex contour of fermionic $p_0$-integrals: eq.~\eqref{eq:cauchy1} or fig.~\ref{fig:contoursplit} (left) for convergent line integrals involving non-differentiated distribution functions, and eq.~\eqref{eq:primecontoursimple} or fig.~\ref{fig:contoursplit} (right) for thermal corrections involving $\delta$-function limits. Their use requires the order of various integrations to be interchangeable and leads to many simplifications as demonstrated in secs.~\ref{sec:IBPintro}--\ref{sec:2-loop}. For the simplicity of presentation, the example calculations we present here are restricted to one- and two-loop calculations involving massless propagators. The proposed method is, however, more general, and we anticipate its true utility to manifest at the three- and four-loop orders. 

We also note that in practical computations, our approach can (and should) be complemented by purely spatial IBP relations, which have been studied in the past both in vacuum and in thermal systems lacking chemical potential. At the two-loop order in particular, the integrands of $(d+1)$-dimensional bubble diagrams with chemical potentials experience factorization in a manner similar to how $d$-dimensional real-valued vacuum bubbles can be expressed as linear combinations of products of their {\em collinear} mass scales, $m_1+m_2 = m_3$ (see \cite{Davydychev:2022dcw} and \newtext{appendix}~\ref{sec:factorizationof2loop}), with complex-valued temporal momenta now playing the role of masses. An explicit example of the application of spatial IBP relations is given in appendix~\ref{app:spatialcheck}, with emphasis on how relations derived at $\mu=0$ may differ from the $\mu \neq 0$ ones due to symmetries broken by a nonzero chemical potential. 

\newtext{Finally, we note that while the spatial operators of eq.~\eqref{eq:ibp:2l:spatial} can be used to derive closed expressions for $\mathcal{S}_{11s}$ for $s \in \{1,2\}$, their $(d+1)$-dimensional counterparts in eqs.~\eqref{eq:sunsetansatz} and \eqref{eq:2loopibp2first} are noticeably simpler. Furthermore, the explicit one-loop factorizations arising from temporal differentiations [eqs.~\eqref{eq:sunsetexplicitIBP}--\eqref{eq:ibp2loopexample} and \eqref{eq:S112temporalderivation}] differ from those obtained with spatial IBPs, cf.\ eqs.~\eqref{eq:S111:T:cross} and \eqref{eq:S112:T:cross}. These features imply that our new framework is capable of producing novel and useful results, a property that we expect to be particularly pronounced at higher loop orders. While formally not a closed group action with respect to differentiation, unlike the purely spatial differential operators, the temporal derivatives are seen to complement the existing IBP framework. }

\subsection{Future directions}

To conclude our presentation, let us make a few remarks on the potential extensions and applications of our main results, including some details of calculations at higher perturbative orders. So far, our discussion has relied on the presence of only one physical scale (the chemical potential $\mu$) in the integrals considered. While this special case often suffices for phenomenological applications in neutron-star physics, it should be noted that our generic method can be straightforwardly generalized to the presence of flavor-dependent chemical potentials, nonzero fermion (or boson) masses, and even to $n$-point functions, i.e.~the presence of external momenta. The complexity of the novel $\delta$-function integrals, however, rapidly increases with the number of scales present, which we demonstrate in appendix~\ref{app:twoscaleIBP} where the one-loop integrals of sec.~\ref{sec:1-loopformulae} are generalized to the case of nonzero masses. While this exercise and its two-loop extension in eq.~\eqref{eq:2loopt0mass} can still be completed in a fairly straightforward manner, introducing external momenta with non-vanishing temporal components or considering topologies more complicated than the sunset one would invalidate all the computational tools relying on collinearity.

At higher loop orders, practical calculations will inevitably rely on a large-scale automation of the IBP reduction algorithm. While we have not yet performed extensive studies in this direction, we expect an automated approach to  be fully compatible with our strategy to generate relations where all new thermal terms are of type~\ref{typeB}, i.e.~contain a maximal number of differentiated distribution functions that reduce to delta functions as $T\to 0$. While the calculations will surely be technically more demanding than in our treatment of the fermionic sunset integral in sec.~\ref{sec:2-loop}, we expect important simplifications to occur from the fact that the extra boundary terms have the form of complexified $d$-dimensional massive integrals that can be evaluated with the standard methods of vacuum field theory.

In an algorithmic implementation of the IBP approach, one faces an infinite, underdetermined system of inhomogeneous linear equations for the unknown integrals.%
\footnote{
 Note that this discussion assumes that all Lorentz, group, and Dirac algebra have been taken care of after the generation of Feynman diagrams and that the resulting integrals are scalarized.
}
In a practical approach, we are typically only interested in a finite number of master integrals, so instead of solving an infinite system at once, it is sufficient to only solve for a finite number of integrals that also constitute a finite linear system of unknowns. This system can in turn be solved via Gaussian elimination -- or a variant of the so-called Laporta algorithm~\cite{Laporta:2000dsw}. For the fermionic sunset considered in sec.~\ref{sec:2-loop}, this implies applying both the diagonal $(d+1)$-dimensional identities~\eqref{eq:ibp:11}--\eqref{eq:ibp:12}, the spatial identities~\eqref{eq:ibp:11:spatial}--\eqref{eq:ibp:12:spatial}, and their off-diagonal variations to a sufficiently large but finite set of integrals such that all initial Feynman diagrams are expressed in terms of the masters. The complexity of this process rapidly increases with the number of loops given that at the $n$-loop level, the IBP-system of vacuum integrals is generated by $2n^2$ diagonal and $2n^2$ off-diagonal identities.

As we want the master integrals to be as simple as possible, we \newtext{propose an ordering}
prescription that penalizes more complicated integrals during the reduction. In contrast to the standard Laporta algorithm~\cite{Laporta:2000dsw}, and basing on our treatment of the fermionic sunset $\mathcal{S}_{\alpha_1 \alpha_2 \alpha_3}^{s_1 s_2}$ (cf.\ eq.~\eqref{eq:S:a123:s12}), we propose the following ordering in a decreasing degree of complexity:
\begin{itemize}
  \item[(i)]
    lowest number of primed distribution functions,
  \item[(ii)]
    highest total power of temporal momenta in the numerator ($s_1+s_2$),
  \item[(iii)]
     highest total number of propagator powers in the denominator ($\alpha_1+\dots+\alpha_3$).
\end{itemize}
With this type of an ordering, it is conceivable to implement a generalized Laporta algorithm that respects the presence of the novel type~\ref{typeB} boundary terms at nonzero $T$ and $\mu$. A practical implementation of the algorithm is beyond the scope of this preparatory study, but we plan to tackle it in the near future as part of the evaluation of the four-loop vacuum diagrams of QCD in the limit of vanishing temperature and nonzero chemical potentials.

%
\section*{Acknowledgements}
The authors thank Saga Säppi and York Schr\"{o}der for enlightening conversations and comments on an early version of this work.
J{\"O} acknowledges financial support from the Vilho, Yrjö and Kalle Väisälä
Foundation of the Finnish Academy of Science and Letters.
J{\"O}, PS, and AV have been supported by
the Academy of Finland grant no.~1322507 as well as by
the European Research Council, grant no.~725369.
PS acknowledges support by the Deutsche Forschungsgemeinschaft (DFG, German Research Foundation) through
the CRC-TR 211 `Strong-interaction matter under extreme conditions' --
project number 315477589 --
TRR 211.

%
\appendix
\renewcommand{\thesection}{\Alph{section}}
\renewcommand{\thesubsection}{\Alph{section}.\arabic{subsection}}
\renewcommand{\theequation}{\Alph{section}.\arabic{equation}}

%
\section{Applying the Cauchy theorem to loop integrals}
\label{sec:Cauchy_appendix}

This appendix collects additional details on the constraints related to the Cauchy theorem as it appears in secs.~\ref{sec:formalism}--\ref{sec:2-loop}. We also further detail the application of the Cauchy theorem at the one-loop level.

\subsection{Single propagator}
\label{app:singleprop}

The simplest integral encountered at the one-loop level is given by eq.~\eqref{eq:fermion1master_new}. As an example, we first revisit this integral, working in the parameter region where it converges:
\begin{equation}
\label{eq:I:alpha:f}
  \mathcal{I}_{\alpha}(\mu) \equiv
  \lim_{T\to 0}
  \mathcal{I}_{\alpha}(\mu,T) =
  \lim_{T\to 0}
  \oint_{p_0}^f
  \int_{\vec{p}}
  \frac{\nFt(p_0)}{[p_0^2+p^2]^{\alpha}}
  < \infty
  \, ,
  \quad
  \text{for}
  \quad
  \Bigl\{
    \frac{1}{2} < \alpha < 1 \,\Big| \, 2\alpha -d > 1
  \Bigr\}
  \, .
\end{equation}

A subset of the above parametric conditions for convergence arises from both the UV behavior and novel $\mu$-specific divergences occurring at $p \sim \mu$ and $p_0 \to 0$. To track down the regularity of the integral, we apply the low-temperature step function limit as in eq.~\eqref{eq:nfbtheta} to only consider the upper line integral. After a change of variables to move to the real axis, we remove the regulator $\eta$, perform the $d$-dimensionally regularized integral, and then generalize the beta function on the complex plane following~\cite{Gorda:2022yex}. As a result, we obtain the intermediate regulated expression
\begin{align}
\label{eq:naive_intermediate}
    \mathcal{I}_{\alpha}(\mu) &=
    \int_{-\infty}^\infty
    \frac{{\rm d}p_0}{2\pi}
    I_{\alpha}(p_0+i\mu)
    =
    \biggl(\frac{e^\gammaE\bmu^2}{4\pi} \biggr)^\frac{3-d}{2}
    \frac{
      \Gamma\bigl(\alpha -\frac{d}{2} \bigr)}{(4\pi)^\frac{d}{2}
      \Gamma\bigl(\alpha\bigr)
      \Gamma\bigl(\frac{d}{2}\bigr)}
      \int_{-\infty}^\infty \frac{{\rm d}p_0}{2\pi}
    \left[(p_0+i \mu)^2 \right]^\frac{d-2\alpha}{2}
    \, .
\end{align}
On the last line, we have clearly separated the UV convergence condition $2 \alpha - d > 0$ arising from Euler's Gamma function and $d+1-2 \alpha < 0$ from the remaining integration.
Conversely, when starting with the $p_0$-integration, the UV convergence requires that
$\alpha > \frac{1}{2}$. The remaining parametric condition arises from eq.\eqref{eq:Iapole}, which occurs at $p = \mu$ and $p_0 \sim 0$  on the first line of eq.~\eqref{eq:naive_intermediate}. Let us reiterate the equation
\begin{equation}
    \int_{-\epsilon}^\epsilon \frac{{\rm d}p_0}{2 \pi} \frac{1}{[p_0 + i\mu + ip]^\alpha [p_0 + i\mu - ip]^\alpha} =
    \int_{-\epsilon}^\epsilon \frac{{\rm d}p_0}{2 \pi} \frac{1}{[p_0 + 2i\mu ]^\alpha [p_0]^\alpha}
    \, ,
\end{equation}
and note that it only converges for $\alpha < 1$.%
\footnote{
  We emphasize that the residue theorem yields {\em physically} correct results at $\alpha = 1$ in agreement with the analytic continuation from the results computed with the Cauchy theorem at $\alpha < 1$ \cite{Gorda:2022yex}. This property does not extend to $\alpha > 1$.
}
Combining these conditions, both the full integral in the region of eq.~\eqref{eq:I:alpha:f} and also the $p_0$-integral converge. The latter allows us to use the Cauchy theorem in the computations.

It is worth noting that it would be possible to apply the Cauchy theorem already for the original contour integral of eq.~\eqref{eq:I:alpha:f}. While this integral does not play a role in our computations involving $\delta$-function entities, it is quite demonstrative, and to this end, we next reverse the order of integration starting with the temporal integral. To start, we, however, need to again remove the lower line integral and replace the distribution function by unity for the upper one. Since everything apart from the above-mentioned occupation-function limit is exponentially suppressed by $e^{- \beta \eta}$ (for $\eta > 0$, see eq.~\eqref{eq:nfbtheta}), we can neglect subleading terms, giving the thermally leading integral the convergent structure
\begin{align}
\label{eq:1loopresfast1}
    \mathcal{I}_{\alpha}(\mu) &=
    \int_{\vec{p}} \left[
        \int_{-\infty + i \mu + i \eta}^{ i \mu + i \eta}
      + \int_{i \mu + i \eta}^{\infty + i \mu + i \eta}
    \right]
  \frac{{\rm d}p_0}{2\pi}
  \frac{1}{[p_0^2+p^2]^\alpha}
  + \mathcal{O} \bigl( e^{- \beta \eta}\bigr)
  \, ,
  \nn &=
    \int_{\vec{p}} \left[
          \int_{ -i \mu }^{ \infty -i \mu }
       +  \int_{i \mu }^{\infty + i \mu }
      \right] \frac{{\rm d}p_0}{2 \pi} \frac{1}{[p_0^2+p^2]^\alpha}
  \, .
\end{align}

Given that the above integral is convergent, the $\eta$-regulator in the line integral could be dropped, followed by changing $p_0 \mapsto - p_0$ in the first integral; this is the setup described in sec.~\ref{sec:formalism}. As the integrand is even in $p_0$, it does not visibly experience the sign change as described in eq.~\eqref{eq:cauchy1}. To find the most suitable alternative representation for both of line integrals, we close the contours along the imaginary and real axes as in fig.~\ref{fig:contoursplit} (left). Neglecting the finite segment at real positive infinity $\re(p_0) = \infty$ is possible due to the convergence for $\alpha > \frac{1}{2}$.
This way, we find the minimal integral expression
\begin{align}
\label{eq:1loopresfast2}
    \mathcal{I}_{\alpha}(\mu) =
     \int_{\vec{p}} \biggl[
        \int_{-\infty}^\infty
      - \int_0^{i\mu}
      - \int_{0}^{-i\mu}\biggr]
    \frac{{\rm d}p_0}{2 \pi}
    \frac{1}{[p_0^2+p^2]^\alpha}
    &=
    -2 \re
      \int_{\vec{p}}
      \int_0^{i\mu} \frac{{\rm d}p_0}{2\pi}
      \frac{1}{[p_0^2+p^2]^\alpha}
    \nn &=
    -2 \re
      \int_{\vec{p}} \theta(\mu-p)
      \int_0^{i\mu}
      \frac{{\rm d}p_0}{2 \pi}
      \frac{1}{[p_0^2+p^2]^\alpha}
    \, .
\end{align}
The real line integral vanishes in dimensional regularization, and we were able to simplify the second line since all values $p > \mu$ explicitly yield imaginary values for the integral inside the brackets. While the result is formally limited to a small subset of dimensions and exponents, we can analytically continue it to almost the full parameter space, as suggested in eq.~\eqref{eq:fermion1master_new}.

At the one-loop level also another class of master integrals can appear
depending if their temporal momentum numerator powers are
even ($\sigma = 0$) or
odd ($\sigma = 1$).
By combining the Cauchy theorem and starting with the spatial integral, we obtain
\begin{align}
\label{eq:I:alpha:s}
  \mathcal{I}_{\alpha}^{2s+\sigma}(\mu) &=
    \lim_{T\to 0}
    \oint_{p_0}^f
    \int_{\vec{p}}
    \frac{p_0^\sigma[p_0^2]^s\,\nFt(p_0)}{[p_0^2 + p^2]^\alpha}
    \nn &=
    \biggl( \frac{e^\gammaE\bmu^2}{4\pi} \biggr)^\frac{3-d}{2}
    \frac{
      \Gamma\bigl(\alpha-\frac{d}{2}\bigr)}{(4\pi)^\frac{d}{2}
      \Gamma\bigl(\alpha\bigr)}
    \biggl[
      -\int_{0}^{i\mu}
      -\int_{0}^{-i\mu}
    \biggr]
      \frac{{\rm d} p_0}{2\pi}\,
      p_0^\sigma [p_0^2]^{\frac{d}{2}-\alpha + s}
    \nn &=
    -i^\sigma
    \biggl( \frac{e^\gammaE\bmu^2}{4\pi} \biggr)^\frac{3-d}{2}
    \frac{
      \Gamma\bigl(\alpha - \frac{d}{2}\bigr)
      }{ \pi (4\pi)^\frac{d}{2}
      \Gamma\bigl(\alpha\bigr)
      }
    \frac{\mu^{d+1-2 \alpha + 2s +\sigma}}{\bigl(d+1-2 \alpha + 2s +\sigma\bigr)}
    \sin\Bigl[\frac{\pi}{2}(d+2-2\alpha+2s)\Bigr]
    \\ &=
    -i^\sigma
    \biggl( \frac{e^\gammaE\bmu^2}{4\pi} \biggr)^\frac{3-d}{2}
    \frac{
      \Gamma\bigl(\alpha - \frac{d}{2}\bigr)
      }{ (4\pi)^\frac{d}{2}
      \Gamma\bigl(\alpha\bigr)
      \Gamma\bigl(\alpha -s - \frac{d}{2}\bigr)
      \Gamma\bigl(\frac{d}{2}+1-\alpha +s \bigr)}
    \frac{\mu^{d+1-2 \alpha + 2s +\sigma}}{\bigl(d+1-2 \alpha + 2s + \sigma\bigr)}
  \;,\nonumber
\end{align}
where $\mathcal{I}^{0}_{\alpha}(\mu) \equiv \mathcal{I}^{ }_{\alpha}(\mu)$.

\subsection{Multiple propagators}
\label{app:multiprop}

As discussed in the main text, the Cauchy theorem approach generalizes to integrals with multiple propagators and loops. In this section, we indeed extract convergence constraints for multi-loop and multi-propagator structures similar to the one-loop case in eq.~\eqref{eq:I:alpha:f}. To this end, we consider an integral with a $(d+1)$-dimensional external momentum $K= (k_0,\vec{k})$ with a complex-valued temporal component, relevant to thermal field theory calculations:
\begin{align}
  \TopoSBtxt(\Lqq,\Agl,\Aqu,,,2,1) =
  \mathcal{I}_{\alpha_{1}\alpha_{2}}(\mu,T) &=
    \oint_{P}^f
    \frac{\nFt(p_0)}{
      [p_0^2+p^2]^{\alpha_{1}}
    }
    \frac{1}{
      [(p_0-k_0)^2+|\vec{p}-\vec{k}|^2]^{\alpha_{2}}
    }
    \, .
\end{align}
The novel problematic divergence should now only occur at $|\vec{p}-\vec{k}|^2 \sim [\mu-\im(k_0)]^2$ for $\alpha_{2} \geq 1$. Therefore, we can even deal with multiple propagators and analytically continue the result by applying the exponent regulation of appendix~\ref{app:singleprop}.

Each loop order introduces one additional contour integral, akin to the two-loop level discussion in sec.~\ref{sec:2-loop}. As at the one-loop level, we study the convergence properties of eq.~\eqref{eq:S:a123:s12}, and write the strict (naive) low-temperature limits in terms of the upper line integrals of the contours:
\begin{align}
\label{eq:S:2l:T0}
&   \lim_{T \to 0} \oint_{p_0, q_0}^f
    \frac{\nFt(p_0)}{
      [p_0^2+p^2]^{\alpha_{1}}
    }
    \frac{\nFt(q_0)}{
      [q_0^2+q^2]^{\alpha_{2}}
    }
    \frac{1}{
      [(p_0-q_0)^2+|\vec{p}-\vec{q}|^2]^{\alpha_{3}}
    }
    \nn
    &=\int_{-\infty + i \mu}^{\infty + i \mu}
    \frac{{\rm d}p_0}{(2\pi)}
    \frac{{\rm d}q_0}{(2\pi)}
    \frac{1}{
      [p_0^2+p^2]^{\alpha_{1}}
      [q_0^2+q^2]^{\alpha_{2}}
      [(p_0-q_0)^2+|\vec{p}-\vec{q}|^2]^{\alpha_{3}}}
    \, .
\end{align}
The first two (fermionic) propagators are similar in their behavior when $p = \mu$ for
$\alpha_{1} \geq 1$ and $q = \mu$ for $\alpha_{2} \geq 1$. The third (bosonic) propagator on the other hand diverges in the hyperplane $|\vec{p}-\vec{q}| = 0$ for $\alpha_{3}\geq 1$, similar to the other two cases but shifted from the position near the origin by the length of $p_0$ or $q_0$ depending on the chosen integration variable. Again, by neglecting these divergences, not only the integral becomes dependent on the integration order but also
the dimensionally regularized loop momenta. Thus, the conditions
$\alpha_{1},\alpha_{2},\alpha_{3}< 1$ are necessary (but not sufficient) to be able to apply
the Cauchy theorem. Further conditions are needed to treat the UV behavior of both zero-components of the $(d+1)$-loop momenta. For this purpose, we use the Feynman parametrization
\begin{equation}
\label{eq:Feynmanparamet}
    \frac{1}{A_1^{\alpha_1} A_2^{\alpha_2}} =
    \frac{\Gamma(\alpha_1+\alpha_2)}{\Gamma(\alpha_1)\Gamma(\alpha_2)}
    \int_0^1 {\rm d}z
    \frac{z^{\alpha_1-1}(1-z)^{\alpha_2-1}}{\left[z A_1 +(1-z)A_2 \right]^{\alpha_1+\alpha_2}}
    \, ,
\end{equation}
which is valid for $\alpha_1, \alpha_2 >0$. Parametrizing the second and third propagator in this fashion yields
\begin{align}
  \eqref{eq:S:2l:T0} &=
   \frac{
     \Gamma\bigl(\alpha_{23})}{
     \Gamma\bigl(\alpha_{2})
     \Gamma\bigl(\alpha_{3})}
   \int_0^1 {\rm d}x\, x^{\alpha_{2}-1} (1-x)^{\alpha_{3}-1}
   \\
   &\hspace{1cm}\times
   \int_{-\infty + i \mu}^{\infty + i \mu}
   \frac{{\rm d}p_0}{(2\pi)}
   \frac{{\rm d}q_0}{(2\pi)}
   \frac{1}{
     [p_0^2+p^2]^{\alpha_{1}}
     [(q_0-x p_0)^2 + x(1-x)p_0^2 + x|\vec{p}-\vec{q}|^2 + (1-x)q^2]^{\alpha_{23}}}
   \, ,\nonumber
\end{align}
where we apply the compact notation $\alpha_{\{i\}} = \sum_{j\in \{i\}} \alpha_{j}$.

The UV behavior of the integral is contained in the $q_0$-integral if $\frac{1}{2}-\alpha_{23}< 0$. To consider this behavior explicitly, we change the integration variables
according to $u(q_0) = q_0-x p_0$ for fixed values of $x$ and $p_0$, and focus on the innermost integral
\begin{equation}
    \int_{-\infty+(1-x) i \mu}^{\infty + (1-x) i \mu}
    \frac{{\rm d}u}{\left[u^2+x(1-x)p_0^2+ x|\vec{p}-\vec{q}|^2+(1-x) q^2 \right]^{\alpha_{2 3}}}
    \, .
\end{equation}
The UV behavior of this expression can be studied easier by moving the line integral to run along the real axis which gives rise to a beta function as in our Cauchy theorem prescription.
Hence, the computation reduces to evaluating
\begin{align}
  2\int_{0}^{\infty } &
  \frac{{\rm d}u}{\left[u^2+x(1-x)p_0^2 +x|\vec{p}-\vec{q}|^2+(1-x) q^2 \right]^{\alpha_{23}}}
  \nn &=
    \frac{
      \Gamma\bigl(\alpha_{23}-\frac{1}{2}\bigr)
      \Gamma\bigl(\frac{1}{2} \bigr)}{
      \Gamma\bigl(\alpha_{23}\bigr)
    }
  \left[ x(1-x)p_0^2 +x|\vec{p}-\vec{q}|^2+(1-x) q^2 \right]^{\frac{1}{2}-\alpha_{23}}
    \, ,
\end{align}
where we used a complex-valued generalization of Euler's beta function to find the given representation (see e.g.\ sec.~III of \cite{Gorda:2022yex}). 

Extracting all of the elements contributing to the remaining $p_0$-integration above, we need to consider another Feynman parametrization:
\begin{align}
    \int_{-\infty+i \mu}^{\infty + i \mu}
    \frac{{\rm d}p_0}{[p_0^2+p^2]^{\alpha_{1}}}
    \frac{1}{\bigl[p_0^2+\frac{1}{1-x}\left|\vec{p}-\vec{q} \right|^2+\frac{q^2}{x}\bigr]^{\alpha_{23}-\frac{1}{2}}}
    &=
    \frac{
      \Gamma\bigl(\alpha_{123}-\frac{1}{2} \bigr)}{
      \Gamma\bigl(\alpha_{1}\bigr)
      \Gamma\bigl(\alpha_{23}-\frac{1}{2} \bigr)}
    \int_0^1 {\rm d}y\,y^{\alpha_{23}-\frac{3}{2}}(1-y)^{\alpha_{1}-1}
    \nn[2mm] &\hspace{-2.5cm}\times
    \int_{-\infty+i \mu}^{\infty + i \mu}
    \frac{{\rm d}p_0}{\bigl[p_0^2+(1-y) p^2+\frac{y}{1-x}|\vec{p}-\vec{q}|^2+\frac{y q^2}{x} \bigr]^{\alpha_{123}-\frac{1}{2}}}
    \, .
\end{align}
The UV behavior of this integral can again be extracted after a reflection of the $p_0$-integral to the real axis and the evaluation of the corresponding beta function. The resulting hypergeometric integral structure
\begin{align}
  \eqref{eq:S:2l:T0} \mapsto
 \frac{
   \Gamma\bigl(\alpha_{123} -1\bigr)
   \Gamma\bigl(\frac{1}{2} \bigr)^2 }{
   \Gamma\bigl(\alpha_1\bigr)
   \Gamma\bigl(\alpha_2\bigr)
   \Gamma\bigl(\alpha_3\bigr)
   \Gamma\bigl(\alpha_{23}\bigr)}
  &
  \int_0^1
  {\rm d}x\,
  {\rm d}y\,
  x^{\alpha_{12}-\frac{3}{2}}
  (1-x)^{\alpha_{13}-\frac{3}{2}}
  y^{\alpha_{23}-\frac{3}{2}}
  (1-y)^{\alpha_{1}-1}
  \nn[2mm] \times &
  \left[x(1-x) (1-y) p^2+ xy |\vec{p}-\vec{q}|^2+(1-x)y q^2 \right]^{1-\alpha_{123}}
  \,,
\end{align}
is well-defined for $\alpha_{123}-1 > 0$, $\alpha_{i < j} > \frac{1}{2}$ and $\alpha_i > 0$ for $\{i,j\} \in \{1,2,3\}$. These constraints are complemented by the above-mentioned conditions $\alpha_i < 1$, associated with the IR behavior of the propagator structure.

Applying total derivatives to any full expression such as eq.~\eqref{eq:S:a123:s12} can organically generate different propagator powers. However, differentiation can not be successfully applied directly to formulae, in which the zero-temperature limit has been previously taken such that the distribution function is fully removed; see eq.~\eqref{eq:1loopresfast2}. Indeed, differentiated distribution functions generate essential corrections arising from the finite-temperature regulation as implied in
eq.~\eqref{eq:nfb_complexdelta}. In the presence of such $\delta$-function limits, the Cauchy theorem does not need to be applied to find a closed-form result. A more convenient prescription involves a clever splitting of the original contour in two (see sec.~\ref{sec:formalism}) and the completion of the two line integral halves. This approach is faithful to the original conditions of the fermionic sum-integral and allows a simple substitution procedure to replace highly non-trivial complex contour computations.

%
\section{One-loop master integral with a mass scale }
\label{sec:massl}
In~\cite{Gorda:2022yex}, two methods were introduced for approaching the computation of
the fermion one-loop master integral involving a single imaginary scale $i \mu$. The integration order, where one starts from the spatial integral, excels in directly treating the (potential) $p \simeq \mu$ divergence for general $\alpha  > 1$ through a generalization of Euler's beta function. Adding another (mass) scale to the propagator would merely shift the location of the  (potential) novel divergence while other properties stay similar, so that in the $T=0$ limit, the massive variant
\begin{equation}
\label{eq:massfiniteT}
  \mathcal{I}_{\alpha}^{s} (\mu,m,T) \equiv
    \oint_{P}^f
    \frac{p_0^s\, \nFt(p_0)}{[p_0^2+p^2+m^2]^\alpha}
    \, ,
\end{equation}
where
$\mathcal{I}_{\alpha}^{ } (\mu,m,T) \equiv \mathcal{I}_{\alpha}^{0} (\mu,m,T)$.
The massive integral has the same convergence properties as the massless integral, with convergence achieved in the triangle-shaped subregion of eq.~\eqref{eq:I:alpha:f}. In the opposing case of $m > \mu$, we on the other hand automatically avoid not only the $p \simeq \mu$ divergence, but also any $\mu$ dependence on the result as will be explicitly demonstrated below.

In this appendix, we explicitly compute the low-temperature limit of eq.~\eqref{eq:massfiniteT} using the regulatory methods of~\cite{Gorda:2022yex} as well as hypergeometric algebra. Subsequently, we construct a generalization of the IBP relation of eq.~\eqref{eq:1loopdeltaformula1} in the low-temperature limit, including the additional mass scale.

\subsection{Direct computation}

In analogy with the zero-mass case of our eq.~\eqref{eq:prime1d} (see also~\cite{Gorda:2022yex}), computing first the $d$-dimensional dimensionally regularized spatial integral in eq.~\eqref{eq:massfiniteT} yields
\begin{eqnarray}
\label{eq:mass1st}
    \mathcal{I}_\alpha (\mu,m, T) &=&
    \biggl( \frac{e^\gammaE\bmu^2}{4\pi} \biggr)^\frac{3-d}{2}
    \frac{\Gamma \left(\alpha-\frac{d}{2}\right)}{(4\pi)^\frac{d}{2} \Gamma(\alpha)}
    \oint_{p_0}^f
    \nFt(p_0)
    \left[p_0^2 + m^2\right]^\frac{d-2\alpha}{2}
    \nn
    &\stackrel{T\to 0}{=}&
    \biggl( \frac{e^\gammaE\bmu^2}{4\pi} \biggr)^\frac{3-d}{2}
    \frac{
      \Gamma\bigl(\alpha-\frac{d}{2}\bigr)}{\pi (4\pi)^\frac{d}{2}
      \Gamma\bigl(\alpha\bigr)}
    \re \int_0^\infty {\rm d}p_0 \left[(p_0+i\mu)^2+m^2 \right]^\frac{d-2\alpha}{2}
    \nn
    &=&
    \biggl( \frac{e^\gammaE\bmu^2}{4\pi} \biggr)^\frac{3-d}{2}
    \frac{
      \Gamma\bigl(\alpha-\frac{d}{2}\bigr)}{
      \Gamma\bigl(\alpha\bigr)}
    \frac{\mu^{d+1-2\alpha}}{\pi(4\pi)^\frac{d}{2}}
    \re \int_0^\infty {\rm d}p_0 \biggl[(p_0+i)^2+\Bigl(\frac{m}{\mu} \Bigr)^2 \biggr]^\frac{d-2\alpha}{2}
    \, ,
    \hspace{1cm}
\end{eqnarray}
wherein we simultaneously scaled out the chemical potential and took the low-temperature limit by discarding the lower part of the contour as it is exponentially suppressed. The remaining integral is most straightforwardly evaluated using the Cauchy theorem, assuming the usual convergence region of eq.~\eqref{eq:I:alpha:f}. To this end, we apply eq.~\eqref{eq:cauchy1} and write
\begin{align}
\label{eq:mass2nd}
    \int_0^\infty {\rm d}p_0 \biggl[(p_0+i)^2+\Bigl(\frac{m}{\mu} \Bigr)^2 \biggr]^\frac{d-2\alpha}{2}
    &=
    \biggl[
        \int_0^\infty
  {\rm d}p_0
      - \int_0^i
  {\rm d}p_0
    \biggr]
    \biggl[p_0^2+\Bigl(\frac{m}{\mu} \Bigr)^2 \biggr]^\frac{d-2\alpha}{2}
    \\[2mm] &=
    \Bigl( \frac{m^2}{\mu^2} \Bigr)^{\frac{d+1-2\alpha}{2}}
    \frac{
        \Gamma\bigl(\frac{3}{2} \bigr)
        \Gamma\bigl(\alpha-\frac{d+1}{2} \bigr)
      }{
        \Gamma\bigl(\alpha-\frac{d}{2} \bigr)}
    - \frac{i}{2} \int_0^1 {\rm d}x\, x^{-\frac{1}{2}}
    \biggl[ \Bigl( \frac{m}{\mu} \Bigr)^2-x \biggr]^{\frac{d-2\alpha}{2}}
    \, .\nonumber
\end{align}

First inspecting the case $m > \mu$, we note that the last term above is purely imaginary. Inserting the real part of the remaining expression into eq.~\eqref{eq:mass1st} yields then
\begin{equation}
\label{eq:massgrtmu}
   \mathcal{I}_\alpha (m > \mu) =
   \int_{P} \frac{1}{[P^2 + m^2]^\alpha} =
   \biggl( \frac{e^\gammaE\bmu^2}{4\pi} \biggr)^\frac{3-d}{2}
   \frac{
    \Gamma\bigl(\alpha-\frac{d+1}{2}\bigr)}{
    \Gamma\bigl(\alpha\bigr)}
   \frac{[m^2]^{\frac{d+1-2\alpha}{2}}}{(4\pi)^\frac{d+1}{2}}
   \, ,
\end{equation}
which agrees with
the massive vacuum integral $I_{\alpha}(m)$ from eq.~\eqref{eq:I:m},
in $(d+1)$ dimensions, with no chemical potential dependence present.

For $m < \mu$, we proceed somewhat differently by first simplifying the last term in
eq.~\eqref{eq:mass2nd} according to
\begin{align}
    \frac{i}{2} \int_0^1 {\rm d}x\, x^{-\frac{1}{2}}
    \biggl[ \Bigl( \frac{m}{\mu} \Bigr)^2-x \biggr]^{\frac{d-2\alpha}{2}} &=
    i \Bigl( \frac{m^2}{\mu^2} \Bigr)^\frac{d-2 \alpha}{2}
    {}_2 F_1 \biggl[\frac{1}{2}, \alpha-\frac{d}{2}; \frac{3}{2}; \frac{\mu^2}{m^2} \biggr]
    \nn &=
    i\Bigl(\frac{m^2}{\mu^2}-1 \Bigr)^\frac{d-2\alpha}{2}
    {}_2 F_1 \biggl[1, \alpha-\frac{d}{2}; \frac{3}{2}; \frac{\mu^2}{\mu^2-m^2}\biggr]
   \, ,
\end{align}
where on the last line we applied the so-called Pfaff transformation of~\cite{abramowitz1964handbook}.%
\footnote{
  This transformation~\cite{abramowitz1964handbook} reads
  $
  {}_2 F_1 \left[a,b;c;z \right] = (1-z)^{-b}{}_2 F_1 \bigl[c-a,b;c;\frac{z}{z-1} \bigr]
  $.
}
The motivation for this transformation stems from the series expansion of the hypergeometric function ${}_2 F_1 \left[a,b;c;z \right]$ only converging when its last argument satisfies $z < 1$.
To obtain an expression satisfying this constraint, we further apply
the transformation~\cite{Barnes1908,Bailey1935GeneralizedHS,abramowitz1964handbook}
\begin{align}
     {}_2 F_1 \left[a,b;c;z \right] &=
     \frac{
       \Gamma(c)
       \Gamma(-a-b+c)}{
       \Gamma(c-a)
       \Gamma(c-b)}
     {}_2 F_1 \left[a,b;a+b-c+1;1-z \right]
     \nn
     &+\frac{
       \Gamma(c)
       \Gamma(a+b-c)}{
       \Gamma(a)
       \Gamma(b)}(1-z)^{c-a-b}
     {}_2 F_1 \left[c-a,c-b;-a-b+c+1;1-z \right]
     \, ,
\end{align}
which allows us to write
\begin{align}
  \frac{i}{2} \int_0^1 {\rm d}x\, x^{-\frac{1}{2}}
  \biggl[ \Bigl( \frac{m}{\mu} \Bigr)^2-x \biggr]^{\frac{d-2\alpha}{2}}
    &=
    \frac{i}{d+1-2\alpha}
    \Bigl(\frac{m^2}{\mu^2}-1 \Bigr)^\frac{d-2\alpha}{2}
    {}_2 F_1 \left[1,\alpha-\frac{d}{2}; \alpha-\frac{d-1}{2}; -\frac{m^2}{\mu^2-m^2} \right]
  \nn[1mm] &
  + i\Bigl( \frac{m^2}{\mu^2} \Bigr)^\frac{d-2\alpha}{2}
  \Bigl(\frac{m^2}{m^2-\mu^2} \Bigr)^{\frac{1}{2}}
  \Bigl(\frac{\mu^2-m^2}{\mu^2}\Bigr)^{\frac{1}{2}}
  \frac{
    \Gamma\bigl(\frac{3}{2} \bigr)
    \Gamma\bigl(\alpha-\frac{d+1}{2} \bigr)
  }{
    \Gamma\bigl(\alpha-\frac{d}{2} \bigr)}
  \\[2mm]
  &\mapsto
  \frac{\left( \Pi^+ \right)^{d+1-2\alpha}}{d+1-2\alpha}
  \Bigl(1 - \frac{m^2}{\mu^2} \Bigr)^\frac{d-2\alpha}{2}
  {}_2 F_1 \left[1,\alpha-\frac{d}{2}; \alpha-\frac{d-1}{2}; -\frac{m^2}{\mu^2-m^2} \right]
  \nn &
  + \left(\Pi^+ \right)\left(\Pi^+ \right)^{-1}
    \Bigl( \frac{m^2}{\mu^2} \Bigr)^\frac{d+1-2\alpha}{2}
    \frac{
      \Gamma\bigl(\frac{3}{2} \bigr)
      \Gamma\bigl(\alpha-\frac{d+1}{2} \bigr)}{
      \Gamma\bigl(\alpha-\frac{d}{2} \bigr)}
  \, .
\end{align}
In the second iteration,
we regulated the complex-valued power functions
using $\Pi^+ \equiv \exp \left[\frac{i\pi}{2}(1+ \kappa) \right]$ as defined in eq.~\eqref{eq:i:reg}.

Taking now the real part of the above expression and the $\kappa$-regulator to vanish,
we can write for $m < \mu$
\begin{align}
  \re \biggl\{ \frac{i}{2} \int_0^1 {\rm d}x\, x^{-\frac{1}{2}}
      \biggl[ \Bigl( \frac{m}{\mu} \Bigr)^2-x \biggr]^{\frac{d-2\alpha}{2}} \biggr\}
  &=
  \frac{\cos \left[ \frac{\pi(d+1-2\alpha)}{2} \right]}{(d+1-2\alpha)}
  \nn &\times
  \Bigl(1-\frac{m^2}{\mu^2} \Bigr)^\frac{d-2\alpha}{2}
  {}_2 F_1 \left[1,\alpha-\frac{d}{2}; \alpha-\frac{d-1}{2}; -\frac{m^2}{\mu^2-m^2} \right]
  \nn[1mm]
  &+ \Bigl( \frac{m^2}{\mu^2} \Bigr)^\frac{d+1-2\alpha}{2}
  \frac{
      \Gamma\bigl(\frac{3}{2} \bigr)
      \Gamma\bigl(\alpha-\frac{d+1}{2} \bigr)
    }{
      \Gamma\bigl(\alpha-\frac{d}{2} \bigr)}
  \, ,
\end{align}
where we extracted the real part via $\re (\Pi^+)^s \sim \cos (\pi(1+\kappa) s/2)$ for
$s \in \mathbb{R}$. As a final step, we use Euler's reflection formula on the trigonometric function and combine this expression with eqs.~\eqref{eq:mass1st}--\eqref{eq:massgrtmu}, which upon further simplifying the result with eq.~\eqref{eq:fermion1master_new} leads to
\begin{equation}
\label{eq:massformulasimple}
  \mathcal{I}_\alpha(m < \mu) =
  \mathcal{I}_\alpha(\mu)
  \Bigl(1 - \frac{m^2}{\mu^2} \Bigr)^\frac{d-2\alpha}{2}
  {}_2 F_1 \left[1,\alpha-\frac{d}{2}; \alpha-\frac{d-1}{2}; -\frac{m^2}{\mu^2-m^2} \right]
  \, .
\end{equation}
This expression allows studying the $m \ll \mu$ regime in detail, and in particular, leads us to verify that the $m \to 0$ limit indeed yields eq.~\eqref{eq:fermion1master_new} and that the result for $\mathcal{I}_{\alpha}(\mu,m,T)$ is continuous in the limit $m \to \mu^{-}$.
The latter follows from the result
\begin{equation}
\label{eq:limitcheck}
    \Bigl( 1 - \frac{m^2}{\mu^2} \Bigr)^\frac{d-2\alpha}{2}
    {}_2 F_1 \left[1,\alpha-\frac{d}{2}; \alpha-\frac{d-1}{2}; -\frac{m^2}{\mu^2-m^2}\right]
    \stackrel{m \rightarrow \mu^{-}}{\longrightarrow}
    \frac{
      \Gamma\bigl(\frac{d}{2} + 1 -\alpha \bigr)
      \Gamma\bigl(\alpha -\frac{d-1}{2} \bigr)}{
      \Gamma\bigl(\frac{1}{2}\bigr)}
    \, ,
\end{equation}
provided convergence is enforced by the parameter constraints in eq.~\eqref{eq:I:alpha:f}.
In combination with eq.~\eqref{eq:massformulasimple},
this shows the continuous limit
$\mathcal{I}_\alpha(m < \mu)  = \lim_{m \to\mu^{-}} \mathcal{I}_\alpha (m > \mu)$.

The above result has a logical interpretation even outside of the parametric values, for which the limit $m \to\mu$ remains formally convergent. Denoting $\chi = \mu^2-m^2$, we recognize that eq.~\eqref{eq:limitcheck} is the $\mathcal{O}(\chi^0)$ term of a series expansion around $\chi = 0$, with  all other terms being of order  $\mathcal{O} \bigl(\chi^k \bigr)$ or $\mathcal{O} \bigl(\chi^{\frac{d-2\alpha}{2}+k}\bigr)$ with $k \in \mathbb{Z}_+$.
Whenever $\mathcal{I}_\alpha (m > \mu)$ is UV-convergent, both subsets only contain positive powers, and similarly to our analytical continuation of the Euler gamma functions in $\mathcal{I}_\alpha (m > \mu)$, we may regulate both the dimension and denominator exponent parameters to remove all terms
except for the one shown in eq.~\eqref{eq:limitcheck}.

\subsection{Integration-by-parts with two scales}
\label{app:twoscaleIBP}

Integration-by-parts relations can also be derived for massive integrals as in eq.~\eqref{eq:massfiniteT}. Given the full solution in eq.~\eqref{eq:massformulasimple},
we expect to  find inhomogeneous equations for master integral structures for $m < \mu$. For eq.~\eqref{eq:massfiniteT}, the IBP approach aligns with the discussion of sec.~\ref{sec:1-loopformulae}.
By acting on the massive master integral with
diagonal spatial and temporal total derivatives, we find
\begin{subequations}
\begin{align}
\label{eq:ibp:1l:m:spatial}
    0 &=
    \Bigl(\frac{\partial}{\partial p_i} \circ p_i\Bigr)
    \mathcal{I}_\alpha^{s} (\mu,m,T)
    \nn[2mm]
    & =
    (d-2 \alpha)
    \mathcal{I}_\alpha^{s} (\mu,m,T)
    +2 \alpha m^2 \mathcal{I}_{\alpha+1}^{s} (\mu, m, T)
    +2 \alpha \mathcal{I}_{\alpha+1}^{s+2} (\mu, m, T)
    \, ,
    \\[2mm]
\label{eq:ibp:1l:m:temporal}
    0 &=
    \Bigl(\frac{\partial}{\partial p_0} \circ p_0\Bigr)
    \mathcal{I}_\alpha^{s} (\mu,m,T)
    \nn[2mm]
    & =
    (1+s)
    \mathcal{I}_\alpha^{s} (\mu,m,T)
    -2 \alpha \mathcal{I}_{\alpha+1}^{s+2} (\mu, m, T)
    +\oint_{P}^f
      \frac{
        p_0^{s+1}\,
        \nFt'(p_0)
      }{[p_0^2+p^2 + m^2]^\alpha}
    \, ,
    \\[2mm]
\label{eq:ibp:1l:m:d+1}
    0 &=
    \Bigl(\frac{\partial}{\partial P_\mu} \circ P_\mu\Bigr)
    \mathcal{I}_\alpha^{s} (\mu,m,T)
    \nn
    & =
    (d+1-2 \alpha + s)
    \mathcal{I}_\alpha^{s} (\mu,m,T)
    +2 \alpha m^2 \mathcal{I}_{\alpha+1}^{s} (\mu, m, T)
    +\oint_{P}^f
      \frac{
        p_0^{s+1}\,
        \nFt'(p_0)
      }{[p_0^2+p^2 + m^2]^\alpha}
    \, ,
\end{align}
\end{subequations}
where in the last line we combined spatial and temporal total derivatives into a
$d+1$ dimensional total derivative bilinear
akin to identity~\eqref{eq:ibp:1l:d+1}.
After setting $s=0$, we again discern two hierarchy regimes:

\paragraph*{\bf Hierarchy $m > \mu$.}
In the low-temperature limit, the right-hand side vanishes such that
\begin{equation}
  \oint_{p_0}^f \int_{\vec{p}} \frac{p_0 \nFt'(p_0)}{[p_0^2+p^2 +m^2]^\alpha}
  \stackrel{T\to 0}{=}
    \biggl( \frac{e^{\gammaE}\bmu^2}{4 \pi} \biggr)^\frac{3-d}{2}
    \frac{
      \Gamma\bigl(\alpha-\frac{d}{2} \bigr)}{\pi (4\pi)^\frac{d}{2}
      \Gamma\bigr(\alpha\bigr)}
    \re \left\{ i \mu [m^2-\mu^2]^\frac{d-2\alpha}{2}  \right\} = 0
    \, ,
\end{equation}
where we have used the results of sec.~\ref{sec:1-loopformulae}. This naturally implies the familiar recursion from zero-temperature scalar field theory, i.e.
\begin{equation}
    \mathcal{I}_{\alpha+1} (m > \mu) = -\frac{d+1-2\alpha}{2\alpha m^2} \mathcal{I}_{\alpha} (m> \mu)
    \, .
\end{equation}

\paragraph*{\bf Hierarchy $m < \mu$.}
In the opposite regime, we regulate the power-law term and the imaginary unit as above to find
\begin{eqnarray}
    \oint_{p_0}^f \int_{\vec{p}} \frac{p_0 \nFt'(p_0)}{[p_0^2+p^2+m^2]^\alpha}
    &\stackrel{T\to 0}{=}&
    \biggl( \frac{e^{\gammaE}\bmu^2}{4 \pi} \biggr)^\frac{3-d}{2}
    \frac{
      \Gamma\bigl(\alpha-\frac{d}{2} \bigr)}{\pi (4\pi)^\frac{d}{2}
      \Gamma\bigl(\alpha\bigr)}
    \re \Bigl\{\mu [\mu^2-m^2]^\frac{d-2\alpha}{2} \left(\Pi^+\right)^\frac{d+1-2 \alpha}{2} \Bigr\}
    \nn[2mm] &=&
    - (d+1-2\alpha)\, \mathcal{I}_\alpha (\mu) \Bigl(1 - \frac{m^2}{\mu^2} \Bigr)^\frac{d-2\alpha}{2}
    \, ,
\end{eqnarray}
where we again used the $\Pi^+$ convention of eq.~\eqref{eq:i:reg}.
This implies the validity of the identity
\begin{eqnarray}
  (d+1-2\alpha)\,\mathcal{I}_\alpha (m <\mu )
  + 2 \alpha m^2 \mathcal{I}_{\alpha+1} (m < \mu ) =
  (d+1-2\alpha)\,\mathcal{I}_\alpha (\mu) \Bigl(1 - \frac{m^2}{\mu^2} \Bigr)^\frac{d-2\alpha}{2}
  \, .
\end{eqnarray}

%
\section{One-loop consistency checks and additional examples}
\label{sec:consistency}

This appendix complements the results of sec.~\ref{sec:1-loopformulae} by providing consistency checks related to known standard properties of dimensionally regularized loop integrals.
In particular, we focus on the one-loop result given in eq.~\eqref{eq:fermion1master_new} as well as its finite-temperature generalization
\begin{align}
\label{eq:I:alpha:f:T}
    \mathcal{I}_\alpha (\mu,T) &=
      \oint_{p_0}^f
      \nFt(p_0)\,
      I_{\alpha}(p_0)
    \stackrel{T\to 0}{=}
    \biggl( \frac{e^\gammaE\bmu^2}{4\pi} \biggr)^\frac{3-d}{2}
    \frac{
      \Gamma\bigl( \alpha-\frac{d}{2} \bigr)}{
      \Gamma\bigl(\alpha\bigr)}
    \oint_{p_0}^f
      \nFt(p_0)
      [p_0^2]^{\frac{d-2\alpha}{2}}
    \, .
\end{align}
Along the way, we also discuss the role of chemical potential in parametric differentiation and
its relation to symmetries at the one-loop order. We further emphasize that our results should be formally considered analytical continuations of mathematically convergent integrals, which in the low-temperature regime again implies parametric constraints akin to eq.~\eqref{eq:I:alpha:f}.

\subsection{Linear algebra and loop-momentum differentiation}

For the sake of completeness, let us first demonstrate that the fermionic one-loop master integral allows the application of various linear decompositions while keeping the result intact. For this purpose, we multiply eq.~\eqref{eq:I:alpha:f:T} by unity $1= (p_0^2 + p^2)/[p_0^2 + p^2]$ and first evaluate the spatial integral, producing
\begin{align}
\label{eq:finitelinsum2}
    \mathcal{I}_\alpha (\mu,T) &=    \oint_{p_0}^f
      p_0^2\,\nFt^{ }(p_0)
    \int_{\vec{p}}
    \frac{1}{[p_0^2+p^2]^{\alpha+1}}
    +\oint_{p_0}^f
    \nFt^{ }(p_0)
    \int_{\vec{p}}
    \frac{p^2}{[p_0^2+p^2]^{\alpha+1}}
    \nn
    &=
    \biggl( \frac{e^\gammaE\bmu^2}{4\pi} \biggr)^\frac{3-d}{2}
    \Biggl[
        \frac{
          \Gamma\bigl(\alpha+1-\frac{d}{2}\bigr)}{(4\pi)^\frac{d}{2}
          \Gamma\bigl(\alpha+1\bigr)}
      + \frac{(2 \pi)^2
          \Gamma\bigl(\frac{d}{2}+1\bigr)
          \Gamma\bigl(\alpha-\frac{d}{2}\bigr)}{\pi (4\pi)^\frac{d+2}{2}
          \Gamma\bigl(\frac{d}{2} \bigr)
          \Gamma\bigl(\alpha+1 \bigr)}
      \Biggr]
    \oint_{p_0}^f
    \nFt^{ }(p_0)
    [p_0^2]^{\frac{d-2\alpha}{2}}
    \nn
    &= \Bigl[ \Bigl(1- \frac{d}{2 \alpha} \Bigr)+\frac{d}{2 \alpha} \Bigr] \mathcal{I}_\alpha (\mu, T)
    \, .
\end{align}
After moving to numerator structures containing exclusively spatial structures, all required symmetries arise from the dimensionally regularized beta function. 

In sec.~\ref{sec:1-loopformulae}, we also discussed the effect of the diagonal total derivative operator $\frac{\partial}{\partial p_0} \circ p_0$, where one simplification was found in the vanishing of total spatial derivatives, or conversely the generated boundary terms. This is a central element in dimensionally regularized integral algebra, and here we apply the one-loop master integral to confirm this result explicitly. By first performing the spatial integration, we retain all differential and linear algebra within the innermost integral. The diagonal operator can be re-written such that
\begin{equation}
\oint_{p_0}^f \nFt(p_0)\int_{\vec{p}} \partial_i p^i =
\oint_{p_0}^f \nFt(p_0)\int_{\vec{p}} \left(d + p^i \partial_i \right)
  \, .
\end{equation}
Upon performing the innermost integral, this operator structure introduces two terms: one proportional to the master integral and another with increased radial power, of which the latter can be geometrically interpreted as a master integral in a higher dimension. To this end, the total \newtext{spatial} derivative \newtext{acting on the integrand of eq.~\eqref{eq:I:alpha:f:T} results in} %
\footnote{
  To simplify the computation, we vary the spatial dimensionality of the master integral,
  $\mathcal{I}_\alpha (d,\mu, T)$, and write it explicitly in the superscript.
}
\begin{align}
\label{eq:spatialIBP2}
  0 &=
  d\,\mathcal{I}_\alpha (d, \mu, T) -(2\alpha)  \frac{\Omega_{d}}{\Omega_{d+2}} (2 \pi)^2 \mathcal{I}_{\alpha+1} (d+2,\mu, T)
  \nn &=
  d\,\mathcal{I}_{\alpha} (d, \mu, T) -4 \pi \cdot \frac{d}{2} \cdot (2 \alpha) \mathcal{I}_{\alpha+1} (d+2, \mu, T)
  \, .
\end{align}
\newtext{We emphasize that a naive application of the residue theorem at $T = 0$ (cf.~\cite{Gorda:2022yex}) would not agree with this relation}.%
\footnote{
  \newtext{A naive application of the residue theorem to eq.~\eqref{eq:fermion1master_new} results in $\mathcal{I}_\alpha^{\rm naive} (d,\mu, T = 0) =
  -\left( \frac{e^{\gamma_{\rm E}} \bmu^2}{4\pi} \right)^\frac{3-d}{2} \frac{\Gamma \left(\alpha-1/2 \right)}{ (4 \pi)^{d/2} \sqrt{\pi}\Gamma(\alpha) \Gamma \left(d/2\right)}\frac{\mu^{d+1-2\alpha}}{d+1-2\alpha}$, which leads to a non-vanishing right-hand side of eq.~\eqref{eq:spatialIBP2}.}}

The results listed above and in sec.~\ref{sec:1-loopformulae} give great insight into expressions with trivial or quadratic numerator structures. To address the subset of cases with numerator structures linear in the zero-components of momenta, let us next discuss the case \newtext{of the temporal} derivative $\frac{\partial}{\partial p_0}$, which acts on the distribution function. Unlike the bilinear operators introduced in sec.~\ref{sec:IBPintro}, such an operator allows for relating primed ($\nF'$) integrals to ones with linear numerator structures that cannot be further simplified using linear algebra from other total derivatives.
The simplest such structure is
the temporal \newtext{integro-differential relation} 
in eq.~\eqref{eq:ibp:1l:temporal}
for \newtext{a fixed numerator parameter} $s=-1$
\begin{align}
\label{eq:1loopextra1}
    \frac{\partial}{\partial p_0} \mathcal{I}_{\alpha}^{ } (\mu,T) &=
      - 2 \alpha \mathcal{I}_{\alpha+1}^{1} (\mu,T)
      + \oint_{P}^f \frac{ \nFt'(p_0)}{[p_0^2+p^2]^\alpha}
      \, ,
\end{align}
which can again be confirmed to vanish by direct computation of both expressions on the right-hand side.
The first integral on the right-hand side is given by eq.~\eqref{eq:I:alpha:s},
which was evaluated using the methods introduced
in sec.~III of \cite{Gorda:2022yex}.
By combining the Cauchy theorem and starting with the spatial integral, we obtain
\begin{align}
\label{eq:demolinear}
    2 \alpha
    \mathcal{I}_{\alpha+1}^{1}(\mu)
    &\stackrel{T \to 0}{=}
   i \biggl( \frac{e^{\gammaE}\bmu^2}{4\pi} \biggr)^\frac{3-d}{2}
      \frac{\Gamma \left(\alpha-\frac{d}{2} \right)}{\pi (4\pi)^\frac{d}{2} \Gamma (\alpha) } \mu^{d-2\alpha}
      \sin \Bigl[ \frac{\pi}{2}\left(d-2\alpha \right) \Bigr]
     \, .
\end{align}
The second expression in eq.~\eqref{eq:1loopextra1} can be computed using methods introduced in secs.~\ref{sec:formalism} and \ref{sec:1-loopformulae}, yielding
\begin{align}
\label{eq:1looplast0}
       \oint_{p_0}^f \int_{\vec{p}}
       \frac{ \nFt' (p_0)}{[p_0^2+p^2]^\alpha}
       &\stackrel{T\to 0}{=}
       i \biggl( \frac{e^{\gammaE}\bmu^2}{4\pi} \biggr)^\frac{3-d}{2}
       \frac{
          \Gamma\bigl(\alpha-\frac{d}{2} \bigr)}{\pi (4\pi)^\frac{d}{2}
          \Gamma\bigl(\alpha\bigr)} \mu^{d-2 \alpha}
       \sin \Bigl[ \frac{\pi}{2}\left(d-2\alpha \right) \Bigr]
       =
       2 \alpha
        \mathcal{I}_{\alpha+1}^{1}(\mu)
      \, ,
\end{align}
which indeed agrees with eq.~\eqref{eq:demolinear}, causing eq.~\eqref{eq:1loopextra1} to vanish as expected. This is another consistency check between the two contour methods introduced
in sec.~\ref{sec:formalism} for the evaluation of the loop integrals in eqs.~\eqref{eq:cauchy1} and \eqref{eq:primecontoursimple}.

\subsection{Parametric differentiation}

\newtext{Parametric differentiation or \emph{Feynman's trick} can also be used to associate most single-scale ($\mu$) one-loop integrals to the master integral $\mathcal{I}_\alpha (\mu)$.} The practical value of such an operation relies heavily on the assumption that the differentiated structure can be dealt with more easily than the non-differentiated one, and that there are no scale-invariant contributions removed by the differentiation. As for the latter concern, we note that the integrals discussed in this work contain two independent scales, $\mu$ and $T$, the latter of which is taken to an unessential limit (and hence should not contribute in a meaningful way to the results we are seeking).
To this end, the $\mu$-differentiation can only miss fully scale-independent, $\mathcal{O}(\mu^0)$, contributions to the integral in question.%
\footnote{
  The situation is different for integrals with, e.g., both mass and chemical potential scales.
  In these cases, each parametric differentiation carries the risk of canceling out a part of the full solution. Such a case can be found in appendix~\ref{sec:massl} for the scale hierarchy $m > \mu$.
}

In the low-temperature limit, the distribution functions bring no scale to the energy dimension of the full integral while their derivatives lower it by one, so that all loop integrals of interest take the form of a power of $\mu$. The $\eta$-regulators are on the other hand suppressed in our explicit results, whose low-temperature limits take forms such as
\begin{eqnarray}
  \lim_{T \to 0} \oint_{p_0}^f \int_{\vec{p}}
  \frac{p_0^{s}\,p^\gamma}{[p_0^2+p^2]^\alpha} \nFt(p_0)  &\sim & \mu^{d+\gamma+s+1-2\alpha}
   \, ,\\
  \lim_{T \to 0} \oint_{p_0}^f \int_{\vec{p}}
  \frac{p_0^{s}\,p^\gamma}{[p_0^2+p^2]^\alpha} \nFt'(p_0) &\sim & \mu^{d+\gamma+s-2\alpha}
  \, .
\end{eqnarray}
Accordingly, the presence of dimensional regularization should be sufficient to avoid any loss of information in parametric differentiation.

At the one-loop level, Feynman's trick can be used both as another consistency check as well as in the derivation of $\mu$-dependent symmetries extending beyond the low-temperature limit. While the full solution is explicitly temperature-dependent, the only parameter we can  differentiate is the chemical potential, which leads to \newtext{non-unique} identities \newtext{involving the null space of $\partial_\mu$}, such as
\begin{equation}
    \frac{\partial}{\partial \mu} \oint_{P}^f \frac{\nFt(p_0)}{[p_0^2+p^2]^\alpha} =
    \frac{\partial}{\partial \mu} \biggl\{
        \oint_{P}^f \frac{\nFt(p_0) }{[p_0^2+p^2]^\alpha}
      - \oint_{P}^{f,\,\mu = 0} \frac{\nF \left[i \beta p_0 \right]}{[p_0^2+p^2]^\alpha}
    \biggr\}
    \, .
\end{equation}
For multi-loop computations and when applied directly, the trick generates differential equations to replace the inhomogeneous recursive equations, corresponding to the power-law behavior of the result in $\mu$.%
\footnote{\newtext{
If multiple distribution functions are present, the $\mu$-differentiation is considerably less useful. The problems can be partially alleviated by introducing an independent chemical potential for each loop momentum, but this complicates the structure of the master integrals.}} While computations involving parametric differentiation are not central to our work, we add a few examples below to demonstrate their connection to the formulae used here and in~\cite{Gorda:2022yex}.

Similarly to many previous computations presented in this work, we need to first keep the temperature nonzero to ensure that we do not miss important contributions to our integrals that might vanish upon differentiation at exactly $T=0$. Without any loss of information, we can then write the partial differentiation of the distribution function as
\begin{equation}
  \oint_{p_0}^f\!
  \nFt' (p_0)  p_0 \int_{\vec{p}} \frac{1}{[p^2+p_0^2]^\alpha} =
  \oint_{p_0}^f i\frac{\partial}{\partial \mu} \left[\nFt (p_0)  \right] \int_{\vec{p}} \frac{i p_0}{[p^2+p_0^2]^\alpha}
  \, ,
  \quad
  \text{with}
  \quad
    \frac{\partial}{\partial \mu} \nFt(p_0) =
  -i\frac{\partial}{\partial p_0} \nFt(p_0)
  \, ,
\end{equation}
and subsequently,
move the differentiation outside the $p_0$-integral due to the closed nature of the integration contour. By differentiating the low-temperature limit corresponding to the rest of the integrand on the right-hand side, we find using methods discussed above
\begin{align}
\label{eq:partiallimits}
    \frac{\partial}{\partial \mu }\biggl\{
      \lim_{T \to 0}
      \oint_{p_0}^f\! \nFt (p_0)
      &
      \int_{\vec{p}}
      \frac{i p_0}{[p^2+p_0^2]^\alpha} \biggr\}
    \nn &=
    -\frac{\partial}{\partial \mu}
    \biggl\{\biggl(\frac{e^{\gammaE}\bmu^2}{4 \pi}\biggr)^{\frac{3-d}{2}}
    \frac{
      \Gamma\bigl(\alpha-\frac{d}{2} \bigr) }{ \pi (4 \pi)^\frac{d}{2}
      \Gamma\bigl(\alpha\bigr) }
    \frac{\mu^{d+2-2\alpha}}{(d+2-2 \alpha)}
      \cos \left[ \frac{\pi}{2}(d+3-2\alpha) \right]\biggr\}
    \nn
    &= -(d+1-2\alpha) \mathcal{I}_\alpha (\mu)
    \nn
    &= \lim_{T \to 0}  \oint_{p_0}^f   \nFt' (p_0)  p_0
    \int_{\vec{p}} \frac{1}{[p^2+p_0^2]^\alpha}
    \, .
\end{align}
As stated above, this implies that the parametric differentiation encompasses all terms containing explicit
$\mu$-dependence.
Furthermore, we note that the order of the zero-temperature limit and parametric differentiation does not affect the result.
Using this flexibility with the limits, we can also confirm the result derived for master integrals equipped with linear $p_0$-structures:
\begin{equation}
\label{eq:crosspartialmu}
    i \frac{\partial}{\partial \mu} \mathcal{I}_\alpha (\mu) =
    -i \biggl( \frac{e^{\gammaE} \bmu^2}{4 \pi}\biggr)^\frac{3-d}{2} \frac{\mu^{d-2\alpha}}{(4 \pi)^\frac{d}{2} 
      \Gamma\bigl(\alpha\bigr)
      \Gamma\bigl(\frac{d}{2} + 1 -\alpha \bigr)}
    \, ,
\end{equation}
which indeed agrees with eq.~\eqref{eq:1looplast0}.

Finally, we note that we can rather easily derive additional formulae for similar master integrals, including e.g.
\begin{equation}
    \lim_{T \to 0} \oint_{p_0}^f \nFt' (p_0)  p_0 \int_{\vec{p}}
    \frac{1}{[p^2+p_0^2]^\alpha} = -\frac{1}{2(\alpha-1)} \frac{\partial^2}{\partial \mu^2} \mathcal{I}_{\alpha-1} (\mu)
    \, ,
\end{equation}
and that we can discern from
eqs.~\eqref{eq:fermion1master_new} and \eqref{eq:1loopdeltaformula1} --
or eq.~\eqref{eq:crosspartialmu} -- that
the $p_0$-numerator in eq.~\eqref{eq:partiallimits} can be replaced by $\mu$ such that
\begin{eqnarray}
    - \mu \frac{\partial}{\partial \mu} \mathcal{I}_\alpha (\mu)
    &\stackrel{{\rm eq.\eqref{eq:crosspartialmu}}}{=}&
    (i\mu) \times  \lim_{T \to 0}\oint_{p_0}^f  \nFt'(p_0)  \int_{\vec{p}}
    \frac{1}{[p^2+p_0^2]^\alpha}
    \nn
    &\stackrel{{\rm eq.\eqref{eq:1loopdeltaformula1}}}{=}&
    \lim_{T \to 0}\oint_{p_0}^f  \nFt'(p_0) p_0 \int_{\vec{p}}
    \frac{1}{[p^2+p_0^2]^\alpha}
    \nn
    &\stackrel{{\eqref{eq:ibp:1l:d+1}}}{=}&
    \Bigl[P_\mu \circ \frac{\partial}{\partial P_\mu}
    +2\alpha \Bigr]\mathcal{I}_\alpha (\mu)
    \, .
\end{eqnarray}
Here, we applied eq.~\eqref{eq:ibp:1l:d+1} on the last line to re-write the relation in terms of
the operator
$P_\mu \circ \frac{\partial}{\partial P_\mu}$ which is reversed to
operators of the set in eq.~\eqref{eq:ibp:set}.
This relation further demonstrates the non-trivial nature of the $p_0 = \pm i \mu$ substitutions arising from eq.~\eqref{eq:primecontoursimple} in the low-temperature limit. The relevant details of the delta function limits and contours have been discussed already in secs.~\ref{sec:formalism} and \ref{sec:1-loopformulae}.

%
\section{Additional details of two-loop IBP computations}
\label{app:examples}

In this appendix, we present several extensions and consistency checks for the computations presented in sec~\ref{sec:2-loop}. In the first part, we introduce conventions for the spatial integrations taking place in eq.~\eqref{eq:resunset1} and similar relations, while in the second subsection, we study in more detail the vanishing integrals that contain propagators with shifted bosonic momenta. The third subsection then evaluates the group of two-loop integrals  $\{\mathcal{S}_{11s}\, |\, s \in \mathbb{N}\}$ at $T = 0$,  using the cutting rule toolkit from~\cite{Ghisoiu:2016swa} and providing a comparison point for our independent evaluation of such diagrams. The last subsection finally utilizes spatial IBP operators from~\cite{Laine:2019uua} to derive IBP relations for the integrals $\mathcal{S}_{111}$ and $\mathcal{S}_{112}$ that explicitly differ from what one would expect at $\mu=0$. These results act as further cross-checks of our full $(d+1)$-dimensional IBP relations~\eqref{eq:ibp2loopexample} and \eqref{eq:S112:T0:fac}.

\subsection{Vacuum integrals with complex mass scales}
\label{sec:vacuum:2l}

Let us begin by considering the one-loop integral of eq.~\eqref{eq:cauchyex1} and the two-loop integrals of eq.~\eqref{eq:resunset1} without explicit replacements. Akin to sec.~\ref{sec:1-loopformulae} of~\cite{Gorda:2022yex}, we may regulate the quadratic scales using $i \mu \mapsto \Pi^+ \mu$ from eq.~\eqref{eq:i:reg}, which allows using the generalized Euler beta functions in the dimensionally regularized radial integrations. At the one- and two-loop levels, the initial spatial momentum integration corresponds to
vacuum integrals of the type
\begin{align}
\label{eq:S:m12}
  S_{\alpha_1\alpha_2\alpha_3}(m_1,m_2,0) &=
      \biggl( \frac{e^\gammaE \bmu^2}{4\pi} \biggr)^{3-d}\frac{
        \Gamma\bigl(\alpha_{123}-d\bigr)
        \Gamma\bigl(\frac{d}{2}-\alpha_{3} \bigr)
        \Gamma\bigl(\alpha_{13}-\frac{d}{2} \bigr)
        \Gamma\bigl(\alpha_{23}-\frac{d}{2} \bigr)}{
        \Gamma\bigl(\alpha_{1}\bigr)
        \Gamma\bigl(\alpha_{2}\bigr)
        \Gamma\bigl(\frac{d}{2} \bigr)
        \Gamma\bigl(\alpha_{123}+\alpha_{3}-d \bigr)
        }
      \frac{[m_1^2]^{d-\alpha_{123}}}{(4\pi)^d}
      \nn &
      \hphantom{\biggl( \frac{e^\gammaE \bmu^2}{4\pi} \biggr)^{3-d}}
      \times {}_2 F_1 \left[
        \alpha_{23}-\frac{d}{2},
        \alpha_{123}-d;
        \alpha_{123}+\alpha_{3}-d;
        1- \frac{m_2^2}{m_1^2}
        \right]
      \,,\\[2mm]
\label{eq:S111:m12}
  S_{111}(m_1,m_2,m_1+m_2) &=
  \frac{1}{2}\frac{(d-2)}{(d-3)}\Bigl[
      \frac{I_{1}(m_1)}{m_1}
      \frac{I_{1}(m_2)}{m_2}
    - \frac{I_{1}(m_1)}{m_1}
      \frac{I_{1}(m_3)}{m_3}
    - \frac{I_{1}(m_2)}{m_2}
      \frac{I_{1}(m_3)}{m_3}
    \Bigr]
  \,,\\[2mm]
\label{eq:S121:m12}
  S_{121}(m_1,m_2,m_1+m_2) &=
  \frac{1}{4}\frac{(d-2)}{(d-5)}\Bigl[
      \frac{I_{1}(m_1)}{m_1^2}
      \frac{I_{1}(m_3)}{m_3^2}
    + \Bigl((d-4)\frac{m_3}{m_2} -1 \Bigr)
      \frac{I_{1}(m_2)}{m_2^2}
      \frac{I_{1}(m_3)}{m_3^2}
    \nn[1mm] &
    \hphantom{{}=
    \frac{1}{4}\frac{(d-2)}{(d-5)}\Bigl[
      \frac{I_{1}(m_1)}{m_1^2}
      \frac{I_{1}(m_3)}{m_3^2}}
    - \Bigl((d-4)\frac{m_1}{m_2} +1 \Bigr)
      \frac{I_{1}(m_1)}{m_1^2}
      \frac{I_{1}(m_2)}{m_2^2}
    \Bigr]
  \,,
\end{align}
where in the first integral one assumes the hierarchy $m_1^2 \geq m_2^2$ and in the latter two cases the collinear relation $m_3 = m_1 + m_2$ that gives rise to the factorization of the vacuum sunset reported in~\cite{Davydychev:1992mt,Davydychev:2022dcw}.

In the evaluation of the maximally primed type~\ref{typeB} integrals of sec.~\ref{sec:maxprime}, we may apply the above collinear results because one of the three complex mass scales in the corresponding $d$-dimensional integrals always vanishes due to the substitutions of $p_0$ and $q_0$ in relations such as eq.~\eqref{eq:resunset1}. The specific mapping between the masses $m_i$ appearing above and the momenta $p_0$, $q_0$, and $p_0\pm q_0$ is determined by the integral in question, typically leading to only one of the three terms in eqs.~(\ref{eq:S111:m12}) and (\ref{eq:S121:m12}) obtaining a nonzero value. To remove the vanishing mass scale from the denominator, we typically use the one-loop massive vacuum IBP relation
$I_{\alpha+1}(m) = -(d-2\alpha)/(2\alpha m^2) I_{\alpha}(m)$ (see \newtext{appendix}~\ref{sec:factorizationof2loop}), which leads to two cases
\begin{align}
\label{eq:repiece1}
     S_{111}(p_0,0,p_0) &= -\frac{2 }{(d-3)(d-2)} \Bigl[p_0 I_2 (p_0) \Bigr]^2
     \,,
     & \text{for}\quad
     p_0-q_0&=0
     \,,\\[2mm]
\label{eq:repiece2}
     S_{111}(p_0,q_0,0) &= \frac{2}{(d-3)(d-2)}
      \Bigl[p_0 I_2 (p_0) \Bigr]
      \Bigl[q_0 I_2 (q_0) \Bigr]
    \,,
    & \text{for}\quad
     p_0+q_0&=0
      \,.
\end{align}
To finally arrive at results such as eq.~\eqref{eq:primesunsetsol1}, we must still connect the temporal momentum components to the substitutions $\mu (\Pi^+)^{\pm 1}$. In this way, eq.~\eqref{eq:repiece1} is associated with the $p_0 = q_0 = \mu \Pi^+$ substitution in the first term on the right-hand side of eq.~\eqref{eq:resunset1}, while eq.~\eqref{eq:repiece2} is associated with the second term there.

One more subtlety, however, remains: for $S_{111}(\mu \Pi^+, \mu / \Pi^+,\mathcal{O}(\kappa))$, a straightforward generalization of the hypergeometric expression~\eqref{eq:S:m12} would provide a complex-valued result for an integral that is by construction its own complex conjugate.  To remedy this, we must return to the defining integral (obtained after shifting $\vec{p} \mapsto \vec{p} + \vec{q}$)
\begin{align}
\label{eq:twomassformula}
  S_{\alpha_1\alpha_2\alpha_3}(p_0,q_0,\mathcal{O}(\kappa)) &=
    \int_{\vec{p},\vec{q}}
    \frac{1}{
        [|\vec{p}+\vec{q}|^2+p_0^2]^{\alpha_{1}}
        [q^2+q_0^2]^{\alpha_{2}}
        [p^2 + \mathcal{O}(\kappa)]^{\alpha_{3}}
      }
      \, ,
\end{align}
where we can use standard Feynman parametrization if $\alpha_1, \alpha_2 >0$. By regulating again the complex-valued mass scales such that $i \mapsto \Pi^+$, we may straightforwardly evaluate the two spatial loop integrations, obtaining a hypergeometric integral of the form
\begin{equation}
\label{eq:intermhypergeom}
S_{\alpha_1\alpha_2\alpha_3}(p_0,q_0,0) =   \biggl(\frac{e^\gammaE \bmu^2}{4 \pi} \biggr)^{3-d}
  \frac{
    \Gamma\bigl(\alpha_{123}-d\bigr)
    \Gamma\bigl(\frac{d}{2}-\alpha_{3} \bigr)
    }{(4 \pi)^d
    \Gamma\bigl(\alpha_{1}\bigr)
    \Gamma\bigl(\alpha_{2}\bigr)
    \Gamma\bigl(\frac{d}{2} \bigr)}
    \int_0^1\!{\rm d}x
    \frac{x^{\alpha_{23}-\frac{d}{2}-1}(1-x)^{\alpha_{13}-\frac{d}{2}-1}}{\left[x q_0^2+(1-x) p_0^2 \right]^{\alpha_{123}-d}},
\end{equation}
where again $\alpha_{\{i\}} = \sum_{j=\{i\}} \alpha_{j}$.

To proceed from here, let us next assume that $\alpha_3 \in \mathbb{N}$. Since the denominator and numerator powers differ by $(\alpha_{23}-\frac{d}{2})+(\alpha_{13}-\frac{d}{2})-(\alpha_{123} - d) = \alpha_3$
for the remaining hypergeometric integral, we can rewrite its integrand by multiplying with unity
\begin{align}
 \frac{[xq_0^2+(1-x)p_0^2]^{\alpha_3}}{[xq_0^2+(1-x)p_0^2]^{\alpha_3}} &=
  \frac{[p_0^2+(q_0^2-p_0^2)x]^{\alpha_3}}{[xq_0^2+(1-x)p_0^2]^{\alpha_3}}=
  \frac{[p_0^2+\mathcal{O}(\kappa)x]^{\alpha_3}}{[xq_0^2+(1-x)p_0^2]^{\alpha_3}}
   \,,
\end{align}
where the subtraction $p_0^2-q_0^2 = \mathcal{O}(\kappa)$ is proportional to the $\kappa$-regulator scale.
Writing the numerator of the multiplier in terms of a binomial expansion, we may then remove the contribution proportional to the regulator, obtaining
 \begin{align}
 \label{eq:binomialstandard}
   \int_0^1\!{\rm d}x
   \frac{x^{\alpha_{23}-\frac{d}{2}-1}(1-x)^{\alpha_{13}-\frac{d}{2}-1}}{\left[x q_0^2+(1-x) p_0^2 \right]^{\alpha_{123}-d}}
   &=
   \sum_{k=0}^{\alpha_3} \binom{\alpha_3}{k} [p_0^2]^k [q_0^2-p_0^2]^{\alpha_3-k}
   \int_0^1\!{\rm d}x
   \frac{x^{\alpha_3+\alpha_{23}-\frac{d}{2}-k-1}(1-x)^{\alpha_{13}-\frac{d}{2}-1}}{\left[x q_0^2+(1-x) p_0^2 \right]^{\alpha_3+\alpha_{123}-d}}
   \nn
   &= \frac{\Gamma \left( \alpha_{23}-\frac{d}{2}\right)\Gamma \left( \alpha_{13}-\frac{d}{2}\right)}{\Gamma\left( \alpha_3+\alpha_{123}-d\right)}
   \frac{[p_0^2]^{\alpha_3}}{
      [p_0^2]^{\alpha_{13}-\frac{d}{2}}
      [q_0^2]^{\alpha_{23}-\frac{d}{2}}}
  + \mathcal{O} (\kappa)
  \,,
\end{align}
where we have reversed the Feynman parametrization of eq.~\eqref{eq:Feynmanparamet} to find the factorization explicitly manifested.

Collecting the above results, we recognize that for $\alpha_3 \in \mathbb{N}$, we can write
 \begin{equation}
\label{eq:final}
S_{\alpha_1\alpha_2\alpha_3}(p_0,q_0,0) =
   \biggl(\frac{e^\gammaE \bmu^2}{4 \pi} \biggr)^{3-d}
  \frac{
    \Gamma\bigl(\alpha_{123}-d\bigr)
    \Gamma\bigl(\frac{d}{2}-\alpha_{3} \bigr)
    }{
    \Gamma\bigl(\alpha_{1}\bigr)
    \Gamma\bigl(\alpha_{2}\bigr)
    \Gamma\bigl(\frac{d}{2}\bigr)}
    \frac{
      \Gamma\bigl( \alpha_{23}-\frac{d}{2}\bigr)
      \Gamma\bigl( \alpha_{13}-\frac{d}{2}\bigr)}{
      \Gamma\bigl( \alpha_3+\alpha_{123}-d\bigr)}
    \frac{
      [p_0^2]^{\frac{d}{2}-\alpha_1}
      [q_0^2]^{\frac{d}{2}-\alpha_{23}}}{(4\pi)^d}
    \,,
\end{equation}
where the binomial expansion in eq.~\eqref{eq:binomialstandard} has clearly broken the symmetry between $p_0$ and $q_0$. A more symmetric result can be reinstated  by writing the solution arising from \eqref{eq:binomialstandard} trigonometrically using $[q_0^2]^n = \cos (\pi n)\mu^{2n}$ for $n \in \mathbb{N}$. With this parametrization, we obtain
\begin{equation}
     [p_0^2]^{\frac{d}{2}-\alpha_1}
      [q_0^2]^{\frac{d}{2}-\alpha_{23}} = \mu^{2(d-\alpha_{123})}
      \Bigl\{
        \cos \bigl[\pi \left(d-\alpha_{123} \right) \bigr]
        \pm i \sin \bigl[ \pi \left(d-\alpha_{123} \right) \bigr]
      \Bigr\}
    \,,
\end{equation}
for $p_0 = q_0 = (\Pi^+)^{\pm 1} \mu$ and
\begin{align}
    [p_0^2]^{\frac{d}{2}-\alpha_1}
    [q_0^2]^{\frac{d}{2}-\alpha_{23}} &=
      \frac{\mu^{2(d-\alpha_{123})}}{2}
      \Bigl\{
        \cos \bigl[\pi \left(\alpha_{23}-\alpha_{1}\right) \bigr]
      + \cos \bigl[\pi \left(\alpha_{2}-\alpha_{13}\right) \bigr]
      \Bigr\}
    \nn &\pm
    i \frac{\mu^{2(d-\alpha_{123})}}{2}
    \Bigl\{
        \sin \bigl[ \pi \left(\alpha_{23}-\alpha_{1}\right) \bigr]
      + \sin \bigl[ \pi \left(\alpha_{2}-\alpha_{13}\right) \bigr]
    \Bigr\}
    \,,
 \end{align}
for $p_0 = (\Pi^+)^{\pm 1} \mu$ and $q_0 = (\Pi^+)^{\mp 1} \mu$.

The result we have arrived at is equivalent to performing the binomial expansion symmetrically in terms of both $p_0$ and $q_0$, but cannot be straightforwardly extended to $\alpha_3 \in \mathbb{R}_+$. This can be seen by carefully studying Newton's extension to the binomial theorem in the context of eq.~\eqref{eq:binomialstandard}. Indeed, for $\alpha_3 \in (0,1)$, $p_0 = \mu \Pi^+$ and $q_0 = \mu (\Pi^+)^{-1}$, we obtain
\begin{align}
 \label{eq:binomialdeluxe}
   \int_0^1\!{\rm d}x&
   \frac{x^{\alpha_{23}-\frac{d}{2}-1}(1-x)^{\alpha_{13}-\frac{d}{2}-1}}{\left[x q_0^2+(1-x) p_0^2 \right]^{\alpha_3+\alpha_{123}-d}}
    \left[x q_0^2+(1-x) p_0^2 \right]^{\alpha_3}
   \nn &=
   \mu^{2 \alpha_3}\int_0^\frac{1}{2}\!{\rm d}x
   \frac{x^{\alpha_{23}-\frac{d}{2}-1}(1-x)^{\alpha_{13}-\frac{d}{2}-1}}{\left[x q_0^2+(1-x) p_0^2 \right]^{\alpha_3+\alpha_{123}-d}}
    \left[-1 + i \pi \kappa (1-2x) + \mathcal{O}(\kappa^2) \right]^{\alpha_3}
   \nn &
   +\mu^{2 \alpha_3}\int_\frac{1}{2}^1\!{\rm d}x
   \frac{x^{\alpha_{23}-\frac{d}{2}-1}(1-x)^{\alpha_{13}-\frac{d}{2}-1}}{\left[x q_0^2+(1-x) p_0^2 \right]^{\alpha_3+\alpha_{123}-d}}
  \left[-1 - i \pi \kappa (2x-1) + \mathcal{O}(\kappa^2) \right]^{\alpha_3}
   \,,
\end{align}
where we have paid close attention to the signs of the $\mathcal{O}(\kappa)$ terms, according to which the integral has been split in two. In the end, we can collect the leading contributions from both integration intervals such that
\begin{align}
\label{eq:binomialdeluxe2}
\eqref{eq:binomialdeluxe} &=
  \mu^{2 \alpha_3}\Biggl\{
   \cos (\pi \alpha_3 )\int_0^1\!{\rm d}x
   + i \sin (\pi \alpha_3)
   \biggl[
    \int_0^\frac{1}{2}\!{\rm d}x
   - \int_\frac{1}{2}^1\!{\rm d}x
   \biggr]
  \Biggr\}
   \frac{x^{\alpha_{23}-\frac{d}{2}-1}(1-x)^{\alpha_{13}-\frac{d}{2}-1}}{\left[x q_0^2+(1-x) p_0^2 \right]^{\alpha_3+\alpha_{123}-d}}
   \,,
\end{align}
which retains the familiar factorizing structure of eq.~\eqref{eq:final} when $\alpha_3 \in \mathbb{N}$.

A natural generalization to eq.~\eqref{eq:final} is found by adding a mass $m \in \mathbb{R}_+$ to the propagators. Such a  scale can be inserted to both \emph{fermionic} propagators of eq.~\eqref{eq:twomassformula} (associated to $p_0$ and $q_0$) while still conserving the binomial expansion condition $(p_0^2+m^2)-(q_0^2+m^2) = \mathcal{O} (\kappa)$. The regulator $\kappa$ can be fitted such that for $\mu > m$ we may replace $\Pi^+ \mu \mapsto \Pi^+\sqrt{\mu^2-m^2}$ in the thermal integral akin to the one-loop example of appendix~\ref{sec:massl}. For $m > \mu$, the result on the other hand trivializes to the classical real-valued sunset with only one scale $p_0^2 = q_0^2  = m^2-\mu^2 > 0$. We emphasize, though, that the $m > \mu$ hierarchy leads to the vanishing of the type \ref{typeB} term due to subtractions akin to eq.~\eqref{eq:2loopmlong}, making the corresponding IBP relation independent of $\mu$.

Next, we apply eq.~\eqref{eq:final} to an explicit evaluation of the sunset $\mathcal{S}_{111} $ with both Fermi-Dirac distributions differentiated. In practice, we write the integrand in terms of the spatial integral $S_{111}(p_0,q_0,0)$ and use the contour deformations from eq.~\eqref{eq:primecontoursimple} to recover the solution to eq.~\eqref{eq:primesunsetsol1},
\begin{eqnarray}
\label{eq:S111:pp:sol}
{\bf D}_{p}
    {\bf D}_{q}\,
    \mathcal{S}_{111} &=&\oint_{p_0, q_0}\! \nFt'(p_0)\nFt'(q_0) S_{111}^{ } (q_0, p_0-q_0, p_0)\nn[1mm]
  &\!\stackrel{T\to 0}{=}\!&
  \frac{d-2}{d-3}\biggl(\frac{e^\gammaE \bmu^2}{4 \pi} \biggr)^{3-d}
  \frac{\mu^{2(d-2)}}{(4\pi)^d}
    \frac{
      \Gamma^2\bigl(1-\frac{d}{2}\bigr)}{(2\pi)^2}\Bigl(\cos[\pi(d-2)]-1 \Bigr)
    \nn &\!=\!&
  -\biggl(\frac{e^\gammaE \bmu^2}{4\pi} \biggr)^{3-d}
  \frac{\mu^{2(d-2)}}{(4\pi)^d}
    \frac{1}{
      \Gamma\bigl(\frac{d}{2} \bigr)
      \Gamma\bigl(\frac{d}{2}-1\bigr) (d-3)}
    \,,
\end{eqnarray}
where standard trigonometric formulae and Euler's reflection formula have been applied. Similarly, it is straightforward to extend the result to the full class of maximally differentiated two-loop vacuum bubbles, for which $\alpha_1, \alpha_2 \in \mathbb{R}_+$ and $\alpha_3 \in \mathbb{N}$. Applying again eq.~\eqref{eq:final}, we obtain
\begin{eqnarray}
\label{eq:finalmathcal}
    {\bf D}_{p}
    {\bf D}_{q}\,
    \mathcal{S}_{\alpha_1 \alpha_2 \alpha_3} &=&
  \oint_{p_0,q_0}\!\nFt'(p_0)\nFt'(q_0) S_{\alpha_2 \alpha_3 \alpha_1}^{ } (q_0, p_0-q_0, p_0)
    \nn
    &\stackrel{T\to 0}{=}&\,
    \biggl(\frac{e^\gammaE \bmu^2}{4 \pi} \biggr)^{3-d}
    \frac{
   \Gamma\bigl(\alpha_{123}-d\bigr)
    \Gamma\bigl(\frac{d}{2}-\alpha_{3} \bigr)
    }{(4\pi)^d
    \Gamma\bigl(\alpha_{1}\bigr)
    \Gamma\bigl(\alpha_{2}\bigr)
    \Gamma\bigl(\frac{d}{2} \bigr)}
    \frac{
      \Gamma\bigl( \alpha_{23}-\frac{d}{2}\bigr)
      \Gamma\bigl( \alpha_{13}-\frac{d}{2}\bigr)}{
      \Gamma\bigl( \alpha_3+\alpha_{123}-d\bigr)}
    \nn
    &&\times (-1)^{\alpha_3}
      \frac{2 \mu^{2(d-\alpha_{123})}}{(2\pi)^2}
      \Bigl(
          \cos \bigl[\pi(d-\alpha_{12}) \bigr]
        - \cos \bigl[\pi(\alpha_1-\alpha_2) \bigr]
      \Bigr)
      \,.
\end{eqnarray}

\newtext{Finally, we note that numerators with integer powers of temporal momentum components $\{p_0\}$ can be implemented through eq.~\eqref{eq:resunset1} and hence do not introduce any further complications. Following the conventions of sec.~\ref{sec:2-loop}, the two type \ref{typeB} terms required in the derivation of $\mathcal{S}_{111}$ read}
\newtext{
\begin{align}
\label{eq:S111typeBterms}
    \mathbf{D}_p \mathbf{D}_q \mathcal{S}_{111}^{11} &\stackrel{\noveltext{T = 0}}{=}\mathbf{D}_p \mathbf{D}_q \mathcal{S}_{111}^{20}
    \stackrel{\noveltext{T = 0}}{=} -\mu^2 \mathbf{D}_p \mathbf{D}_q \mathcal{S}_{111}
    \stackrel{\noveltext{T = 0}}{=} \frac{2(d-3)}{(d-2)} \mu^2\,\mathcal{I}_2 (\mu) \mathcal{I}_2 (\mu)
    \,.
\end{align}
}
\subsection{Momentum-shifted bosonic propagators}
\label{app:momentumshift}

Next, we take a closer look at the zero-temperature limits of the integrals $\mathcal{S}_{102}$ and $\mathcal{S}_{012}$, appearing in eq.~\eqref{eq:S111:2}. Following the convention of eq.~\eqref{eq:fermionsum}, we may write
\begin{align}
\label{eq:I:1l:fb}
  \ToptVE(\Aqu,\Agl) =
  \mathcal{S}_{\alpha_1 0\,\alpha_2}
  &=
  \oint_{P,Q}^f
  \frac{\nFt (p_0)\nFt(q_0)}{
    [p_0^2+p^2]^{\alpha_{1}}
    [(p_0-q_0)^2+|\vec{p}-\vec{q}|^2]^{\alpha_{2}}}
  \nn
  &=
  \Tint{\{P\}}
  \frac{1}{[p_n^2 + p^2]^{\alpha_{1}}}
  \Tint{\{Q\}}
  \frac{1}{[(p_n - q_m)^2 + |\vec{p}-\vec{q}|^2]^{\alpha_{2}}}
  \nn &=
  \Tint{\{P\}}
  \frac{1}{[p_n^2 + p^2]^{\alpha_{1}}}
  \Tint{R}
  \frac{1}{[r_m^2 + r^2]^{\alpha_{2}}}
  =
  \oint_{P}^f
  \frac{\nFt(p_0)}{[p_0^2+p^2]^{\alpha_{1}}}
  \oint_{R}^b
  \frac{\nB \left[ i \beta r_0 \right]}{[r_0^2+r^2]^{\alpha_{2}}}
  \, ,
\end{align}
where we have used the sum-integral definitions of eq.~\eqref{eq:fermionsum} and noted the bosonic nature of the momentum  $R\equiv Q-P$. To proceed to the low-temperature limit, we first perform the spatial integral and then compute the remaining zero-component integral through the step function limit of $\nB$ in eq.~\eqref{eq:nfbtheta} (cf.\ sec.~III of \cite{Gorda:2022yex}):
\begin{align}
\label{eq:vanishingboson}
   \oint_{p_0}^b
    \nB \left[i\beta p_0 \right]
    I_{\alpha}(p_0)
   &= 2
   \biggl( \frac{e^\gammaE \bmu^2}{4\pi} \biggr)^\frac{3-d}{2}
   \frac{
      \Gamma\bigl(\alpha-\frac{d}{2} \bigr)}{(4\pi)^\frac{d}{2}
      \Gamma\bigl(\alpha\bigr)}
     \re \int_{i \eta}^{\infty+i\eta}
     \frac{{\rm d}p_0}{2\pi} [p_0^2]^{\frac{d}{2}-\alpha}
   \nn
   &=-
   \biggr( \frac{e^\gammaE \bmu^2}{4 \pi} \biggr)^\frac{3-d}{2}
   \frac{1}{(4\pi)^\frac{d}{2}
      \Gamma\bigl(\alpha\bigr)
      \Gamma\bigl(\frac{d}{2}+1-\alpha\bigr)
      }
      \frac{\eta^{d+1-2\alpha}}{(d+1-2\alpha)}
  = \mathcal{I}_\alpha(\mu)\Bigr|_{\mu \to \eta}
      \, .
\end{align}

The above expression formally corresponds to the fermionic one-loop integral of eq.~\eqref{eq:fermion1master_new}. Its dependence on the powers of the regulator scale $\eta$ (from the contour) indicates the factorizing of the $R$-integral in eq.~\eqref{eq:I:1l:fb}, leading to the entire diagram vanishing in dimensional regularization%
\footnote{\newtext{Taking the limit $\eta \rightarrow 0$ before letting the dimensional regulator vanish results in a vanishing scaleless integral as expected for $T=0$ bosonic integrals.}}. This result is retained also with a numerator linear in $q_0$ after splitting the computation into two upon writing $q_0 = r_0 + p_0$. An integral with a $p_0$-numerator also follows the same steps as above while an integral with an $r_0$-numerator involves an imaginary part taken during the remaining $r_0$-integral, similar to eq.~\eqref{eq:demolinear}.

Next, we inspect the closely related integral of eq.~\eqref{eq:sunsetexplicitIBP}, which contains one primed and one unprimed distribution function, \noveltext{marking our first example of a simple type \ref{typeC} integral}.
Using the formulae of sec.~\ref{sec:2-loop}, we obtain for it
\begin{align}
  \oint_{p_0, q_0}^f
  &
  \int_{\vec{p},\vec{q}}
  \frac{\nFt(p_0)\, q_0 \nFt'(q_0) }{
    [p_0^2+p^2]^{\alpha_{1}}
    [(p_0-q_0)^2+|\vec{p}-\vec{q}|^2]^{\alpha_{2}}
  }
  =
    \oint_{p_0}^f \nFt(p_0) I_{\alpha}(p_0)
    \oint_{q_0}^f q_0 \nFt'(q_0) I_{\alpha}(p_0 - q_0)
  \, ,
\end{align}
where the right-hand side follows after performing the $d$-dimensional spatial integrals. Focusing next on the remaining two integrals, we change the integration variable of the $q_0$-integral via $q_0 \mapsto r_0 = q_0-p_0$ and also split the $p_0$-integral using the line integral description. This leads to
\begin{align}
\label{eq:listofmixedfermions}
  \oint_{p_0}^f [p_0^2]^{\frac{d}{2}-\alpha_{1}}\nFt(p_0)
  &
  \oint_{q_0}^f q_0 [(p_0-q_0)^2]^{\frac{d}{2}-\alpha_{2}} \nFt'(q_0)
  =
  \nn &
   -\int_{-\infty + i \mu+i \eta_p}^{\infty + i \mu+i \eta_p}
  \frac{{\rm d}p_0}{2 \pi}
  [p_0^2]^{\frac{d}{2}-\alpha_{1}}\nFt(p_0)
  \oint_{r_0}^{b,-} (r_0+p_0)  [r_0^2]^{\frac{d}{2}-\alpha_{2}}
   \nFt'(r_0+p_0)
   \nn &
  + \int_{-\infty + i \mu-i \eta_p}^{\infty + i \mu-i \eta_p} \frac{{\rm d}p_0}{2 \pi}
   [p_0^2]^{\frac{d}{2}-\alpha_{1}} \nFt(p_0)
   \oint_{r_0}^{b,+} (r_0+p_0)  [r_0^2]^{\frac{d}{2}-\alpha_{2}}
   \nFt'(r_0+p_0)
   \, ,
\end{align}
where
$\eta_p$ is the regulator on the $p$-integral contour and the $\pm$ symbols correspond to the $\pm i \eta_p$ shifts of the bosonic contour along the imaginary axis. \noveltext{In the $T \to 0$ limit, the leading behaviour of distribution functions is solely determined by the imaginary part of loop momenta [cf.~eq.~\eqref{eq:nfb_complexdelta}],
\begin{equation}
    \nFt'(r_0+p_0) =
    i \beta \nF'\bigl\{i \beta[ \re(r_0+p_0) + i\im (r_0)\pm i \eta_p ]\bigr\}
    \stackrel{T \to 0}{=}
    -i \delta [\text{Im} (r_0) \pm \eta_p]
    \,,
\end{equation}
which we can be immediately applied given that all the momentum dependence outside of distribution functions is of the form of generalized monomials}\footnote{\noveltext{Monomials do not introduce problematic poles and allow a straightforward taking of the naive $T \to 0$ limit in both differentiated and non-differentiated distributions.}}
\noveltext{and where the $p_0$ line integrals fix the value of $\im(p_0)$.
Thus, the $r_0$ integration akin to eq.~\eqref{eq:primecontoursimple} becomes
\begin{align}
   \oint_{r_0}^{b,+} [r_0^2]^{\frac{d}{2}-\alpha_{2}}
   \nFt'(r_0+p_0)\Bigr|_{\im(p_0) = i\mu - i \eta_p}
   \to\,&
     i \int_{-i \eta_q+i \eta_p}^{i \eta_q + i \eta_p} \frac{{\rm d} r_0}{2 \pi}[r_0^2]^{\frac{d}{2}-\alpha_2}  \delta [\im (r_0) - \eta_p]
   \nn
   +\,&i \int_{-i \eta_q-i \eta_p}^{i \eta_q -i \eta_p} \frac{{\rm d} r_0}{2 \pi}[r_0^2]^{\frac{d}{2}-\alpha_2}  \delta [-\im (r_0) - \eta_p]
   \,,
\end{align}
which shows the proprtionality to the regulator in the form $\propto \eta_p^{d-2\alpha_2}$. Similarly, each term in eq.~\eqref{eq:listofmixedfermions}, through the evaluation of the $r_0$ integrals, contains a factorizing dependence on the regulator powers.} As in the previous example \noveltext{of eq.~\eqref{eq:vanishingboson}}, this indicates that the full integral vanishes in dimensional regularization -- a property that is trivially retained for unit-valued numerators. Should we take the additional regulator $\eta_p$ to zero beforehand, the integral containing $\delta$-sequence expressions \noveltext{in the $T \to 0$ limit} would on the other hand be identical to one with a bosonic $\delta$-sequence (as generated by differentiation).

\subsection{Evaluation of the sunset integral with cutting rules}
\label{app:cuttingrules}

For the sake of completeness, we present next an alternative evaluation of the standard sunset integral at $T = 0$ using the so-called cutting rules of~\cite{Ghisoiu:2016swa}. By setting $\alpha_1 = \alpha_2 = 1$ and $\alpha_3 = s \in \mathbb{N}$ in eq.~\eqref{eq:S:a123:s12}, we are indeed able to take the zero-temperature limit and compute the remaining integral using
the residue theorem~\cite{Ghisoiu:2016swa,Gorda:2022yex}, evaluating multiple integrals at once. We emphasize, however, that raising either of the other two exponent parameters $\alpha_1,\alpha_2$ to values higher than unity would require the introduction of corrections akin to the differentiated distribution functions discussed in sec.~\ref{sec:formalism}.

In the application of the cutting rules, we write the residues as cuts along the fermionic propagators, obtaining thereby the expression
\begin{align}
    \mathcal{S}_{11s} &\stackrel{\noveltext{T \to 0}}{=}
    \int_{\vec{p},\vec{q}} \int_{-\infty }^{\infty}
    \frac{{\rm d}p_0}{(2\pi)}
    \frac{{\rm d}q_0}{(2\pi)}
    \frac{1}{
      [(p_0+i \mu)^2 +p^2]
      [(q_0+i \mu)^2 +q^2]
      [(p_0-q_0)^2 +|\vec{p}-\vec{q}|^2]^s}
    \nn[2mm] &=
      \int_{P,Q} \frac{1}{P^2 Q^2[(P-Q)^2]^{s}}
    - 2 \int_{\vec{p}} \frac{\theta(\mu-p)}{2p} \left[\int_Q  \frac{1}{Q^2 [(Q-P)^2]^{s}} \right]_{p_0 \mapsto i p}
    \nn &\hphantom{{}=\int_{P,Q} \frac{1}{P^2 Q^2[(P-Q)^2]^{s}}}
    + \int_{\vec{p},\vec{q}} \frac{\theta(\mu-p) \theta (\mu-q)}{4 pq} \frac{1}{[(P-Q)^2]^{s}} \biggr|_{{\scriptsize
      \begin{aligned}
        p_0 &\mapsto ip\\[-2mm]
        q_0 &\mapsto iq
      \end{aligned}
      }}\; ,
\end{align}
where we again use capital letters to signify Euclidean $(d+1)$-momenta integrated over $\mathbb{R}^{d+1}$ using dimensional regularization. The first of the three terms on the right-hand side of this expression is said to correspond to a 0-cut, the second to a 1-cut, and the third to a 2-cut contribution to the Feynman integral.

Given that all scaleless integrals vanish in dimensional regularization, the 0-cut contribution on the second line above clearly vanishes, and the same can be seen to be true for the 1-cut term given that the substitution enforces $P^2=0$.  For a general exponent, $s$, the remaining integral and also the full result therefore becomes
\begin{align}
 \int_{\vec{p},\vec{q}}
  \frac{\theta(\mu-p) \theta (\mu-q)}{4pq} \frac{1}{[(P-Q)^2]^{s}} \biggr|_{{\scriptsize
    \begin{aligned}
      p_0 &\mapsto ip\\[-2mm]
      q_0 &\mapsto iq
    \end{aligned}
    }} &=
    \int_{\vec{p},\vec{q}}
    \frac{\theta (\mu-p) \theta (\mu-q)}{2^{2+s} p q \left(p q - \vec{p} \cdot \vec{q} \right)^{s} }
    \nn
    &=
   \biggl(\frac{e^\gammaE \bmu^2}{4\pi} \biggr)^{3-d}
   \frac{1}{2^{2+s}(2\pi)^{2d}}
   \biggl[\frac{2 \pi^\frac{d}{2}}{\Gamma \left( \frac{d}{2} \right)} \biggr]
   \biggl[\frac{2 \pi^\frac{d-1}{2}}{\Gamma \left( \frac{d-1}{2} \right)} \biggr]
   \\ &
   \hphantom{{}=\biggl(\frac{e^\gammaE \bmu^2}{4\pi} \biggr)^{3-d}}
   \times
   \biggl[\int_0^\mu {\rm d}q\, q^{d-2-s}\biggr]^2
   \biggl[\int_{-1}^1 {\rm d}z \frac{(1-z^2)^\frac{d-3}{2}}{(1-z)^s} \biggr]
   \, ,
   \nonumber
\end{align}
where the angular parameter in the dot product reads $z = \vec{p} \cdot \vec{q}/(pq)$. To evaluate the $z$-integral, we next change the integration variable as $z \mapsto y = (1+z)/2$ to obtain a scaled Euler's beta function
\begin{equation}
  \int_{-1}^1 {\rm d}z \frac{(1-z^2)^\frac{d-3}{2}}{(1-z)^{s}} =
  \frac{2^{d-2-s}
  \Gamma\bigl(\frac{d-1-2s}{2} \bigr)
  \Gamma\bigl(\frac{d-1}{2} \bigr)}{
  \Gamma\bigl(d-1-s\bigr)}
  \,,
\end{equation}
again regularized in $d$ dimensions. After finally utilizing  Legendre's duplication formula, we can substitute $2^{d-2-2s}\Gamma \left( \frac{d-1-2s}{2}\right) = \Gamma \left(\frac{1}{2} \right)\Gamma (d-1-2s) / \Gamma \left(\frac{d}{2}-s \right)$
and obtain for the value of the zero-temperature fermionic sunset in $d$ dimensions
\begin{align}
\label{eq:cuttingcheck}
  \mathcal{S}_{11s}
  &\stackrel{\noveltext{T \to 0}}{=} \frac{(d-1)}{(d-1-s)} \frac{\Gamma (s+1) \Gamma(d-2s)}{\Gamma (d-s)} \mathcal{I}_1 (\mu) \mathcal{I}_{s+1} (\mu)
    \, ,
\end{align}
where we applied the one-loop result of eq.~\eqref{eq:fermion1master_new}. It is easy to verify that the full result indeed reproduces our earlier expressions for $\mathcal{S}_{111}$ in eq.~\eqref{eq:ibp2loopexample} and for $\mathcal{S}_{112}$ in eq.~\eqref{eq:S112:T0:fac}.

\subsection{Sunset integral and spatial IBP relations}
\label{app:spatialcheck}

In sec.~\ref{sec:ibp2loopsec}, we derived factorizing low-temperature results for the fermionic sunset integrals $\mathcal{S}_{111}$ and $\mathcal{S}_{112}$ in eqs.~\eqref{eq:ibp2loopexample} and \eqref{eq:S112:T0:fac}, respectively. At nonzero temperature but vanishing chemical potentials, the sunset $\mathcal{S}_{111}$ is on the other hand known to vanish identically, as demonstrated in~\cite{Nishimura:2012ee,Laine:2019uua} using spatial IBP relations and initially argued up to $\mathcal{O}(\epsilon)$ in~\cite{Arnold:1994eb}. Given the non-vanishing nature of our corresponding $T = 0, \mu \neq 0$ results, it is interesting to compare these calculations and attempt to pinpoint the reason for the observed difference.

To gain further insights into the sunset integral, we apply
the diagonal spatial IBP identities from the set~\eqref{eq:ibp:set} using an in-house Laporta-type reduction~\cite{Laporta:2000dsw}, implemented in {\tt FORM}~\cite{Ruijl:2017dtg}. The relations are homogeneous and can be written compactly in operator form
\newtext{(again omitting the two loop master integral $\mathcal{S}_{\alpha_1 \alpha_2 \alpha_3}^{s_1 s_2}$ on which each operator acts)}
as
\setcounter{dummy}{\value{equation}}
\renewcommand{\theequation}{ibp.2}
\begin{subequations}
\setcounter{equation}{2}
\begin{align}
\label{eq:ibp:11:spatial}
  (d-2\alpha_{1} - \alpha_{3})
  &
  + \alpha_{3} {\bf 3_{+}}({\bf 2_{-}} - {\bf 1_{-}})
  \nn &
  +2\alpha_{1} {\bf 1_{+}}{\bf 1^{+}}{\bf 1^{+}}
  +2\alpha_{3} {\bf 3_{+}}({\bf 1^{+}}{\bf 1^{+}} - {\bf 1^{+}}{\bf 2^{+}})
  \equiv
\noveltext{
  \Bigl(\frac{\partial}{\partial p_i}\circ p_i\Bigr)
  \,,
}
  \\
\label{eq:ibp:12:spatial}
  (\alpha_{3} - \alpha_{1})
  + \alpha_{1} {\bf 1_{+}}({\bf 3_{-}} - {\bf 2_{-}})
  &
  + \alpha_{3} {\bf 3_{+}}({\bf 2_{-}} - {\bf 1_{-}})
  \nn &
  + 2\alpha_{1} {\bf 1_{+}}{\bf 1^{+}}{\bf 2^{+}}
  - 2\alpha_{3} {\bf 3_{+}}({\bf 2^{+}}{\bf 2^{+}} - {\bf 1^{+}}{\bf 2^{+}})
  \equiv
\noveltext{
  \Bigl(\frac{\partial}{\partial p_i}\circ q_i\Bigr)
  \,,
}
\end{align}%
\end{subequations}
\renewcommand{\theequation}{\Alph{section}.\arabic{equation}}%
\setcounter{equation}{\value{dummy}}%
\newtext{where the two remaining bilinear combinations can be found via the substitution
${\bf 1_{ }}\leftrightarrow {\bf 2_{ }}$
corresponding to $p_i \leftrightarrow q_i$.}
These relations can be combined to yield
\noveltext{operator ansätze leading to an explicit one-loop factorization of two-loop expressions with collinear temporal scales. The ansatz we choose reads} (cf.~\cite{Laine:2019uua})
\noveltext{
\begin{align}
\label{eq:ibp:2l:spatial}
  0 =
  \biggl\{&
  \Bigl(\frac{\partial}{\partial p_i} \circ p_i\Bigr)
  \Bigl[(d-2s)
    - \mathbf{2}_{+}\bigl(
        2 \cdot \mathbf{1}^{+} \mathbf{2}^{+}
      + \mathbf{3}_{-}
      - \mathbf{1}_{-}
    \bigr)
  \Bigr]
  + \Bigl(\frac{\partial}{\partial p_i} \circ q_i\Bigr)
  \Bigr[
    2 \cdot\mathbf{1}^{+}\mathbf{1}^{+}
    - \mathbf{1}_{-}
  \Bigl]
  \mathbf{2}_{+}
  \nn[1mm]
  +&\Bigl(\frac{\partial}{\partial q_i} \circ p_i\Bigr)
  \Bigl[(d-2s)
    +\mathbf{1}_{+}\bigl(
        2\cdot\mathbf{1}^{+} \mathbf{2}^{+}
      - (s-1) \mathbf{3}_{-}
    \bigr)
  \Bigr]
  - \Bigl(\frac{\partial}{\partial q_i} \circ q_i\Bigr)
  \Bigl[2 \cdot \mathbf{1}^{+} \mathbf{1}^{+}
  \Bigr]
  \mathbf{1}_{+}
  \biggr\}\,
  \mathcal{S}_{11s}
  \,.
\end{align}}
Choosing now $s = 1$ and $s = 2$, this result reduces to the identities%
\footnote{\newtext{The quadratic dependence on the dimension arises from contractions of the form
$\int_{\vec{p}} \frac{\partial}{\partial p_k} p_k = \int_{\vec{p}} d$.}}
\begin{align}
\label{eq:S111:T:cross}
    (d-2)(d-3) \mathcal{S}_{111}&=
    - 2 \Bigl[ \mathcal{I}_{2}^{1}(\mu,T) \Bigr]^2
    \,,\\
\label{eq:S112:T:cross}
    (d-2)(d-5) \mathcal{S}_{112}&=
      \Bigl[ \mathcal{I}_{2}^{ }(\mu,T) \Bigr]^2
    - 2\mathcal{I}_{2}^{ }(\mu,T)\mathcal{I}_{2}^{b}(T)
    \,,
\end{align}
where we dubbed bosonic integrals with the superscript $b$ and
utilized the master integrals of eq.~\eqref{eq:I:alpha:s} and where the fully bosonic integrals $\mathcal{I}^{b}_{\alpha}(T)$ vanish here in the $T\to 0$ limit.

By inspecting eqs.~\eqref{eq:S111:T:cross} and \eqref{eq:S112:T:cross}, we immediately recover the high-temperature fermionic result of~\cite{Laine:2019uua} using the fact that sum-integrals of the form
\begin{equation}
  \oint_{P}^f \frac{p_0^s\,\nFt(p_0)}{[P^2]^\alpha} \equiv
  \Tint{\{P\}} \frac{p_n^s}{\left[p_n^2+p^2 \right]^\alpha}
\end{equation}
identically  vanish for odd values of $s$ at $\mu = 0$. At $T=0, \mu \neq 0$, such integrals, however,
produce nonzero results (see eqs.~\eqref{eq:1loopextra1}--\eqref{eq:demolinear}), which reflects the breaking of the charge conjugation symmetry by nonzero chemical potentials. This highlights the fact that IBP relations derived in vacuum or at $T\neq 0$ but $\mu=0$ cannot be directly applied to nonzero densities.

Lastly, we note that we may use the IBP relation~\eqref{eq:ibp:2l:spatial} \noveltext{(enacted on $\mathcal{S}_{111}$)} as a further cross-check of our results. Indeed, a straightforward calculation utilizing eq.~\eqref{eq:crosspartialmu} produces
\begin{align}
  \mathcal{S}_{111} &=
  -\frac{2}{(d-2)(d-3)} \left[\oint_{P}^f \frac{p_0\nFt (p_0)}{P^4} \right]^2
  \nn &\newtext{\stackrel{T \rightarrow 0}{=}}
    -\frac{2}{(d-2)(d-3)} \left[\frac{i}{2} \partial_\mu \mathcal{I}_1 (\mu) \right]^2
    =
     \frac{(d-1)}{(d-2)^2}
      \mathcal{I}_{1}^{ }(\mu)
      \mathcal{I}_{2}^{ }(\mu)
      \,,
\end{align}
where the final line can be seen to fully agree with our earlier results given in
eqs.~\eqref{eq:ibp2loopexample} and \eqref{eq:cuttingcheck}.

%
\section{Two-loop factorization at finite density}
\label{sec:factorizationof2loop}

\newtext{In sec. \ref{sec:2-loop}, we encountered the partially unexpected factorization of two-loop thermal vacuum integrals into a sum of products of one-loop integrals, the immediate consequences of which we will now investigate further. Such a property has been known to hold for $d$-dimensional $T=\mu=0$ sunset integrals with a collinear mass signature~\cite{Davydychev:2022dcw,Davydychev:1992mt},
\begin{align}
  S_{\alpha_1 \alpha_2 \alpha_3} (m_1,m_2,m_3) &= \sum_{i<j} c_{ij} m_i^{f_{ij}}m_j^{g_{ij}}
  \,,
  & \text{for collinear masses}\quad
  m_3&=  m_1+m_2
\,.
\end{align}
The coefficients, $c_{ij}$, and exponents $\{f_{ij}, g_{ij}\}$ have recently been determined in a closed form for integer-valued exponents $\alpha_n \in \mathbb{N}$ in~\cite{Davydychev:2022dcw} and agree with values previously suggested by recursion formulae such as two-loop vacuum IBP reduction. In our present context, it is a nontrivial and very interesting question, to which extent such a factorization extends to the description of the thermal sunset $\mathcal{S}_{\alpha_1 \alpha_2 \alpha_3}(\mu,T)$ in terms of products of the one-loop integrals
$\mathcal{I}_\alpha (\mu, T)$. In the following, we present such a generalization using the formalism of
secs.~\ref{sec:formalism} and \ref{sec:IBPintro}.}

\newtext{For integer-valued exponents of $\mathcal{S}_{\alpha_1\alpha_2\alpha_3}$ in eq.~\eqref{eq:S:a123:s12}, the corresponding vacuum sunset~\eqref{eq:S:m123} features exactly the factorizing collinear mass signature of~\cite{Davydychev:2022dcw}. To apply such a formula in the context of complex scales and for $q_0 > p_0 >0$ (cf.\ eqs.~\eqref{eq:I:m} and \eqref{eq:S:m123}), we want to recast their results in a form where no temporal momenta appear in the denominator. This can be achieved through identities such as
\begin{align}
\label{eq:factorizationsunsetspat0}
    S_{111}(p_0, q_0-p_0, q_0)
    &= 
    \frac{2}{(d-3)(d-2)} \Bigl[
      -p_0 q_0 I_2(p_0) I_2 (q_0)
      -q_0 (q_0-p_0)I_2(q_0-p_0) I_2 (q_0)
    \nn &\hphantom{{}= \frac{2}{(d-3)(d-2)} \biggl[}
      +p_0 (q_0-p_0)I_2(p_0) I_2 (q_0-p_0)
    \Bigl]
    \;,\\
\label{eq:S112:fac:spat:T0}
  S_{121}(p_0, q_0-p_0, q_0) &=
   \frac{1}{(d-5)(d-2)}
   \Bigl[ I_2(p_0) I_2(q_0)- I_2(p_0)I_2(q_0-p_0)- I_2(q_0)I_2(q_0-p_0) \Bigr]
   \nn&
   - \frac{4}{(d-5)(d-2)}
    \Bigl[(q_0-p_0) I_3 (q_0-p_0) \Bigr]
    \Bigl[q_0 I_2 (q_0) - p_0 I_2 (p_0) \Bigr]
    \,,
\end{align}
which can be derived using the $d$-dimensional one-loop massive vacuum IBP relation $I_{\alpha+1}(m) = -(d-2\alpha)/(2\alpha m^2) I_{\alpha}(m)$ on
eq.~\eqref{eq:S111:m12} for $S_{111}$ and
eq.~\eqref{eq:S121:m12} for $S_{121}$.}

\newtext{Starting from the factorization observed in eq.~\eqref{eq:factorizationsunsetspat0}, we can analytically continue such expressions to complex values following the work of~\cite{Gorda:2022yex}. In particular, we assume that the linear numerator terms can be analytically continued to complex values, keeping the even/odd property of $\mathcal{S}_{\alpha_1\alpha_2\alpha_3}^{s_1 s_2}$ for general values of $s_1,s_2$. To demonstrate the validity of this assumption, we can compute the $T \to 0$ limit of the sunset integrals $\mathcal{S}_{111}$ and $\mathcal{S}_{112}$ using eq.~\eqref{eq:factorizationsunsetspat0}, obtaining
\begin{align}
\label{eq:S111:fac:I1I2}
  \mathcal{S}_{111} &=
   -\frac{2}{(d-3)(d-2)}
   \Bigl[ \mathcal{I}_{2}^{1}(\mu,T)\Bigr]^2
    \,\stackrel{T \to 0}{=}\,
     \frac{(d-1)}{(d-2)^2}
      \mathcal{I}_{1}^{ }(\mu)
      \mathcal{I}_{2}^{ }(\mu)
    \,,\\
  \mathcal{S}_{112} &=
   +\frac{1}{(d-5)(d-2)} \Bigl(
   \Bigl[ \mathcal{I}_{2}^{ }(\mu,T)\Bigr]^2-2 \mathcal{I}_2 (\mu, T) \mathcal{I}_2^b (T) \Bigr)
    \,\stackrel{T \to 0}{=}\,
    \frac{1}{(d-5)(d-2)}
   \Bigl[ \mathcal{I}_{2}^{ }(\mu)\Bigr]^2 
   \,.
\end{align}
Given the antisymmetry between $p_0 \leftrightarrow q_0$, all terms of the underlying spatial integrand~\eqref{eq:factorizationsunsetspat0} and the lower row of eq.~\eqref{eq:S112:fac:spat:T0}, that contain odd bosonic one-loop integrals, can be seen to vanish.
The low-temperature limit $\lim_{T\to 0} \mathcal{I}_{2}^{1}(\mu,T) = \mathcal{I}_{2}^{1}(\mu)$ can furthermore be evaluated using eq.~\eqref{eq:S:a12:s12:odd:ibp} or directly as in eq.~\eqref{eq:I:alpha:s}, while the remaining $\lim_{T\to 0} \mathcal{I}_{\alpha}^{ }(\mu,T) = \mathcal{I}_{\alpha}^{ }(\mu)$ follows from eq.~\eqref{eq:fermion1master_new}. Both relations were of course already confirmed using only IBP relations in eqs.~\eqref{eq:ibp2loopexample} and \eqref{eq:S112:T}, respectively.}

\newtext{Finally, we note that due to the factorization (via collinear scales) of the $d$-dimensional vacuum sunset $S_{\alpha_1\alpha_2\alpha_3}$ for $\alpha_{1},...\alpha_{3} \in \mathbb{N}$~\cite{Davydychev:2022dcw}, it is in fact all thermal two-loop bubbles  $\mathcal{S}_{\alpha_1\alpha_2\alpha_3}$ and not just the above special cases that factorize, and this appears to happen for arbitrary values of  $T$ and $\mu$. Specific examples of this generic result include the low-temperature regime of $T\to 0$ but $\mu \neq 0$, where the factorization emerges explicitly as observed in sec.~\ref{sec:maxprime}, as well as the high-temperature regime $T \neq 0$ but $\mu = 0$, where the factorization was demonstrated already in~\cite{Ghisoiu:2012yk,Nishimura:2012ee} for a few special cases. The factorization  becomes particularly transparent in the IBP approach, where it is automatically built in the relations~\eqref{eq:ibp:11}--\eqref{eq:ibp:12:spatial}; see sec.~\ref{sec:ibp2loopsec} and the appendix \ref{sec:vacuum:2l}. We emphasize, however, that it is not a universal property of multi-loop thermal integrals nor do we expect it to generalize to higher loop orders or massive propagators. Even in these cases, the maximally primed type~\ref{typeB} integrals and the corresponding $\delta$-functions nevertheless remain tractable at low temperatures, thus simplifying the corresponding computations.}

\newtext{One example of possible extensions of the above results is the two-loop fermionic sunset with equal positive mass scales $m^2$ and chemical potential $\mu$ in the fermionic propagators,
\begin{equation}
\label{eq:S:a123:m}
  \mathcal{S}_{\alpha_1\alpha_2\alpha_3}(\mu,m_f,T) =
    \oint_{P,Q}^f
    \frac{\nFt(p_0)}{[p_0^2+p^2+\noveltext{m_f}^2]^{\alpha_1}}
    \frac{\nFt(q_0)}{[q_0^2+q^2+\noveltext{m_f}^2]^{\alpha_2}}
    \frac{1}{[(p_0-q_0)^2 + |\vec{p}-\vec{q}|^2]^{\alpha_3}}
    \,.
\end{equation}
Evaluating this integral directly using factorization is not possible, since the $d$-dimensional integral is no longer collinear. Its maximally primed expression, however, still has this property and can thus be evaluated directly using Feynman parametrization and the corresponding hypergeometric integral~\eqref{eq:final} or
using the factorization observed in~\cite{Davydychev:2022dcw}. The result can be given readily for $\mu > m$ in the form
\begin{align}
\label{eq:2loopt0mass}
    {\bf D}_{p}
    {\bf D}_{q}\,
    \mathcal{S}_{111}(\mu,m,T)
    \stackrel{T\to 0}{=}
    - \frac{2(d-3)}{(d-2)} \left[ \mathcal{I}_2 \left(\sqrt{\mu^2-\noveltext{m_f}^2} \right) \right]^2
    \,,
\end{align}
while for $m > \mu$ it is observed to vanish; see appendices \ref{sec:massl} and \ref{sec:vacuum:2l} for more context and details. Such differentiated integrals are again seen to appear as a part of $(d+1)$-dimensional IBP reduction akin to eq.~\eqref{eq:2loopibp2first} or \eqref{eq:2loopibp:ops}, thus contributing to the determination of the full massive thermal integral.}

{\small
%
\bibliographystyle{utphys}
\bibliography{ref}
}
\end{document}